\documentclass[preprint,12pt,authoryear]{elsarticle}
\usepackage[margin=2.5cm]{geometry}
\usepackage[utf8]{inputenc}
\usepackage{xcolor}

\usepackage[linesnumbered,ruled]{algorithm2e}
\usepackage{natbib}
\usepackage{bm}
\usepackage{graphicx}
\usepackage{verbatim}
\usepackage{soul}

\usepackage{appendix}
\usepackage{subcaption}
\usepackage{setspace}
\usepackage{booktabs}
\usepackage{amsmath}
\numberwithin{equation}{section}
\usepackage[makeroom]{cancel}
\newcommand{\stkout}[1]{\ifmmode\text{\sout{\ensuremath{#1}}}\else\sout{#1}\fi}
\usepackage{amssymb}
\usepackage[colorlinks]{hyperref} 
\hypersetup{ 
    colorlinks=true,       
    linkcolor=red,          
    citecolor=blue,        
    filecolor=magenta,      
    urlcolor=cyan           
}

\usepackage{tikz}
\usepackage{stmaryrd}

\newcommand{\jump}[1]{\llbracket #1 \rrbracket}
\newcommand{\bphi}{\bar \eta}

\newcommand{\fref}[1]{Fig.~\ref{#1}}
\newcommand{\tref}[1]{Tab.~\ref{#1}}

\newcommand{\frefs}[2]{Figs.~\ref{#1}--\ref{#2}}
\newcommand{\ftref}[1]{footnote~\ref{#1}}
\newcommand{\eref}[1]{(\ref{#1})}
\newcommand{\erefs}[2]{(\ref{#1})--(\ref{#2})}
\newcommand{\sref}[1]{Section~\ref{#1}}

\newcommand{\blue}[1]{\color{blue}}{}

\journal{Journal of Computational Materials Science}

\begin{document}

\begin{frontmatter}

\title{A crystal symmetry-invariant Kobayashi--Warren--Carter grain boundary
model and its implementation using a thresholding algorithm}

\author[mechseaddress]{Jaekwang Kim}
\author[mathaddress]{Matt Jacobs}
\author[mathaddress]{Stanley Osher}
\author[mechseaddress]{Nikhil Chandra Admal\corref{mycorrespondingauthor}}
\cortext[mycorrespondingauthor]{Corresponding author}
\ead{admal@illinois.edu}
\address[mechseaddress]{Department of Mechanical Science and Engineering, 
                        University of Illinois Urbana--Champaign}
\address[mathaddress]{Department of Mathematics, University of California Los Angeles}

\begin{abstract}
One of the most important aims of grain boundary modeling 
is to predict the evolution of a large collection of grains in phenomena
such as abnormal grain growth, coupled grain boundary motion, and
recrystallization that occur under extreme thermomechanical loads.
A unified framework to study the coevolution of grain boundaries 
with bulk plasticity has recently been developed by \citet{Admal:2018},
which is based on modeling grain boundaries as continuum dislocations governed
by an energy based on the Kobayashi--Warren--Carter (KWC) model \citep{KWC:1998,KWC:2000}. 
While the resulting unified model demonstrates coupled grain boundary motion 
and polygonization (seen in recrystallization), it is restricted to grain
boundary energies of the Read--Shockley type, which applies only to small
misorientation angles. In addition, the implementation of the unified model using finite
elements inherits the computational challenges of the KWC model 
that originate from the singular diffusive nature of its governing equations.  
The main goal of this study is to generalize the KWC functional to grain
boundary energies beyond the Read—Shockley-type that respect the
bicrystallography of grain boundaries. The computational challenges of the
KWC model are addressed by developing a thresholding method that relies on a primal dual algorithm and
the fast marching method, resulting in an $\mathcal{O}(N \log N)$ algorithm, where $N$ is the number of grid points.
We validate the model by demonstrating the Herring angle and the von Neumann--Mullins relations, 
followed by a study of the grain microstructure evolution in a two-dimensional face-centered cubic copper polycrystal with crystal symmetry-invariant grain boundary 
energy data obtained from the covariance grain boundary model of \citet{Runnels:2016_1,Runnels:2016_2}.
\end{abstract}
\begin{keyword}
    A. phase field model \sep microstructures \sep
    B. motion by curvature \sep constitutive behavior \sep polycrystalline
    materials
\end{keyword}
\end{frontmatter}

\newpage
\section{Introduction}
Most metals and ceramics exist as polycrystals, which are aggregates of single
crystal grains stacked together along grain boundaries. The microstructure of a polycrystal is often
characterized by the orientation distribution of its grains commonly referred to
as texture. The macroscopic properties of polycrystals, which
include yield strength, resistance to creep, fatigue, thermal and magnetic
properties, are strongly influenced by texture. 
Grain boundary engineering refers to the strategy of enhancing
the properties of polycrystalline materials by transforming the 
grain boundary character distribution 
to a desired state using thermomechanical processes \citep{Watanabe:2011}.
Mapping the microstructure-property relationship, and modeling the evolution of
microstructure under various manufacturing processes are fundamental open
problems relevant to grain boundary engineering.

The evolution of grain boundaries is driven by a long list of thermodynamic
forces, of which surface tension plays a central role. For instance, grain boundaries in the
isotropic Mullins' model \citep{Mullins} are driven by their excess surface
energy resulting in motion by curvature with velocity given by
\begin{equation}
v = -m \gamma \kappa,
\label{eqn:motion_by_curvature}
\end{equation}
where $\kappa$ and $\gamma$ denote curvature and 
misorientation-dependent energy density of the grain boundary respectively, while $m$
represents a constant mobility. More
generally, an anisotropic grain boundary evolution arises from
the dependence of $m$ and $\gamma$ on the grain boundary character defined by
the five macroscopic degrees of freedom, which
represent the misorientation and the inclination of the grain
boundary.\footnote{Under an anisotropic energy density $\gamma$ that depends on
    the inclination $n$ of a grain boundary, \eref{eqn:motion_by_curvature}
transforms as $v=-m\kappa(\gamma+ \partial^2 \gamma/\partial n^2)$.}
Recent advances in the development of accurate interatomic
potentials have enabled us to build and refine the grain boundary energy and
mobility landscapes as functions of the grain boundary character 
\citep{Srolovitz:2020_2, Runnels:2016_1, Runnels:2016_2, Olmsted:2009,Bulatov:2013}.
The grain boundary energy landscape reflects the symmetry of a bicrystal,
and understanding the role of crystal symmetry 
contributes enormously towards characterizing grain microstructure
evolution. Moreover, precisely identifying
the grain boundary character distribution responsible for
phenomena such as abnormal grain growth and recrystallization remains an open
problem in materials science. This motivates us to undertake
simulations of large ensembles of appropriately sampled polycrystals
to discover lower-order statistical models for grain microstructure evolution.
\emph{The goal of this paper is to develop a lightweight model for motion by
curvature in the presence of a misorientation-dependent grain boundary energy
density, that can be implemented using an ultrafast algorithm.}

While motion by curvature is the simplest description of grain boundary evolution,
experiments \citep{Rollett:2018,Barmak:2013} and atomistic simulations \citep{Srolovitz:1998,Foiles:2006} demonstrate that surface tension alone
is not the dominant force. Molecular dynamics (MD) simulations have revealed that as grain
boundaries evolve, they plastically deform the underlying material resulting in
lattice distortions that give rise to additional forces on the grain boundary.  
In order to include the effect of grain boundary plasticity, recent mesoscale
models \citep{Admal:2018} have focused on the coevolution grain boundaries and deformation.
Evidently, such models subsume motion by curvature as a special case, and are
computationally more expensive, which is another motivation for us to seek an
ultrafast algorithm. While the focus of this paper is motion by curvature, 
we ensure that our model is amenable to generalizations that include grain boundary plasticity.

Models for grain microstructure evolution can be broadly classified into three
categories: probabilistic, diffuse-interface, and sharp-interface models.
An example of a probabilistic grain growth model
is the Monte-Carlo Potts model \citep{Monte1,Monte2,Monte3,Monte4,Monte5},
wherein a polycrystal is described using points in a lattice, which are allocated
to different grains. A grain boundary is implicitly defined by adjacent
lattice points that belong to different grains. 
Evolution of the microstructure is carried out stochastically 
through random jumps of boundaries in thermodynamically favorable directions. 
While the advantage of the Mote-Carlo Potts model lies in the simplicity of its
implementation, it relies on heuristic rules that do not have a thermodynamic basis. 

In sharp interface models \citep{Mullins,Hillert,AllenCahn}, grain boundaries
are modeled as surfaces that evolve according to motion by curvature given
in \eref{eqn:motion_by_curvature}. Methods to implement
\eref{eqn:motion_by_curvature} rely on either implicitly or explicitly tracking
the moving grain boundaries. For instance, front tracking methods
\citep{FrontTrack1,FrontTrack2,Kinderlehrer2004,Kinderlehrer2006} describe
grain boundaries in two dimensions as line segments along with their connectivity. 
Such a description breaks down at critical events including disappearance of
shrinking grains and topological changes due to merging of grain boundaries.
Therefore, front tracking methods are supplied with additional rules to redefine
the connectivity of line segments to describe such critical events. 
The level set method \citep{LevelSet1,LevelSet2} addresses the above limitation
using an implicit representation. Each grain is described by a function that
is positive within, and negative outside the grain, which implies the zero-valued
isosurface describes the interface surrounding the grain.
An implicit representation can describe topological changes without any additional rules. 
Yet, the main disadvantage of the level set method is that it does not extend to handle 
surfaces with self intersection and junctions, 
which occur in polycrystals. 
More recently, a thresholding method commonly referred to as the
Merriman--Bence--Osher (MBO) \citep{MBO} scheme and its generalization
by \cite{Esedoglu:2015} have been shown to simulate grain kinetics very
efficiently in addition to its ability to predict grain nucleation. The level
set and the MBO methods are memory intensive as they use as many functions as the
number of grains in order to describe a polycrystal. For example, a description of a 3D
polycrystal with $10,000$ grains on a 256$\times$256$\times$256 grid requires
$\sim 1$ TB of memory. 
Thus, for a large scale simulation, 
an additional numerical technique is devised to employ a level set function 
that approximates a large subset of spatially separated grains
\citep{Esedoglu:2009,Esedoglu:2011}. The thresholding method of
\citet{Esedoglu:2015}, which is
first-order accurate in time, has recently been extended by \citet{ZAITZEFF2020109404} 
to a second-order method that is unconditionally energy stable.
In general, all sharp-interface models can be
incorporated with misorientation dependent grain boundary energy densities and mobilities. 
However, including inclination dependence is more challenging, and recent works by 
\citet{Gupta:2014,Bulatov:2019,Gupta:2020} have addressed this challenge.

In diffuse interface models \citep{DiffuseInterface}, 
a polycrystal is defined using functions called phase fields, which are constant in the interior of the grains. 
The regions where the gradients of phase fields are non-zero 
are identified as diffused grain boundaries, which have a characteristic width.
A numerical implementation of a diffuse
interface model requires a grid that is refined enough to resolve the width of
the grain boundary. Therefore, diffuse interface models are computationally more
expensive than their sharp-interface counterparts such as the level set and MBO
methods. The \textit{multi phase field} (MPF) model \citep{MF1,MF2,MF3}, and the
Kobayashi--Warren--Carter (KWC) model \citep{KWC:1998,KWC:2000,KWC:2003} are two
examples of diffuse-interface models for grain boundaries.

The main advantage of the MPF model lies in the simplicity of its construction to include
misorientation dependent grain boundary energies and mobilities. 
Similar to the level set and the MBO methods, 
a naive implementation of the MPF method would use as many phase fields as the 
number of grains, resulting in an excessive use of computational memory.
Since, at any point in the domain, only a few order parameters would be
non-zero, recent implementations \citep{MPFRemap2,MPFRemap} of the MPF model
allow multiple grains which do not share a common boundary to share the same
order parameter. A \emph{grain remapping} algorithm is used to strategically
remap an order parameter shared by two distant grains when they approach close to
each other. Recent advances \citep{Runnels:2019,Wollants,BJLEE:2014} in MPF models 
explore the full anisotropy of grain
boundary energy which consists of misorientation and inclination
dependence.\footnote{Grain boundary energy as a function of inclination is
    typically non-convex. For grain boundary models that incorporate inclination
    dependence to be well-posed, they must include curvature-dependent energy
    densities resulting in a higher-order model which adds to the
computationally intensive nature of the MPF model.} 

In contrast, the KWC model describes a two-dimensional polycrystal using
only two order parameters - one for structural order 
$\eta$ ranging from 0 (disordered phase) to 1 (crystalline state), and the other
for crystal orientation field $\theta$. The elegance of the KWC model is offset
by the severe restriction it imposes on the grain boundary energy. The energy
functional of the KWC model limits the dependence of grain boundary energy on
misorientation angle to a Read--Shockley-type \citep{Read} that does not
respect the crystal symmetry. In addition, the singular diffusive nature
of the KWC model results in stiff equations that are computationally
expensive to solve.

Recognizing the elegance of the KWC model, we formulate a
generalization of the KWC model that can incorporate arbitrary
misorientation-dependent grain boundary energies. In addition, we design a
thresholding method that addresses the challenge of solving the singular diffusive equation
of the KWC model. The resulting model inherits the memory efficiency of the
original KWC model, while having significantly more computational efficiency 
compared to conventional numerical methods such as finite element and finite
difference.

The paper is structured as follows.
In \sref{sec:KWCModel}, we discuss the role of crystal symmetry on the
anisotropy of grain boundary energies, and review the original KWC model and its
limitations.
Then, we propose a generalization of the KWC model to 
incorporate grain boundary energies beyond the Read--Shockley type.
In \sref{sec:thresholding_method}, we design a thresholding algorithm to
implement the generalized KWC model.
Numerical experiments to validate and demonstrate the computational efficiency
of the thresholding method are discussed in \sref{sec:numeric_experiment}.
Finally, we summarize and conclude with a description of future directions.


\section{Grain boundary energy and the Kobayashi--Warren--Carter model}
\label{sec:KWCModel}
\begin{figure}[t]
\begin{center}
\includegraphics[width=0.7\textwidth]{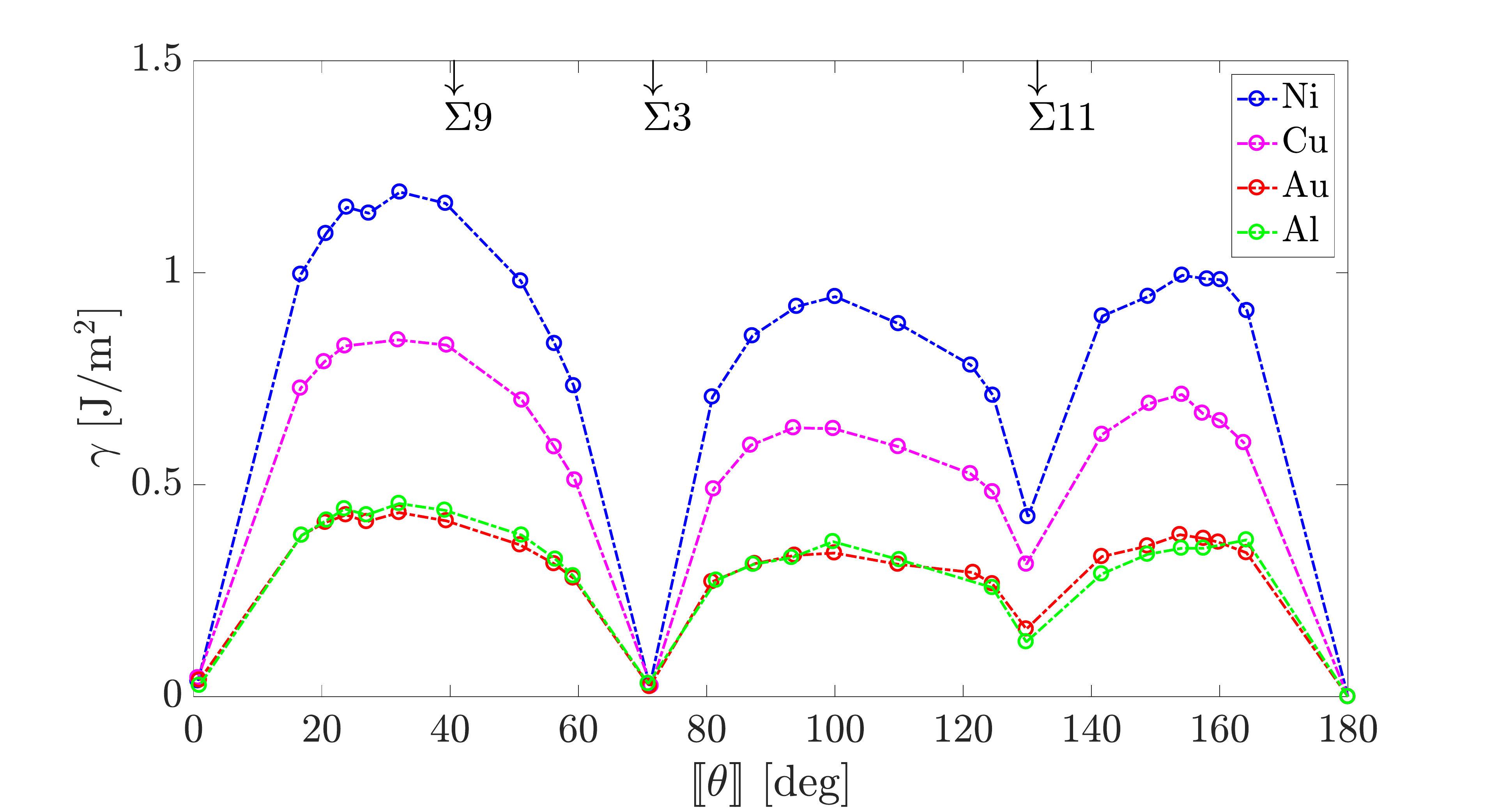}
\end{center}
\caption{A plot of grain boundary energy density as a function of
    misorientation angle of a $[110]$ symmetric-tilt grain boundary in fcc
    copper, computed
    using molecular dynamics \cite{Holm:2010,Bulatov:2014}. Misorientations
corresponding to low energy $\Sigma$ boundaries are marked on the upper axis.}
\label{fig:MDdataset}
\end{figure}

A grain boundary is characterized by five macroscopic degrees of freedom where three degrees
represent a rotation associated with the misorientation between the two
grains, and the remaining two degrees correspond to the inclination of the grain
boundary. More precisely, the grain boundary character space is
given by the topological space $\mathcal T = SO(3)\times SO(3)/SO(2)$, where
$SO(n)$ is the special orthogonal group in $n$ dimensions. Grain
boundaries are equipped with a surface energy density, which is defined as a
function on $\mathcal T$. An energy density that is constant is referred to as
an \emph{isotropic}, and \emph{anisotropic} otherwise. \fref{fig:MDdataset} shows a plot of
grain boundary energy density as a function of misorientation angle for a
$[110]$ symmetric-tilt grain boundary in face-centered cubic (fcc) copper, calculated
using molecular dynamics simulations \citep{Holm:2010,Bulatov:2014}.
Since the symmetry of an fcc lattice ensures that the energy of a
$[110]$ symmetric-tilt grain boundary is symmetric about the $180^{\circ}$
misorientation angle, \fref{fig:MDdataset} shows a plot of energy vs
misorientation angles up to $180^\circ$.
In addition, $\gamma$
exhibits local minima at certain misorientations, marked as $\Sigma_3$ and
$\Sigma_{11}$ in \fref{fig:MDdataset}, due to an enhanced lattice matching
\citep{Runnels:2016_1,Runnels:2016_2,Wolf:1990} between the two adjoining grains. 
Recent efforts \citep{mason2019basis,Bulatov:2013,Runnels:2016_1,Runnels:2016_2,Olmsted:2009,Kim:2014} by
materials scientists in characterizing the grain boundary character space, and parametrizing
grain boundary energy using data from atomistic simulations and experiments,
bring us closer to developing an atomistically informed mesoscale model for
grain boundaries.

The motion of grain boundaries driven by surface tension to decrease the
interfacial energy is a defining characteristic of various grain microstructure models such as the
Mullins model \citep{Mullins}, and its diffuse-interface counterparts such as the
multiphase field and the KWC models. The resulting grain boundary motion, commonly referred to as \emph{motion by curvature}, in the presence of an anisotropic energy density has been shown to have a considerable effect on grain statistics \citep{Barmak:2013}
leading to changes in the macroscopic properties of materials. 
In order to explore the structure-property relationship, recent research efforts have
focused on developing ultrafast algorithms to simulate grain boundary evolution in
large polycrystals in the presence of atomistically derived anisotropic grain
boundary energies. 


In this paper, we recognize the simplicity of the phase
field model of Kobayashi, Warren and Carter, and show that it can be generalized
and implemented using a thresholding method resulting in an ultrafast algorithm for grain boundary
evolution.

\subsection{The KWC model}
\label{sec:kwc}

\begin{figure}[t]
     \centering
     \begin{subfigure}[b]{0.45\textwidth}
         \centering
         \includegraphics[width=\textwidth]{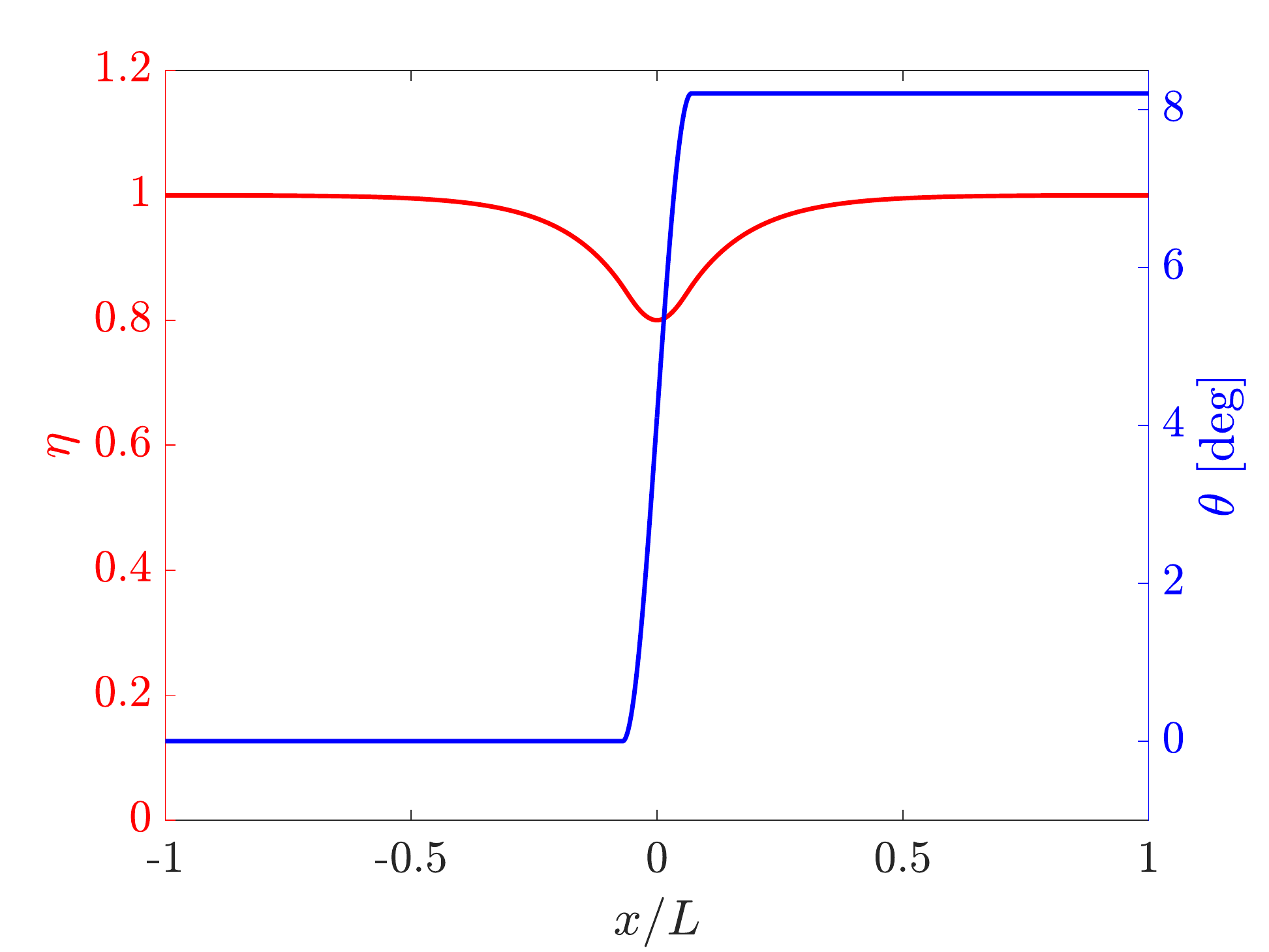}
         \caption{}
         \label{fig:analytic_order}
     \end{subfigure}
     \hfill
     \begin{subfigure}[b]{0.45\textwidth}
         \centering
         \includegraphics[width=\textwidth]{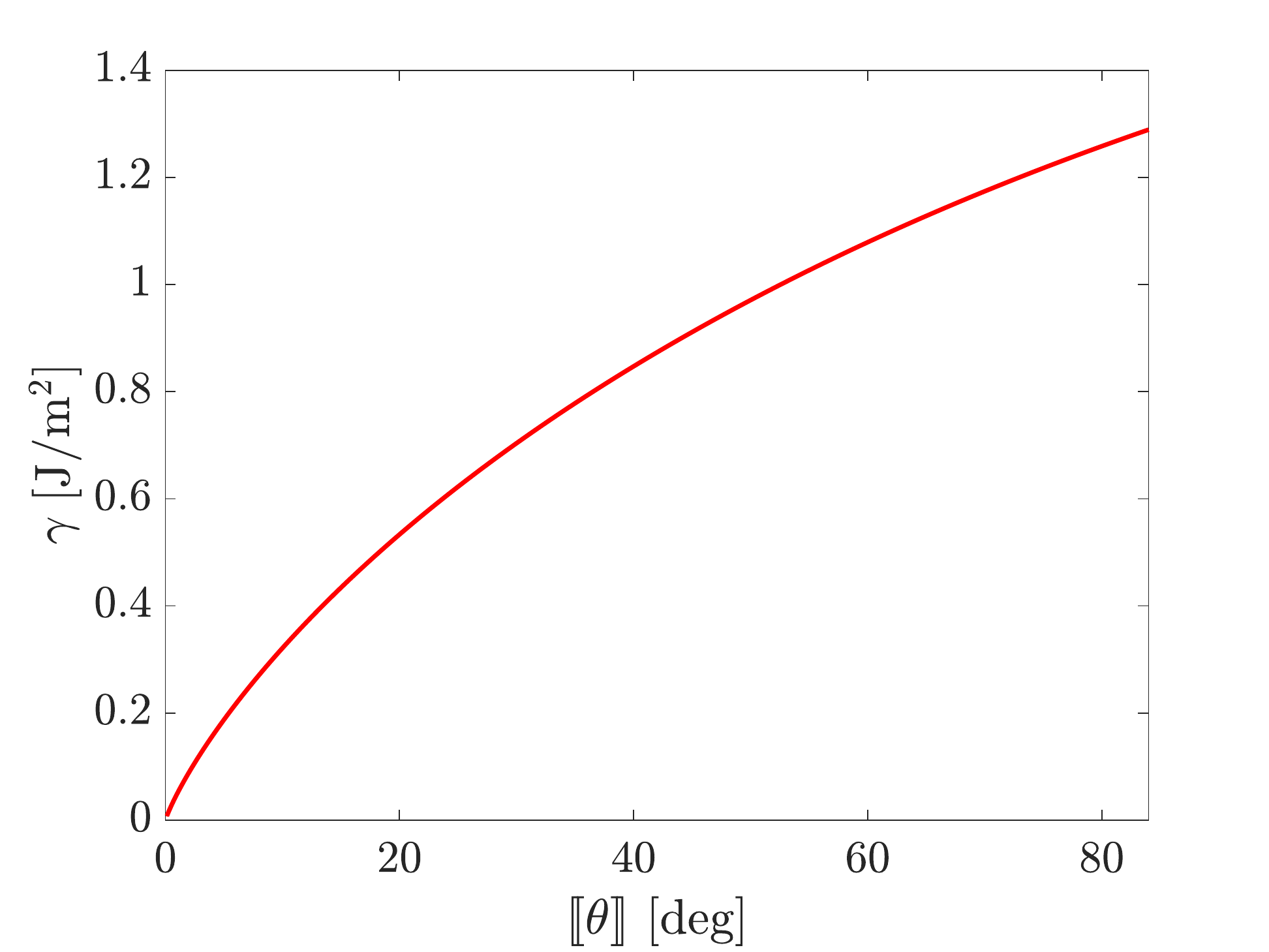}
         \caption{}
         \label{fig:analytic_energy}
     \end{subfigure}
     \caption{Results on the one-dimensional steady state solution of the original KWC model
         describing a flat grain boundary.
         \protect\subref{fig:analytic_order}) A 
         steady state analytical solution of the KWC model for a given
         misorientation. \protect\subref{fig:analytic_energy}) Variation of the
         grain boundary energy as a function of misorientation in the KWC model.
     }
     \label{fig:KWC1D_analytic}
\end{figure}

The Kobayashi--Warren--Carter (KWC) model \citep{KWC:1998,KWC:2000,KWC:2003}
is a dual-phase field model to study grain evolution in polycrystalline materials. 
In this model, an arbitrary polycrystal in two dimensions is described using
only two order parameters $\eta$ and $\theta$. This is one of the main
advantages of employing the KWC model as opposed to the multiphase field model,
which uses as many order parameter as the number of grains.
The order parameter $\eta$ ranges from $0$,
which signifies disorder, to $1$ that describes crystalline order. On other
hand, the order parameter $\theta$ describes the orientation of the
grains.\footnote{In three dimensions, the order parameter $\theta$ is replaced by a rotation
tensor.}

The KWC model is governed by a free energy functional given by
\begin{equation}
\mathcal W[\eta,\theta] = \int_{\Omega}
\left[ \frac{f(\eta)}{\epsilon} + \frac{\epsilon}{2}|\nabla \eta|^2
+ g(\eta) |\nabla \theta| + \frac{\epsilon}{2}  |\nabla \theta|^2
\right]\; dV, 
\label{eqn:KWC_energy}
\end{equation}
where
\begin{equation}
    f=\frac{(1-\eta)^2}{2}
\end{equation}
is a single-well potential with minimum at $\eta=1$, and 
\begin{equation}
g=-\ln(1-\eta)
\label{eqn:g_form}
\end{equation}
is an increasing function.\footnote{
The choice of the logarithmic function for $g$ is supported by the
work of \citet{Alicandro}, which shows that 
the KWC functional converges (in the sense of $\Gamma$-convergence) 
to a surface energy function when $g(1)=\infty$ (See Theorem 4.1 in \cite{Alicandro}).
} The functional in \eref{eqn:KWC_energy} is defined for all functions $\eta$ and
$\theta$ in the Hilbert
space $H^1(\Omega)$.\footnote{$H^1(\Omega)$ denotes the set of all functions on
    $\Omega$ whose first derivatives are square integrable.} The constant $\epsilon>0$ is a dimensionless scaling parameter that 
determines the thickness of the grain boundary region \citep{KWC:2001}. The
evolution equations for the order parameters, assuming a gradient descent (with
respect to the $L^2$-norm) of $\mathcal W$,
are obtained as
\begin{subequations}
    \begin{align}
        \epsilon b_{\eta}\dot{\eta}
        &=   \epsilon \Delta \eta - \frac{f'(\eta)}{\epsilon}
        - g'(\eta) |\nabla \theta|,
        \label{eqn:phi_evolution_orig}\\
        \epsilon b_{\theta} \dot{\theta}
        &= \nabla \cdot \left[ 
                \epsilon \nabla \theta + 
                 g(\eta) \frac{\nabla \theta}{|\nabla \theta|}
            \right],
        \label{eqn:theta_evolution_orig}
    \end{align}
    \label{eqn:evolution}
\end{subequations}
where $b_{\eta}$ and $b_{\theta}$ are the inverse mobilities corresponding to
respective order parameters.\footnote{The KWC model was originally developed to
    simultaneously model grain rotation \emph{and} grain boundary motion. The model can be specialized to demonstrate \emph{only} grain boundary motion by enforcing zero mobility
for $\theta$ in the grain interior. This can be achieved by a
constant $b^\phi$, and a $\phi$-dependent $b^\theta$ \citep{Dorr:2010}.}

A one-dimensional steady state solution of
\eref{eqn:evolution} under Dirichlet boundary conditions 
is plotted in \fref{fig:analytic_order}. 
The value of $\eta<1$ in a neighborhood of the grain boundary
suggests a loss of crystalline order, and the orientation $\theta$
is constant in the interior of the grains, and has a non-zero gradient in a
finite thickness around the grain
boundary. In the limit $\epsilon\to 0$, \citet{KWC:2001} have shown that the
evolution equations in \eref{eqn:evolution} result in the shrinking of the grain
boundary thickness converging to the Mullins model.

Below, we briefly summarize the role of each term appearing in the KWC
functional. We refer the reader to \citet{KWC:2003} for a more
detailed description, and \citet{Admal:2019} for a generalization of the KWC
model to three dimensions. 
The function $f$ drives $\eta(\bm x)$ towards $1$, while
the coupled term $ g(\eta)|\nabla \theta|$ tends to decrease $\eta$ in a
neighborhood of the grain boundary. In addition, the coupled term
tends to localize the jump in $\theta$, while
$|\nabla \theta|^2$ has a tendency to diffuse it, resulting in a 
regularized step function for $\theta$.
It is interesting to note that in the absence of the $|\nabla\theta^2|$ term,
the steady state solution for $\theta$ is a pure step function,
resulting in a model with a blend of sharp- and diffuse-interface
characteristics, i.e. while $\theta$ is sharp, $\eta$ is diffused. Moreover,
grain boundaries cease to evolve in the absence of
$|\nabla\theta|^2$ term \citep{KWC:2001}. In other words, $|\nabla\theta|^2$ in the KWC model has
a dual role of not only regularizing $\theta$ but also rendering non-zero
mobility to the grain boundaries.

The grain boundary energy $\gamma$, as a function of misorientation
angle, predicted by the
KWC model is of the Read--Shockley-type, as shown in
\fref{fig:analytic_energy}.
This is in contrast to the experimentally observed grain
boundary energies shown in \fref{fig:MDdataset}. Despite the elegance of the KWC
model in describing polycrystals with only two order parameters, its restriction 
to Read--Shockley type grain boundary energies is a major limitation compared to
the flexibility of incorporating arbitrary grain boundary energies into the
multiphase field model. The above-stated limitation is one of the main motivation
for us to seek a new formulation of the KWC model to incorporate
arbitrary misorientation-dependent grain boundary energies that respect the bicrystallography of grain boundaries.

\subsection{A crystal symmetry-invariant KWC model}
\label{subsec:alternateKWC}
In this section, we formulate a new KWC model that can 
incorporate arbitrary misorientation-dependent grain boundary energies. 
 
We begin with the KWC functional without the $|\nabla \theta|^2$ term.
From \sref{sec:kwc}, recall that in the absence of the $|\nabla \theta|^2$
term, the steady state solution for
$\theta$ is a step function with the discontinuity occurring at the grain
boundary. In one-dimension, since a discontinuous $\theta$ is not in
$H^1(\Omega)$, the minimizer of $\mathcal W$, with $|\nabla
\theta|^2$ absent, is not attained. This observation motivates us to redefine the
domain of the modified KWC functional such that $\theta$ belongs to the space
of piecewise constant functions, as opposed to $H^1(\Omega)$, and this
enables us to simplify the functional as
\begin{equation}
    \mathcal W[\eta,\theta] = \int_{\Omega}
    \left[ \frac{(1-\eta)^2}{2\epsilon} + \frac{\epsilon}{2 }|\nabla \eta|^2
    \right]\, dV
    - \int_{\mathcal S} \ln{(1-\bar \eta)} [\![ \theta ]\!] \, dS,
    \label{eqn:kwc_sharp}
\end{equation}
where $\bar\eta:\mathcal S \to \mathbb R$ is the restriction of $\eta$ to the
jump set $\mathcal S$ of $\theta$, which represents the union of all grain
boundaries. The steady state solution, given in \eref{eqn:eta_analytical},
corresponding to a one-dimensional bicrystal governed by \eref{eqn:kwc_sharp}, and the
resulting grain boundary energy as a function of the misorientation,
\begin{equation}
    \gamma(\jump\theta) = \frac{\jump\theta}{2} \left( 1- 2\ln \left[ \frac{\jump\theta}{2} \right] \right)
    \label{eqn:gbenergy}
\end{equation}
are derived in \ref{app:analytic}. From \eref{eqn:gbenergy}, it is clear that
the grain boundary energy is of a Read--Shockley-type, which does not respect
the crystal symmetry. 

The above observation leads us to the following generalization of the KWC functional 
\begin{equation}
    \mathcal{W}^{\rm G}[\eta,\theta] = \int_{\Omega}
    \left[ \frac{(1-\eta)^2}{2\epsilon} + 
    \frac{\epsilon}{2 }|\nabla \eta|^2
    \right]\, dV
    + \int_\mathcal S g(\bar \eta) \mathcal{J} \left([\![ \theta ]\!] \right) \,
    dS,
    \label{eqn:KWC_energy_alternate}
\end{equation}
which is defined for all $\eta \in H^1(\Omega)$, and
piecewise constant functions $\theta$;  and $\mathcal J$ is an even function of the
jump in orientation. Under this new formulation, the grain boundary energy function modifies as
\begin{equation}
\begin{aligned}
\gamma^{\rm G}([\![ \theta ]\!])
&=(1-\overline{\eta})^2 -\ln{(1-\overline{\eta}) } \mathcal{J}([\![ \theta]\!]) \\
&= \frac{\mathcal{J}(\jump\theta)}{2} \left( 1- 2\ln \left[ \frac{\mathcal J(\jump\theta) }{2} \right] \right),
\end{aligned}
\label{eqn:KWC_bicrystal_integral}
\end{equation}
where $\bar\eta$ is the value of the stead-state solution on the grain boundary,
given implicitly in terms of $\mathcal J(\jump\theta)$ as\footnote{
    The analog of \eref{eqn:jump_condition} in the original KWC model is
    \eref{eqn:app_jump_condition}, whose derivation is shown in
    \ref{app:analytic}.
}                                                                                
\begin{equation}
2(1-\overline{\eta})^2 = \mathcal{J}([\![ \theta ]\!]).
\label{eqn:jump_condition}
\end{equation}        
Inspired from the terminology in dislocations, we refer to $\mathcal{J}(\jump\theta)$ as the
\textit{core energy}.

\begin{figure}[t]
\begin{center}
\includegraphics[width=0.65\textwidth]{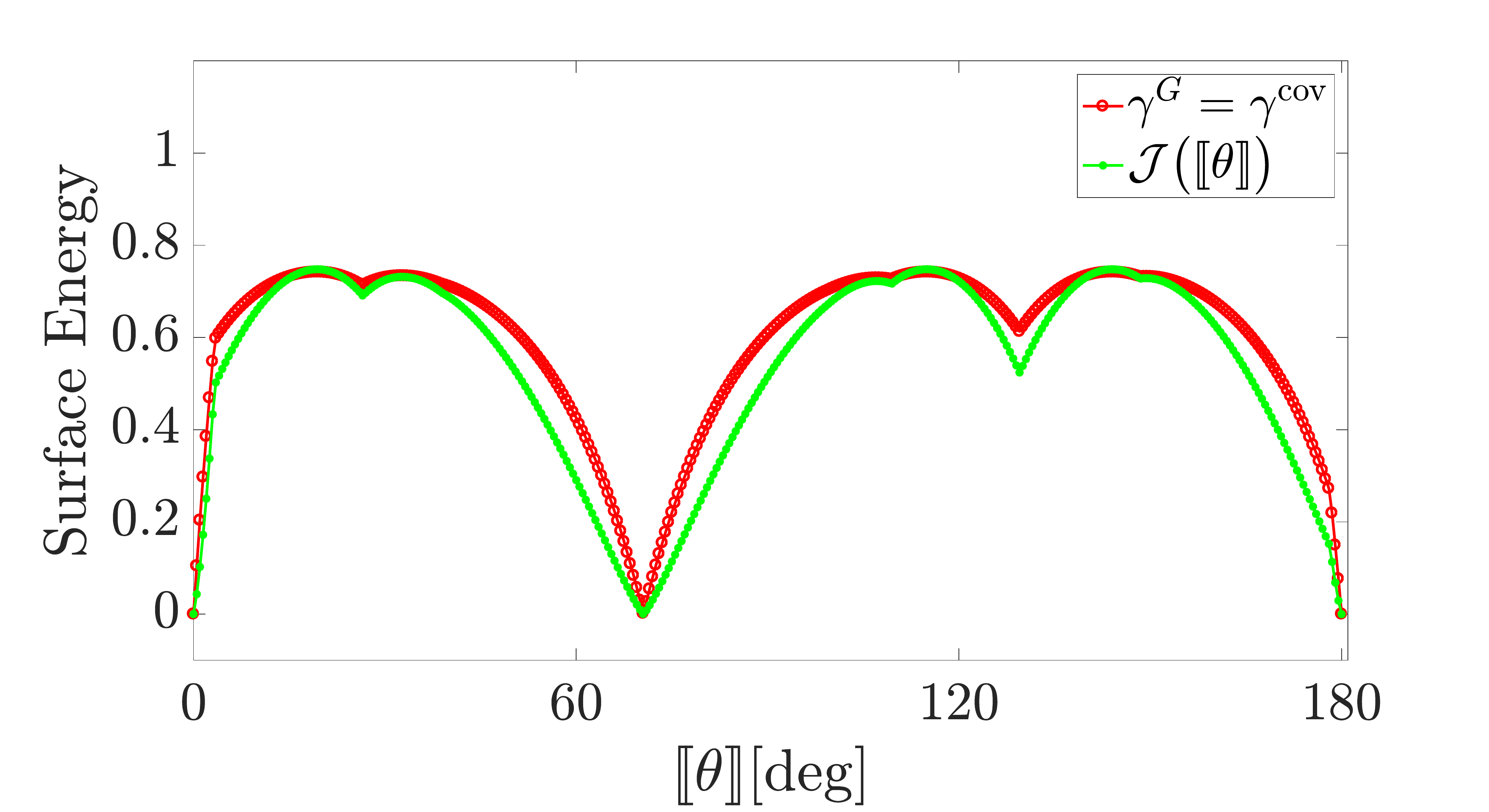}
\end{center}
\caption{A plot (in green) of the core energy $\mathcal J(\jump\theta)$ 
    calculated to match $\gamma^{G}$ \eref{eqn:KWC_bicrystal_integral} 
    to $\gamma^{\rm cov}$ \eref{eqn:gamma_cov} (in red) for $[110]$
    symmetric-tilt grain boundaries in fcc copper. $\gamma^{G}$ is identical to $\gamma^{\rm cov}$
    by construction. 
}
\label{fig:cu_110_covariance}
\end{figure}


From \eref{eqn:KWC_bicrystal_integral}, it is clear that by appropriately
constructing the core energy $\mathcal J$, we can arrive at a $\gamma^{\rm G}$ that
faithfully represents the grain boundary energy and symmetry of the bicrystal.
In other words, the crystal symmetry of the new KWC model is inherited from the
core energy. For illustration, we consider the energy $\gamma^{\rm cov}$ of a $[110]$ symmetric
tilt grain boundary in face-centered cubic (fcc) copper, shown as red data points in
\fref{fig:cu_110_covariance}. $\gamma^{\rm cov}$ is computed using the \textit{covariance model}
developed by \citet{Runnels:2016_1,Runnels:2016_2}, wherein it
is defined as the covariance
of the two lattices adjoining the grain boundary. For
completeness, in \ref{app:covariance_model}, we describe the covariance model of
grain boundary energy, the procedure to compute it, and list the parameters
used to arrive at the data
plotted in \fref{fig:cu_110_covariance}. 
We solve for $\mathcal{J}$ in \eref{eqn:KWC_bicrystal_integral} using the
Newton's method such that the resulting $\gamma^{\rm G}=\gamma^{\rm cov}$ at each
data point. 
In other words, the grain boundary energy of the resulting KWC model
is identical to $\gamma^{\rm cov}$ by construction. 
\fref{fig:cu_110_covariance} shows a plot of the solution $\mathcal J$ in
green, and highlights the common positions of the local minimizers of
$\mathcal J$ and $\gamma^{\rm cov}$.
Typically, the number of Newton iterations for convergence error (absolute value
of the difference between $\gamma^{\rm G}$ and $\gamma^{\rm cov}$) of $10^{-6}
\rm [J/m^2]$ is less than 20.
While the modification of the KWC functional from \eref{eqn:kwc_sharp} to
\eref{eqn:KWC_energy_alternate} allows us to model arbitrary misorientation-dependent
grain boundary energies, the absence of
$|\nabla\theta|^2$ term in \eref{eqn:KWC_energy_alternate} renders the grain boundaries
immobile as mentioned in \sref{sec:kwc}.\footnote{The evolution of $\theta$ by the gradient descent of
    $\mathcal W^{\rm G}$ results in pure rotation while the position of the
grain boundaries remains fixed.} In what follows, we address this
shortcoming by devising a thresholding method to move grain boundaries
by evolving the piecewise-constant $\theta$.

\section{Grain boundary motion in the new KWC model}
\label{sec:thresholding_method}
In this section, we present our approach in which we alternate between
evolving $\eta$ and $\theta$ to evolve a polycrystal governed by
$\mathcal W^{\rm G}$. The order parameter $\eta$ is solved in the following minimization
problem for a given
$\theta$
\begin{equation}
\eta^{*} = 
\operatornamewithlimits{arg\; min}_{\substack{\eta \in H^1(\Omega) \\
\partial \eta/\partial n|_{\partial \Omega}=0}}
\mathcal W^{\rm G}[\eta,\theta].
\label{eqn:eta_sub}
\end{equation}
Next, the orientation field $\theta$ is evolved using a thresholding rule
described in the next section.

In order to solve for $\eta$ in \eref{eqn:eta_sub}, we
note that since $g(\eta)\to \infty$ as $\eta
\to 1$, the functional $W^{\rm G}$ is non-smooth in $\eta$, which makes the
Newton's method not viable. Therefore, we use a primal-dual method recently developed by
\citet{Jacobs:2019}, which has a $\mathcal O(\frac{1}{e}N \log N)$ complexity, 
where $e$ is the error in the numerical solution to \eref{eqn:eta_sub},
and $N$ is the grid size. See \ref{app:primaldual} for a more
detailed description of the primal dual method. 

Next, we develop a thresholding rule to evolve
$\theta$ for a fixed $\eta^*$ obtained in \eref{eqn:eta_sub}. 
The alternate use of the
primal dual method and the thresholding rule at every time step constitutes our
approach to evolving the grain boundaries.

\subsection{The thresholding rule}
\label{sec:rule}

A thresholding method is a sequence of simple rules, executed every time step to reinitialize
the order parameter, such that its evolution describes the motion of a grain
boundary. 

The original idea of using a thresholding method to evolve grain boundaries
goes back to the work \citep{MBO} of Merriman, Bence and Osher (MBO) wherein, similar to the multiphase field model, grains in a polycrystal are described using as many order parameters,
with the caveat that the order parameters are piecewise-constant implying a
sharp interface. An order parameter in the MBO method is evolved based on a two-step thresholding
scheme --- a convolution of the order parameter with a Gaussian kernel followed by
a trivial thresholding --- resulting in motion by curvature.

The MBO method has recently been generalized by \citet{Esedoglu:2015} to 
a variational model, referred to as the \emph{Gaussian kernel
method}. In the MBO and the Gaussian kernel methods,
there are as many characteristic functions as the number of distinct grains.
While the end goal of the KWC model is also to describe motion by curvature,
it is markedly different from the Gaussian kernel method as it uses only two
order parameters to represent a
polycrystal. Therefore, it does not require additional techniques to address the
memory intensive nature of a naive implementation of the MBO/Gaussian kernel
methods.
More importantly, the KWC model lends itself to further generalizations which include
the modeling of grain rotation. Therefore, the goal here is to seek a
thresholding algorithm to implement the KWC model.

In this section, we design a thresholding rule for $\theta$ that results in
motion by curvature. We first recall that $\theta$ is a piecewise-constant field
with a finite range of orientations. This implies, that a thresholding rule for
$\theta$ reassigns $\theta(\bm x)$, for each point $\bm x \in \Omega$, 
to one of the possible orientations. Our thresholding rule originates from the
observation that \emph{the asymmetry of $\eta$ in the neighborhood of a grain boundary
characterizes its curvature.} Below, we explicitly identify this asymmetry
before describing our thresholding rule.

\begin{figure}[t]
\begin{center}
\includegraphics[width=0.50\textwidth]{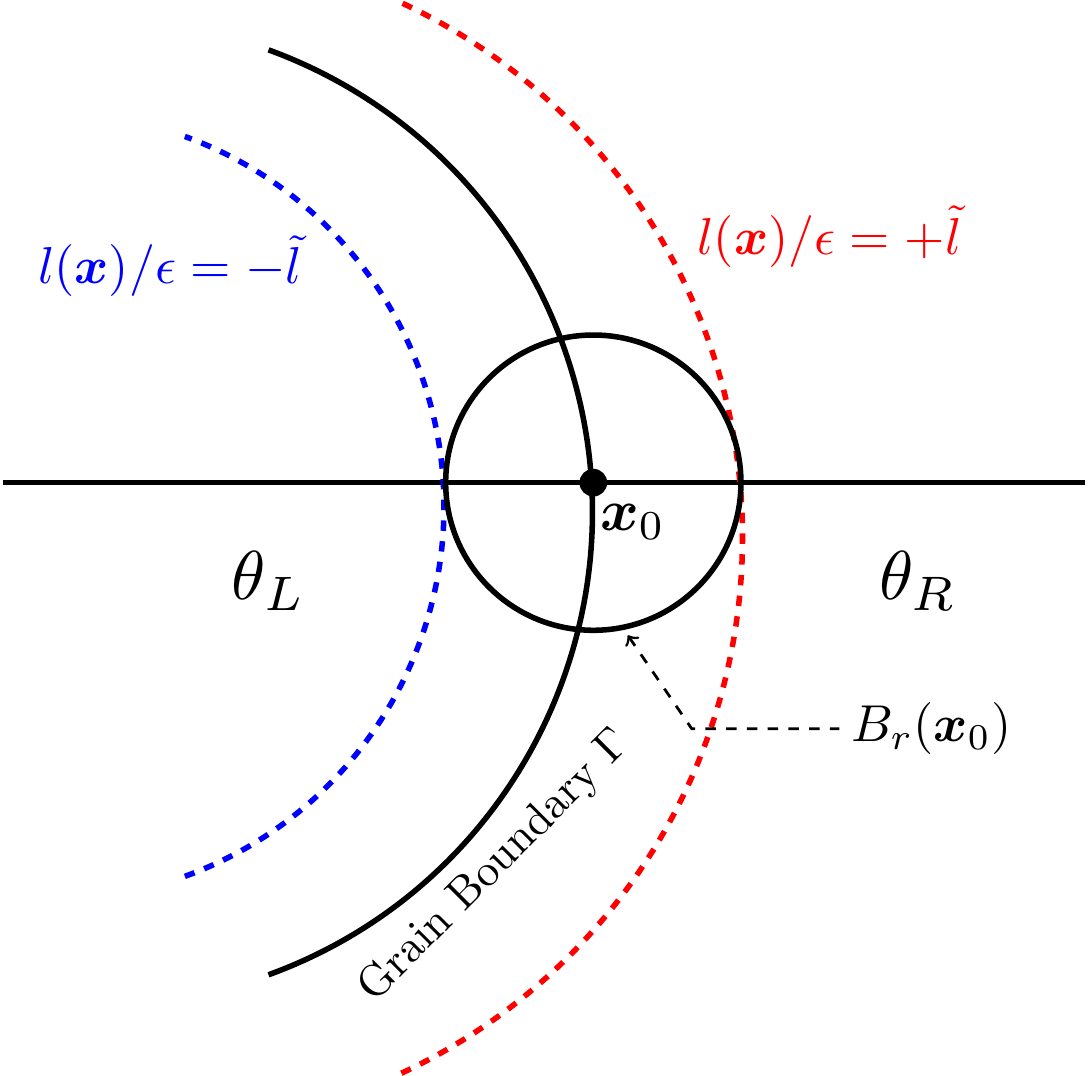}
\end{center}
\caption{Level sets of the distance function $l(\bm x)$ in a neighborhood of
    $\bm x_0$.  The grain boundary $\Gamma$ is depicted as a solid curve, and the dashed curves correspond to
the level sets $l(x)/\epsilon= \pm \tilde l$.
}
\label{fig:level_sets}
\end{figure}
First, we note that the steady-state solution for $\eta$, derived in
\eref{eqn:eta_analytical} for a flat interface (zero curvature), is symmetric about the grain boundary. Next, we derive an
approximate form for $\eta$ in the presence of a non-zero curvature, and
show the dependence of asymmetry on the curvature. Let $\Gamma$ denote a grain
boundary with a non-zero curvature that separates two grains with orientations
$\theta_{\rm L}$ and $\theta_{\rm R}$. 
We postulate that for a $\Gamma$ with a small curvature such that $\epsilon \kappa
\ll 1$, and for a $\bm x_0 \in \Gamma$ away from a triple junction, the solution
$\eta^*$ to \eref{eqn:phi_evolution_orig} is approximated in a small neighborhood of $\bm x_0$ as
\begin{equation}
\eta^*(\bm x) \approx u \left( \frac{l( \bm x)}{\epsilon} \right),
\label{eqn:eta_approx}
\end{equation}
where $l(\bm x)$ is the signed distance function from $\Gamma$ to $\bm x$, as
shown in \fref{fig:level_sets}. In
other words, we assume that $\eta$ only depends on the radial coordinate.
Away from the grain boundary, the solution to the minimization problem in \eref{eqn:eta_sub} satisfies the
equation
\begin{subequations}
    \begin{equation}
        \epsilon \Delta \eta^* - \frac{(\eta^*-1)}{\epsilon}=0.
        \label{eqn:phi_evolution}
    \end{equation}
    \label{eqn:evolution_new}
\end{subequations}
The above equation can be simplified by using a local coordinate system $\bm
x=(\tilde l,s)$, where $\tilde l=l(\bm x)/\epsilon$ is
the scaled radial coordinate, and $s$ is
the distance measured along $\Gamma$ between $\bm x_0$ and the perpendicular projection of $\bm x$ on
$\Gamma$. In this coordinate system, we note that $\triangle \eta(\bm x) = u''/\epsilon^2 + \kappa
u'/\epsilon$, where $\kappa(\tilde l)$ is the curvature of the coordinate line $\{\bm x
\in B_r(\bm x_0) : l(\bm x) = \tilde l\epsilon\}$.
Therefore, \eref{eqn:evolution_new} simplifies as

\begin{equation}
u''(\tilde l)+\epsilon \kappa u'(\tilde l) -u(\tilde l)+1=0. 
\label{eqn:derivationOde}
\end{equation}
Assuming $\kappa(\tilde l) = \kappa(0)$, we obtain the following closed form
solution to \eref{eqn:derivationOde}:
\begin{equation}
u(\tilde l)=
1+C_1 \exp{\left[-\tilde l\left(\frac{\epsilon \kappa + \sqrt{4+\epsilon^2 \kappa^2}}{2}\right)\right]}
+C_2 \exp{\left[-\tilde l\left(\frac{\epsilon \kappa - \sqrt{4+\epsilon^2
\kappa^2}}{2}\right)\right]},
\label{eqn:general_sol}
\end{equation}
where the constants $C_1$ and $C_2$ are determined using the boundary conditions
$u(\pm\infty)=1$.\footnote{The boundary conditions are interpreted in the limit
$\epsilon \to 0$, which results in the boundary conditions $l/\epsilon=\pm
\infty$ for the scaled radial coordinate.\label{ft:bc}}
Subsequently, the solution can be further approximated\footnote{Here, we use the
    approximation $\sqrt{4+\epsilon^2 \kappa^2}\approx 2 +
    \mathcal{O}(\epsilon^2\kappa^2)$.}
under the assumption that
both $\epsilon$ and $\epsilon\kappa$ are small, resulting in
\begin{equation}
    u(\tilde l)=
    \begin{cases}
        1+ (u(0) - 1 ) e^{-(1+0.5 \epsilon \kappa) \tilde l} & \text{ if }
        \tilde l>0, \\
        1+ (u(0) - 1 ) e^{(1-0.5\epsilon \kappa)\tilde l} & \text{ otherwise.}
    \end{cases}
    \label{eqn:eta_profile}
\end{equation}
The asymmetry of $u$ is apparent from \eref{eqn:eta_profile} by noting that in
the presence of a positive curvature, 
the rate at which $u \to 1$ as $\tilde l\to +\infty$ is greater than when 
$\tilde l\to -\infty$.

\begin{figure}[t]
\begin{center}
\includegraphics[width=0.65\textwidth]{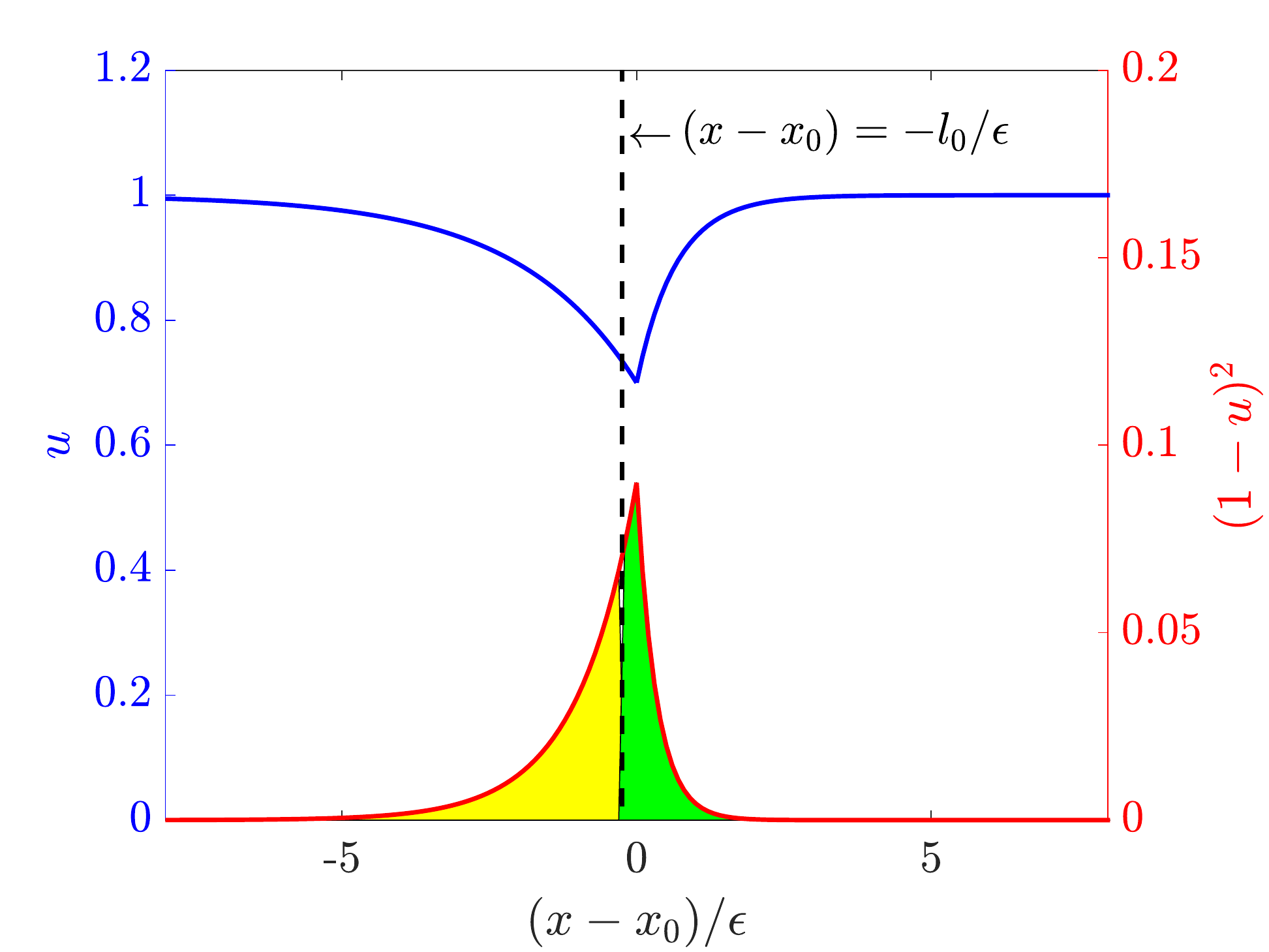}
\end{center}
\caption{ A plot of $\eta^{*}$ in a small neighborhood of $x_0$ (see
\fref{fig:level_sets}) is shown in blue, while $(1-\eta^*)^2$ is shown in red.
The asymmetry of $u$ around $x_0$ due to curvature $\kappa$ is characterized
by the position $x$ at which the two areas shown in yellow and green regions are
equal. The position $x$ is given in terms of $l_0$ ($(x-x_0)=l_0/\epsilon$), which is the
solution of \eref{eqn:suggested_metric}.}
\label{fig:thresholding_observation}
\end{figure}
The asymmetry of $u$ forms the foundation of our thresholding scheme, which is
designed to reassign the values of $\theta$ in the neighborhood of the grain
boundaries resulting in a motion by curvature. To design a thresholding rule, we
identify a unique $l=l_0$, such that 
\begin{equation}
\int^{l_0}_{-\infty} (1-u(l/\epsilon))^2 \, dl
=\int^{+\infty}_{l_0} (1-u(l/\epsilon))^2 \, dl. 
\label{eqn:suggested_metric}
\end{equation}
The two integrals in \eref{eqn:suggested_metric} are depicted as equal areas
under the yellow and green regions in \fref{fig:thresholding_observation}, which
clearly shows that in the presence of a non-zero curvature, the asymmetry of $u$
results in $l_0 \ne 0$.
A straightforward but tedious calculation (see \ref{app:derivThreshold} for details) 
shows that 
\begin{equation}
l_0 = -\frac{\epsilon^2}{4}\kappa + \mathcal{O}(\epsilon^{3}). 
\label{eqn:metric_mean_curvature}
\end{equation}
By reinitializing the orientations of all $\bm x$ with $l(\bm x)<l_0$ to 
$\theta_{\rm L}$, and  to $\theta_{\rm R}$ when $l(\bm x)>l_0$,
we have a thresholding rule that moves the grain boundary by 
$\epsilon^2 \kappa$ in one time step
$dt = \mathsf{t}\epsilon^2/4$, where $\mathsf{t}=1$ is a unit conversion factor. 
Alternating between the $\eta$-update using the primal-dual method, and
the $\theta$-update using the thresholding rule, results in a grain boundary motion by
curvature with mobility equal to the inverse of the grain boundary
energy.\footnote{In this case, the \emph{reduced mobility}
\citep{Esedoglu:2019,Esedoglu:2019_2}, which is defined as the product of grain
boundary energy and mobility, is equal to $1$ for all grain boundaries.}
Although this is a severe restriction on the mobility, we postulate that this can be overcome
by modifying the thresholding rule \eref{eqn:suggested_metric}, and this will be
addressed in a future work. 
The efficiency of the thresholding rule described above rests on
the computation of  $l_0$ in \eref{eqn:suggested_metric}. In the next section,
we use the \emph{fast marching method} to not only compute $l_0$ in an $\mathcal O(N
\log N)$ algorithm, but also generalize the above strategy to an arbitrary
polycrystal.

\subsection{Thresholding dynamics via the fast marching method}
\label{sec:fmm}
The fast marching method (FMM), developed by \citet{tsitsiklis1995efficient}, 
is an algorithm to evolve a surface with a
spatially varying normal velocity. 
A general description of FMM with a stand-alone example is given 
in \ref{app:fastmarching}.
Here, we focus on using the fast marching method to implement the
thresholding algorithm described in \sref{sec:rule} by solving for $l_0$ in
\eref{eqn:suggested_metric}.

\begin{figure}[t]
     \centering
     \begin{subfigure}[b]{0.45\textwidth}
         \centering
         \includegraphics[width=\textwidth]{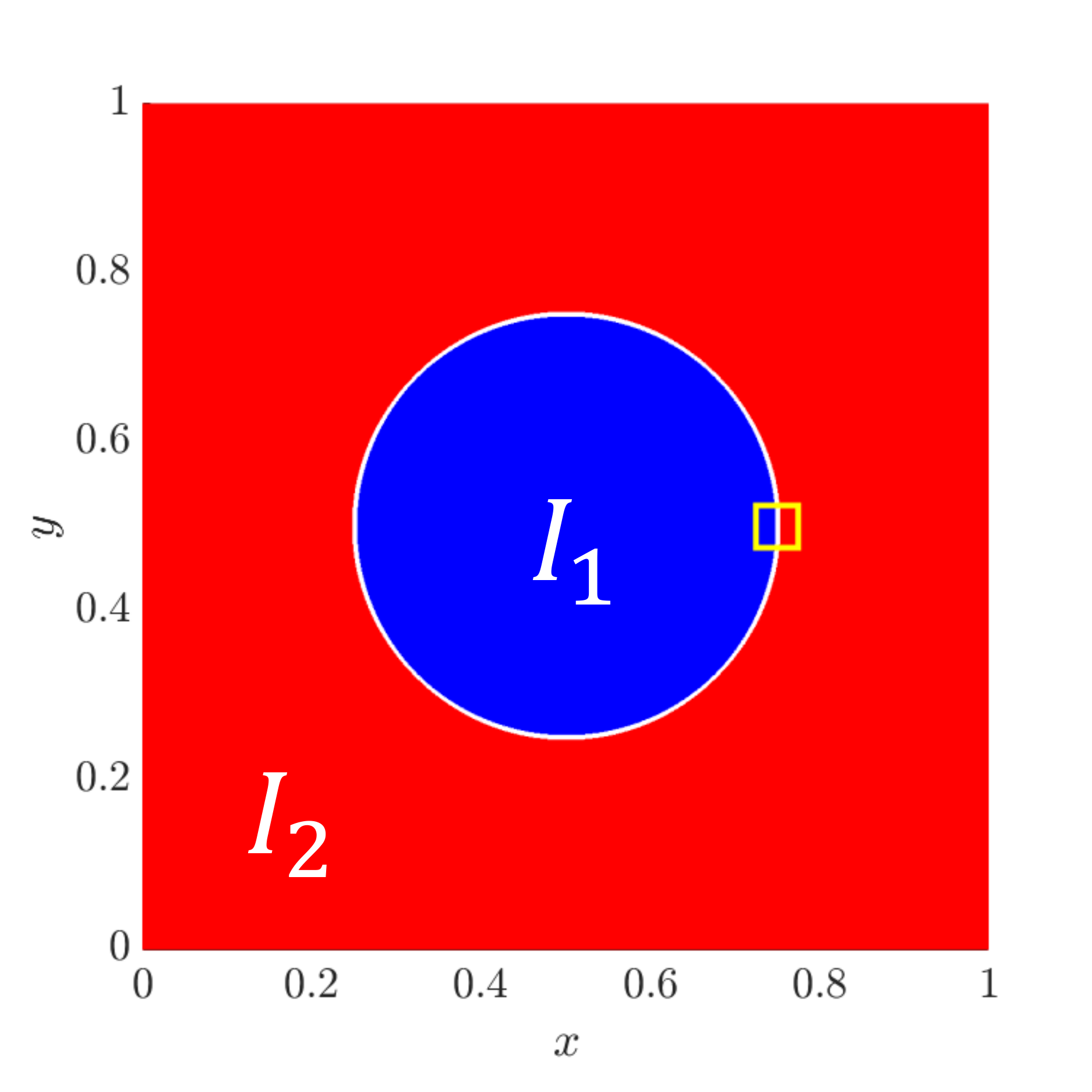}
         \caption{Two interior regions of a bicrystal.}
         \label{fig:circular_grain}
     \end{subfigure}
     \hfill
     \begin{subfigure}[b]{0.45\textwidth}
         \centering
         \includegraphics[width=\textwidth]{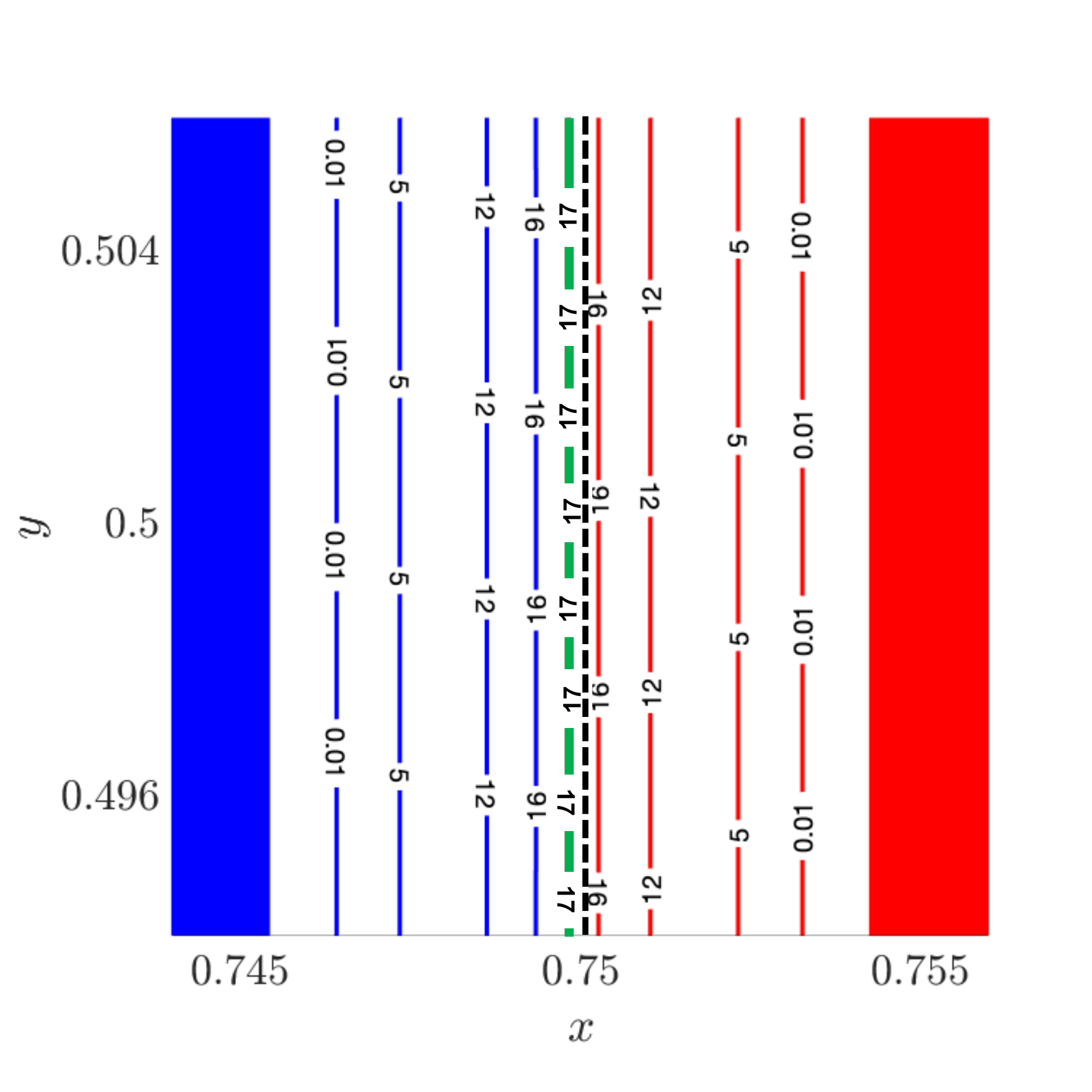}
         \caption{The evolution of $\partial I_1$ and $\partial I_2$ solved
         using the fast marching method.}
         \label{fig:contours}
     \end{subfigure}
     \caption{Shrinking of a circular grain simulated using the thresholding method. 
         \protect\subref{fig:circular_grain}) Two interior 
     regions (red and blue) $I_1$ and $I_2$ are grown towards the grain boundary 
     with a speed $1/(1-\eta^{*})^2$ using the fast marching method. \protect\subref{fig:contours}) 
     A closeup of a rectangular region around the grain boundary, depicted in
     (\protect\subref{fig:circular_grain}), shows the contour lines of the
     fast marching method, which describe the time it takes for $\partial I_1$ or
     $\partial I_2$ to arrive at a grid point. Therefore, the original grain
     boundary, shown as a dashed black line in (\protect\subref{fig:contours}), moves to a new position (solid
 green line) where the two grain interiors meet.}
     \label{fig:theory_example}
\end{figure}

We begin with a description of our implementation of the thresholding scheme for a bicrystal
consisting of a circular grain, followed by its generalization to a polycrystal.
We first recall that the boundary conditions $u(\pm \infty)=1$ used to arrive at
\erefs{eqn:eta_profile}{eqn:suggested_metric} apply only in the limit
$\epsilon \to 0$ as noted in \ftref{ft:bc}. In practice, we choose a finite
limit $l_{\rm b}>0$, and modify \eref{eqn:suggested_metric} as 
\begin{equation}
    \text{Find $l_0$ such that }
    \int^{l_0}_{-l_{\rm b}} (1-u(l/\epsilon))^2 \, dl
    =\int^{l_{\rm b}}_{l_0} (1-u(l/\epsilon))^2 \, dl. 
    \label{eqn:suggested_metric_new}
\end{equation}
In \ref{app:derivThreshold}, we show that the error in $l_0$ due to
the introduction of $l_{\rm b}$ exponentially decreases as $\epsilon \to 0$.
In order to use FMM
to compute $l_0$, we interpret the integrand $(1-u(l/\epsilon))^2$ in
\eref{eqn:suggested_metric_new} as an inverse of the
normal velocity of a surface $\mathcal S_l:=\{\bm x: l(\bm x) =
l\}$ traveling towards the grain boundary.
Under this interpretation, the integrals in \eref{eqn:suggested_metric_new} are a
measure of the time it takes for two initial surfaces $\mathcal S_{-l_{\rm{b}}}$ and $\mathcal
S_{l_{\rm b}}$ on either side of the grain boundary, to meet at $l=l_0$.
We use the fast marching method to evolve the surfaces $\mathcal S_{-l_{\rm{b}}}$ and $\mathcal
S_{l_{\rm b}}$, and implement the thresholding rule described in \sref{sec:rule} by
reassigning the orientation of any point $\bm x$ in the region $\{\bm x \in \Omega:
    |l(\bm x)|<l_{\rm b}\}$  to $\theta_{\rm L}$ if it first encounters
the evolving surface $\mathcal S_{-l_{\rm b}}$, and to $\theta_{\rm R}$ otherwise.

In practice, however, we do not have access to the signed distance
function $l(\bm x)$ to identify the surfaces $\mathcal S_{-l_{\rm b}}$ and
$\mathcal S_{l_{\rm b}}$. 
Instead, we first identify the grain interiors $I_p$ defined as 
\begin{equation}
I_p=\{\bm  x  \in \Omega: \theta(\bm x)=\theta_p, \; \eta(\bm x) > 1-\xi \},
\label{eqn:init_threshold}
\end{equation}
where $\xi>0$ is some fixed small value.
 \fref{fig:circular_grain} shows the grain interiors $I_1$ and
$I_2$ in a bicrystal, and \fref{fig:contours} is a closeup of a rectangular region, 
marked in yellow, around the grain boundary. The original grain boundary is
marked as a black dashed line in \fref{fig:contours}. By construction, the two
surfaces $\partial I_1$ and $\partial I_2$ are
equidistant, up to $\mathcal{O}(\epsilon)$, from the grain boundary, and serve
as substitutes for $\mathcal S_{-l_{\rm b}}$ and $\mathcal S_{l_{\rm b}}$.
The grain interiors are grown in the outward direction with a velocity
$(1-u(l/\epsilon))^{-2}$ using the fast marching
method, and the surface where they meet is the new grain boundary, shown as a green
dashed line in \fref{fig:contours}.

\begin{figure}[t]
\begin{center}
\includegraphics[width=0.5\textwidth]{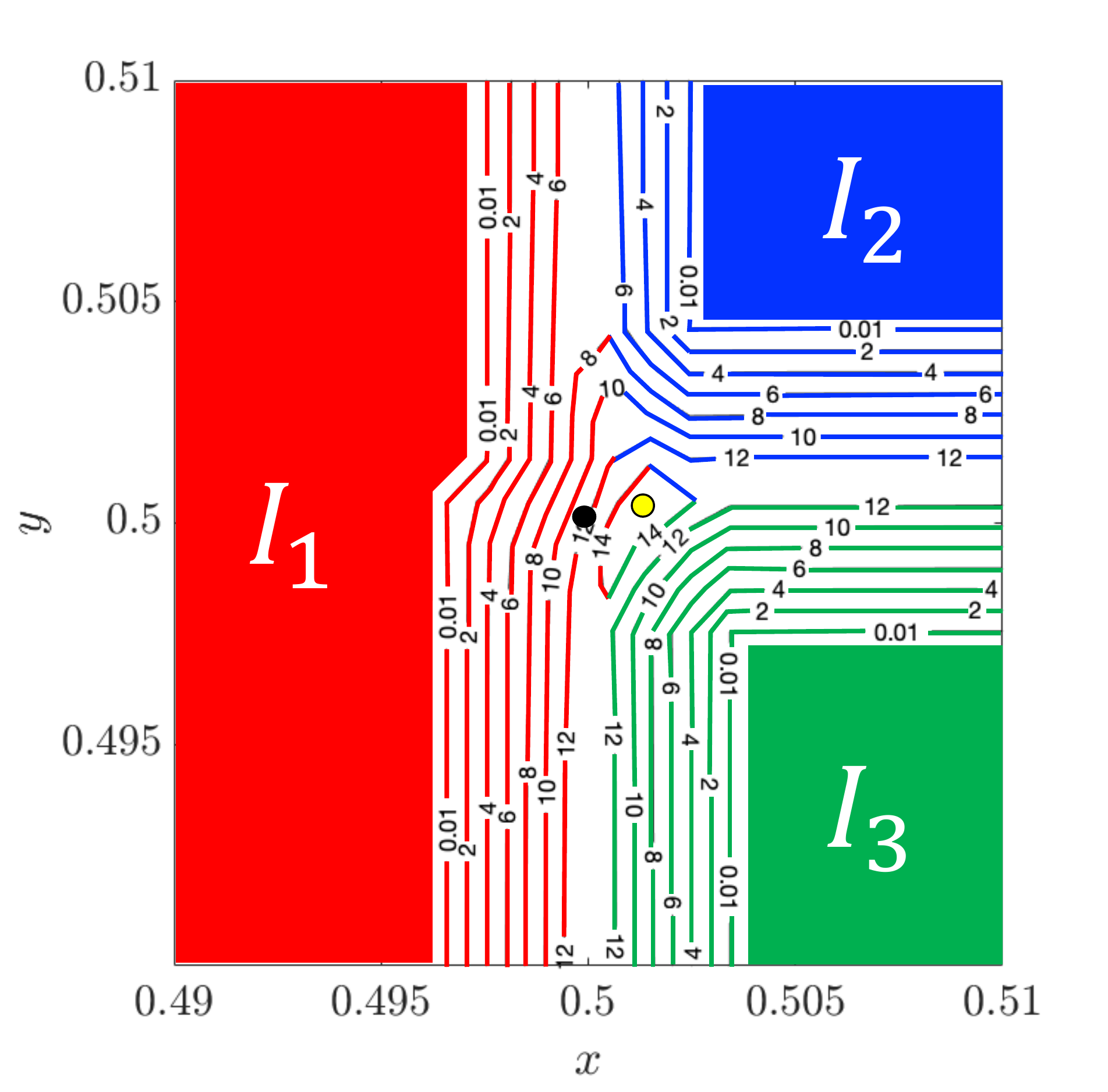}
\end{center}
\caption{Movement of a triple junction according to the thresholding algorithm. 
The triple junction initially at $(x,y)=(0.5,0.5)$ (black filled circle)
moves to a new position (yellow filled circle) 
where the three grain interiors, evolved using the fast marching method, meet at the same time.}
\label{fig:fastmarching_triplepoint}
\end{figure}

We will now generalize the above implementation to an arbitrary
polycrystal consisting of $\mathcal N$ grains, described using a piecewise constant
$\theta$ with values in $\{\theta_1,\dots,\theta_{\mathcal N}\}$. 
Using \eref{eqn:init_threshold}, we identify the $\mathcal N$ grain interiors, and define $I$ as
their union. Next, we grow the grain interiors in their outward unit normal directions until
every point (in the almost everywhere sense) in the domain is in precisely one
grain. We implement this by first collecting all the boundaries of the interior
regions in $\partial I = \partial I_1 \cup \dots \cup \partial I_n$,
and \emph{simultaneously} evolving them in
the outward normal direction with a speed of $1/(1-\eta^{*}(x))^2$ using the fast
marching method.\footnote{Note that the fast marching method is used to evolve
    \emph{all} grain interiors in unison as opposed to
evolving them individually.} As the grain interiors grow, a point $\bm x \in
\Omega-I$ 
is reinitialized to an orientation $\theta_q$ if it encounters $\partial I_q \subset
\partial I$. 
At the end of the fast marching method, all points in $\Omega - I$
have been reinitialized resulting in an updated polycrystal at the end of a
time step. \fref{fig:fastmarching_triplepoint} shows the implementation of the thresholding rule in a
tricrystal. Dirichlet boundary conditions on $\theta$ are imposed by including
all $\bm x \in \partial \Omega$ in the grain interiors. On the other
hand, periodic boundary conditions are achieved by periodically reinitializing
$\theta$ for $\bm x \in \partial \Omega$ during the fast marching step.

\begin{algorithm}[t]
    \setstretch{1.2}
    \LinesNotNumbered
    \SetKwInOut{Input}{Input}
    \SetKwInOut{Output}{Output}
    \Input{A polycrystal with $\mathcal N$ grains with orientations
        $\theta_1,\dots,\theta_{\mathcal N}$, grain boundary core energies
        $\mathcal J(\jump \theta)$; parameters: $\epsilon$, $\xi$, total time $T$,
    and tolerance $\mathsf e$.}
    \Output{Time evolution of the polycrystal}
    Construct the core energy function $\mathcal{J}( [\![ \theta ]\!])$ from
    grain boundary energy data\\
    Initialize $t=0$, and the orientation field $\theta(\bm x,0)$ \\
    \While{ $t < T$} {
    Compute the discrete jump fields $\jump \theta(\bm x,t)$ and
    $\bar{\mathcal J}:= \mathcal J(\jump \theta(\bm x,t))$ on $\Omega$\\
    Regularize  the jump field: $\mathcal J^\star = G * \bar{\mathcal
    J}$, where $G(\bm x)=(1/2\pi \epsilon^2) e^{-\frac{\bm
        |\bm{x}|^2}{2\epsilon^4}}$\\
        \tcp{Solve for $\eta(\bm x,t)$ using the primal-dual algorithm}
    
    Initialize $\eta$ and the dual field $\psi$: $\eta_0(\bm x)=0,\;
    \psi_0( \bm x)=0$, and $n=0$\\
    \SetKwRepeat{Do}{do}{while}
    \Do{  $\|\eta_{n+1}-\eta_{n}\|_{\infty} \leq \mathsf{e}$\quad }
    {
        $n=n+1$\\
        Calculate $\eta_n$ using $\psi_{n-1}$  (\ref{eqn:PD_my})\\
        Calculate $\psi_{n}$ using $\eta_{n}$  (\ref{eqn:psi_update_form})\\
    }
    $\eta(\bm x,t) = \eta_{n+1}$\\
    \tcp{Threshold/update the orientation field}
    Identify interiors of grains: $I_p=\{ \bm{x}  \in \Omega:
    \theta(\bm{x},t)=\theta_p,\, \bar{\mathcal J}(\bm x) < \xi \}$, and set $I =
    \cup_{p=1}^n I_p$\\
    Evolve $I$ with speed $1/(1-\eta(\bm x,t))^2$ using the fast marching method
    and update/threshold the orientations at each point $\bm{x}\in
    \Omega-I$\\
    $t=t+0.25 \epsilon^2$ (\ref{eqn:metric_mean_curvature})
    }
    \caption{Thresholding algorithm for the new KWC model}
    \label{algorithm:KWC}
\end{algorithm}

The primal-dual and the fast marching methods are
implemented on a regular grid of resolution, say $\delta x $. From
\eref{eqn:metric_mean_curvature},
we know that the resolution of the grid should be large enough to resolve a grain
boundary movement of $\epsilon^2 \kappa$ in each time step, i.e.
\begin{equation}
\delta x \ll \epsilon^2 \kappa,
\label{eqn:spacing}
\end{equation}
which is a common requirement of other thresholding methods \citep{Esedoglu:2015,MBO}.
If this condition is not satisfied, grain boundaries would stagnate. 
Since grain boundary evolution results in an overall decrease in
curvature, \eref{eqn:spacing} may cease to hold as the simulation progresses.
Therefore, we adaptively increase $\epsilon$ when a grain boundary stagnates, and as
a consequence, we obtain a time adaptive algorithm since $dt \propto
\epsilon^2$.
On the other hand, an extremely small $\epsilon$ will increase the computational cost of the thresholding
method.

\begin{table}[t]
\centering
\begin{tabular}{c|c|c|}
\cline{2-3}
\multicolumn{1}{l|}{}      & \multicolumn{2}{c|}{Grid Size} \\ \hline
\multicolumn{1}{|c|}{$\xi$}   & $1024\times1024$    & $2048\times2048$    \\ \hline
\multicolumn{1}{|c|}{0.15} & 9.77 \%          & 2.32   \%        \\ \hline
\multicolumn{1}{|c|}{0.10} & 7.74  \%        & 1.24  \%         \\ \hline
\multicolumn{1}{|c|}{0.05} & 3.39   \%       & 0.71  \%         \\ \hline
\multicolumn{1}{|c|}{0.02} & 2.58    \%      & 0.07 \%          \\ \hline
\end{tabular}
\caption{The effect of parameter $\xi$ on deviations from the expected motion by
    curvature. We note that for a $2048\times2048$ grid, $\xi=0.05$ is small enough to achieve an error less than 1\%. }
\label{tab:metricError}
\end{table}

Finally, we explore the effect of $\xi$, introduced in
\eref{eqn:init_threshold}, on
the extent to which \eref{eqn:metric_mean_curvature} is satisfied. Recall that
$\xi$ was introduced in \eref{eqn:init_threshold} to identify grain interiors.
In the case of a circular grain (see \fref{fig:circular_grain}),  \eref{eqn:metric_mean_curvature} implies the
rate of change of radius is given by
\begin{equation}
\dot{R}(t) =-\frac{\epsilon^2}{4R(t)}. 
\end{equation} 
To test if the above equation is satisfied, 
we executed the thresholding algorithm
using the $\eta$-solution from the primal dual algorithm with $\epsilon=0.01$,  
and measured $\dot{R}(t)$.
Relative \% errors in shrinking-rate $\dot{R}$ at different values of $\xi$ 
are summarized in \tref{tab:metricError}.
It is confirmed that for a sufficiently small grid, 
$\xi=0.05$ is small enough to achieve an error less than 1\%.

Algorithm~\ref{algorithm:KWC} summarizes our approach. 
The core energy data $\mathcal{J}$ (e.g., \fref{fig:cu_110_covariance}) is
computed separately using the procedure described in \sref{subsec:alternateKWC},
and used as an input to our method. 
The algorithm alternates between the primal-dual and the fast marching methods resulting in motion by curvature. 
A \texttt{C++} template library that implements
Algorithm~\ref{algorithm:KWC} is available at 
\url{ https://github.com/admal-research-group/GBthresholding}.

  Finally, we remark on the computation of $\mathcal J(\bm x)$ 
on a grid. Since $\jump \theta$, calculated at a grid point $ij$ in either $x$- or
$y$-directions using centered-difference, is shared between two grid points, a
factor of $1/2$ appears in
the following expression used to compute the total jump:
\begin{equation}
    \jump\theta_{ij} = \frac{1}{2} \sqrt{(\theta_{i+1,j}
    -\theta_{i-1,j})^2+(\theta_{i,j-1} -\theta_{i,j+1})^2}.
\end{equation}

\section{Numerical results}
\label{sec:numeric_experiment}
In this section, we present examples that explore various features of grain
boundary evolution predicted by our model. 

\begin{figure}[t]
     \centering
     \begin{subfigure}[b]{0.45\textwidth}
         \centering
         \includegraphics[width=\textwidth]{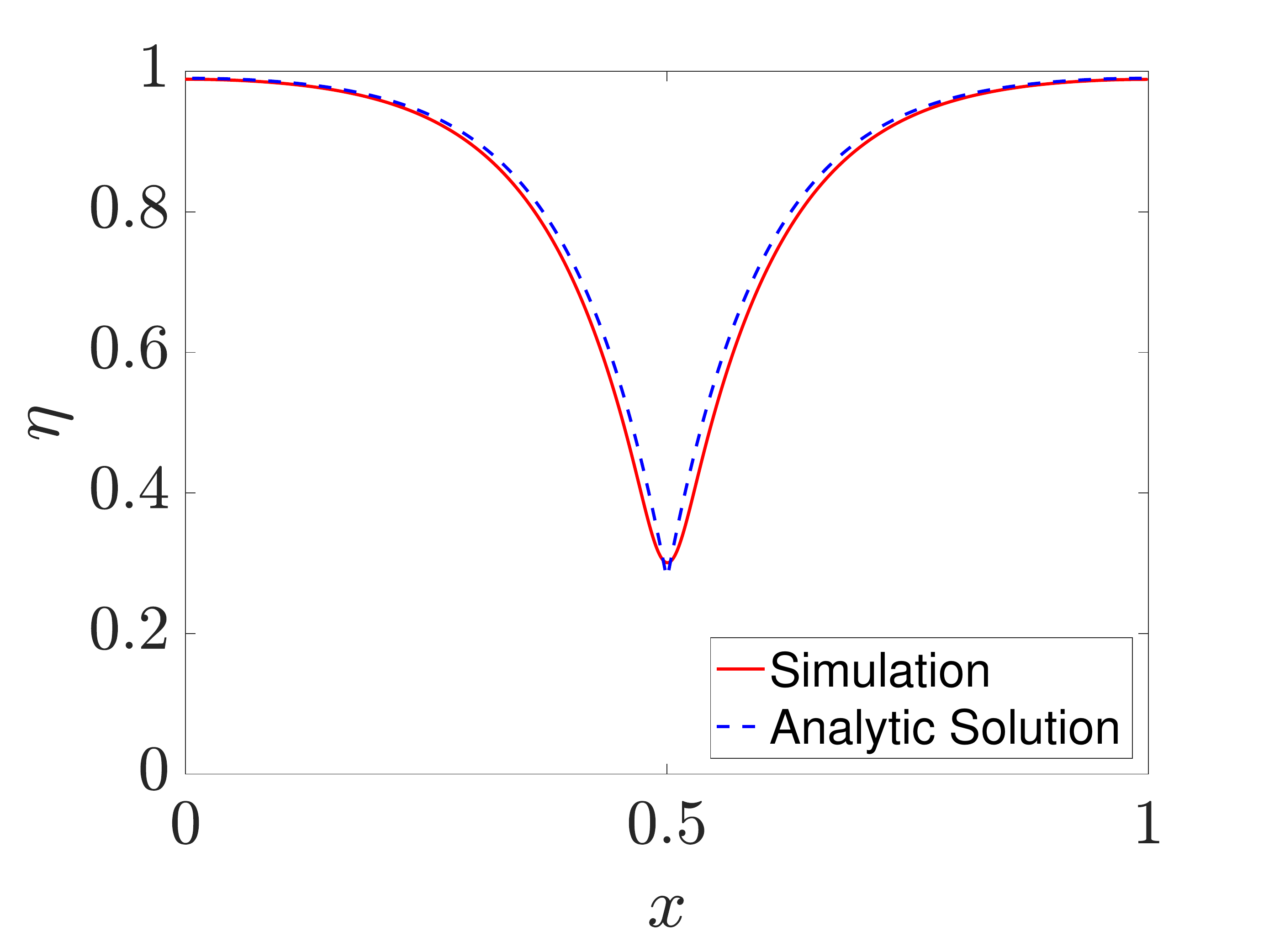}
         \caption{}
         \label{fig:eta_bicrystal}
     \end{subfigure}
     \begin{subfigure}[b]{0.45\textwidth}
         \centering
         \includegraphics[width=\textwidth]{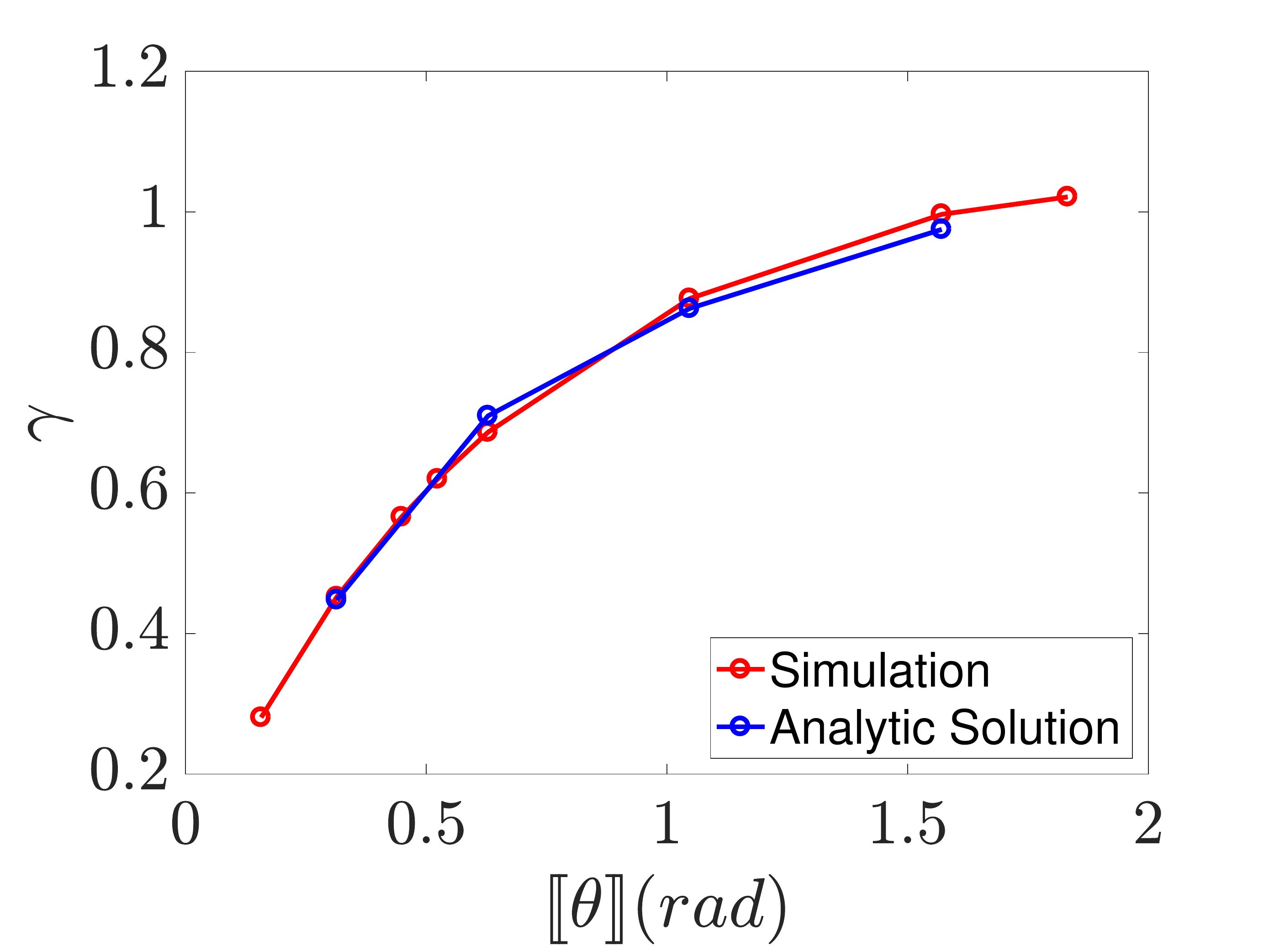}
         \caption{}
         \label{fig:energy_bicrystal}
     \end{subfigure}
     \caption{A comparison of the numerical solution resulting from the
         thresholding algorithm, implemented with $\epsilon=0.1$ on a $512\times
         512$ grid, with the analytical solution. Plots of \protect\subref{fig:eta_bicrystal}) the steady-state
         solution $\eta$, and \protect\subref{fig:energy_bicrystal}) grain
     boundary energy as a function of misorientation.}
     \label{fig:1d_Bicrystal}
\end{figure}
We begin with a simulation of a one-dimensional bicrystal $\Omega=[0,1]$ with a grain
boundary at $x=0.5$, and
$\mathcal J(\jump\theta) = \jump\theta$. The purpose of this simulation is to
ensure that the results of the primal dual algorithm are consistent with the
analytical model described in \ref{app:analytic}. A Neumann boundary condition
$d\eta/dx = 0$ is enforced at the two ends.
In the absence of a curvature, we expect
the grain boundary to remain at $x=0.5$, and $\eta$ reach its steady state.
The tolerance $\mathsf e$ of the primal dual algorithm \eref{eqn:primal_tol} is set to $10^{-6}$.
\fref{fig:eta_bicrystal} confirms the agreement between $\eta$ obtained from the
primal dual algorithm and the analytical form given in \eref{eqn:eta_analytical}.
Furthermore, \fref{fig:energy_bicrystal}
shows that the grain boundary energies predicted by the primal-dual algorithm for various
misorientation angles are in agreement with the analytical
result in \eref{eqn:energy_sol}.

\subsection{Equilibrium of a triple junction}
\label{subsec:herring}

\begin{figure}[t]
     \centering
     \begin{subfigure}[t]{0.3\textwidth}
         \centering
         \includegraphics[trim= 0 0 0 0 clip=true, width=\textwidth]{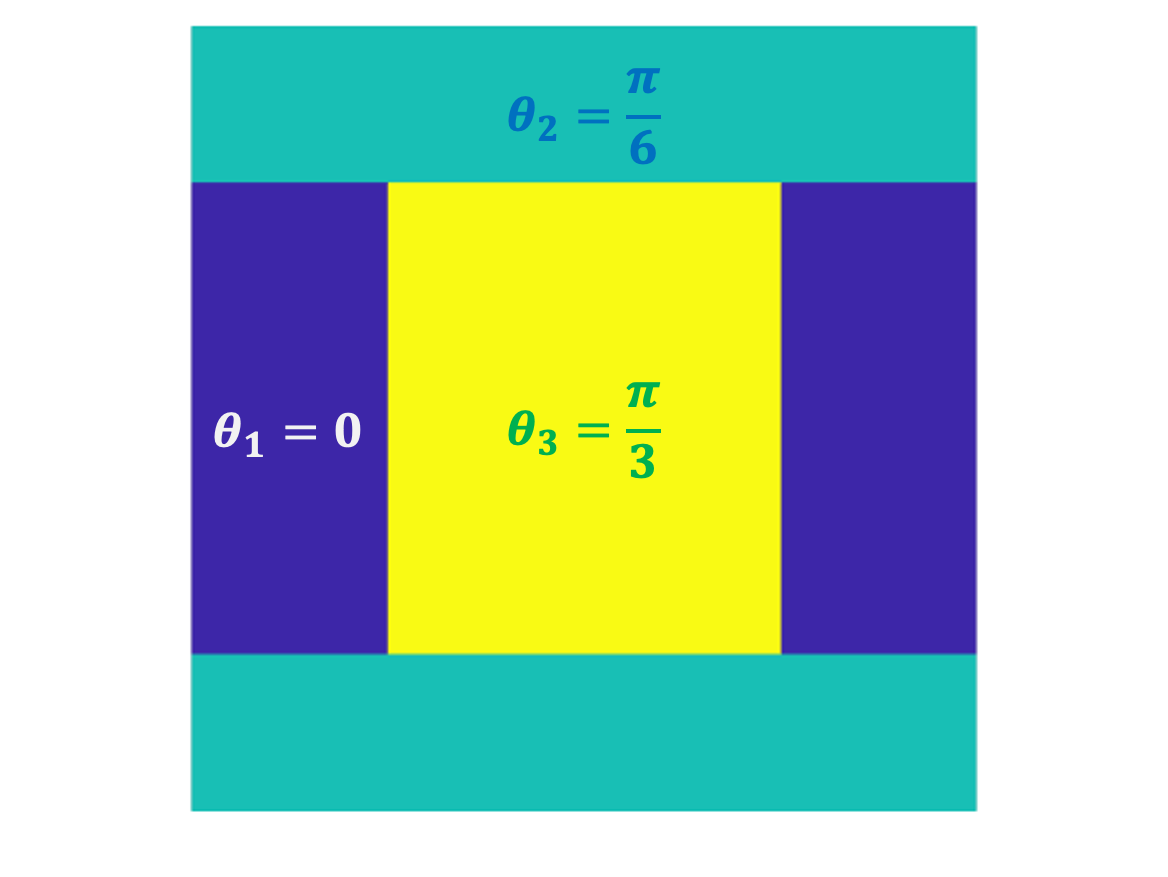}
         \caption{Initial Condition}
         \label{subfig:Tricrystal_init}
     \end{subfigure}
     \begin{subfigure}[t]{0.3\textwidth}
         \centering
         \includegraphics[trim= 0 0 0 0 clip=true, width=\textwidth]{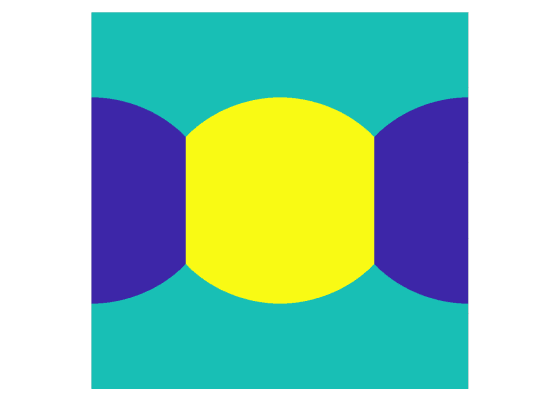}
         \caption{Gaussian Kernel Method}
         \label{subfig:Tricrystal_Thresholding}
     \end{subfigure}
      \begin{subfigure}[t]{0.3\textwidth}
         \centering
         \includegraphics[trim= 0 0 0 0 clip=true, width=\textwidth]{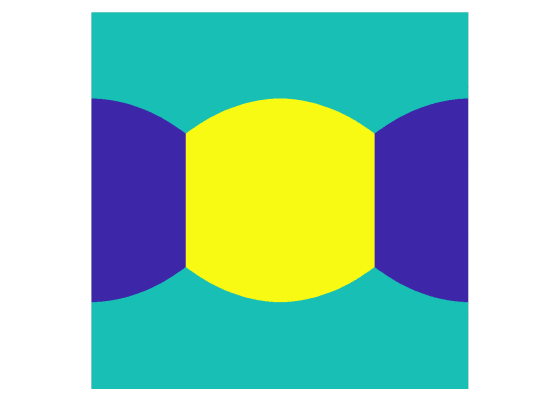}
         \caption{Current method}
         \label{subfig:Tricrystal_Thresholding}
     \end{subfigure}
        \caption{A comparison of the evolutions of a tricrystal under
         periodic boundary conditions obtained using the Gaussian
     kernel method and the generalized KWC model with $\epsilon=0.01$, implemented using
 our method. The dihedral angles predicted by the Gaussian kernel method and our
 method are $(93^\circ,133.5^\circ,133.5^\circ)$ and
 $(106^\circ,127^\circ,127^\circ)$ respectively, while the theoretical values
 are $(90.89^\circ,134.56^\circ,134.56^\circ)$. In \fref{fig:Herring}, we
 demonstrate that the error in the dihedral angles predicted by the generalized KWC
 model decreases as $\epsilon\to 0$.} 
 \label{fig:Tricrystal}
\end{figure}

\begin{figure}[t]
     \centering
     \begin{subfigure}[t]{0.45\textwidth}
         \centering
         \includegraphics[trim= 0 29 0 0 clip=true, width=\textwidth]{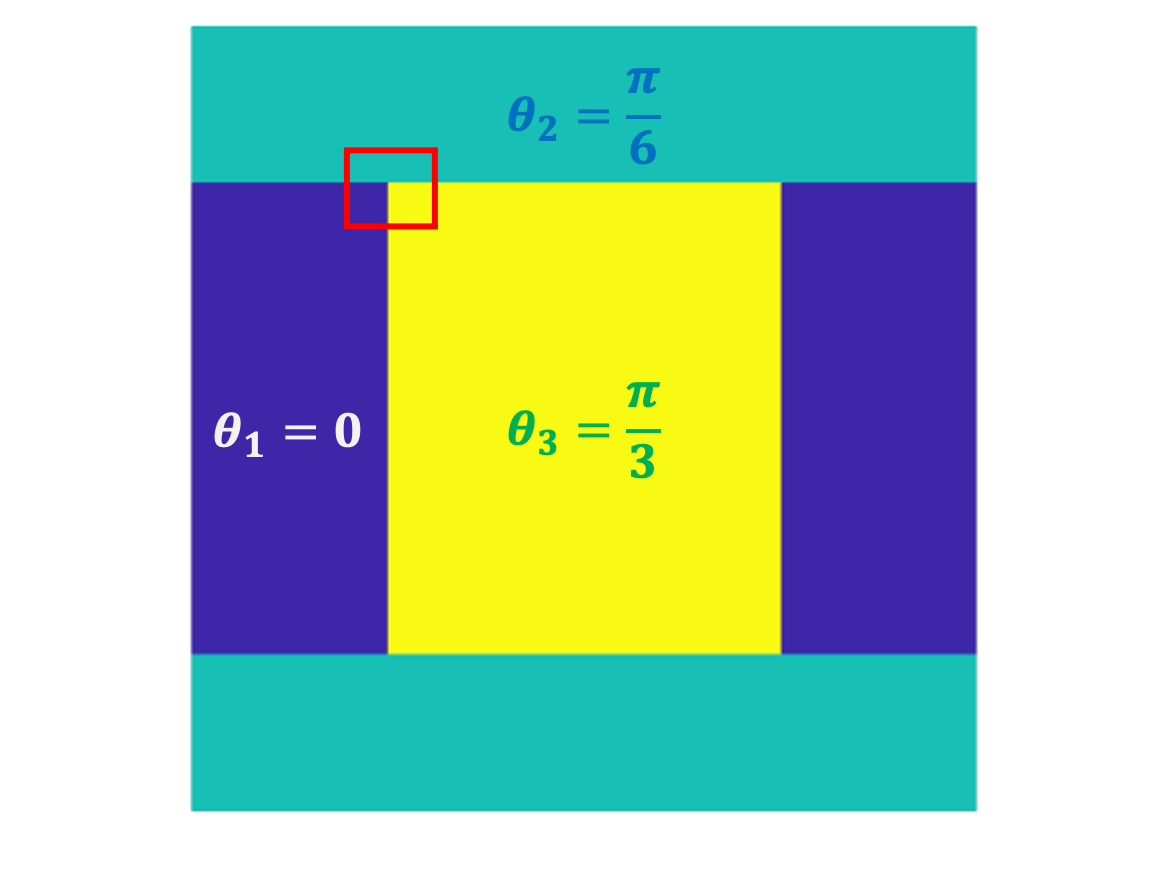}
         \caption{Initial Condition}
         \label{subfig:Herring_init}
     \end{subfigure}
     \begin{subfigure}[t]{0.45\textwidth}
         \centering
         \includegraphics[trim= 0 35 0 0 clip=true, width=\textwidth]{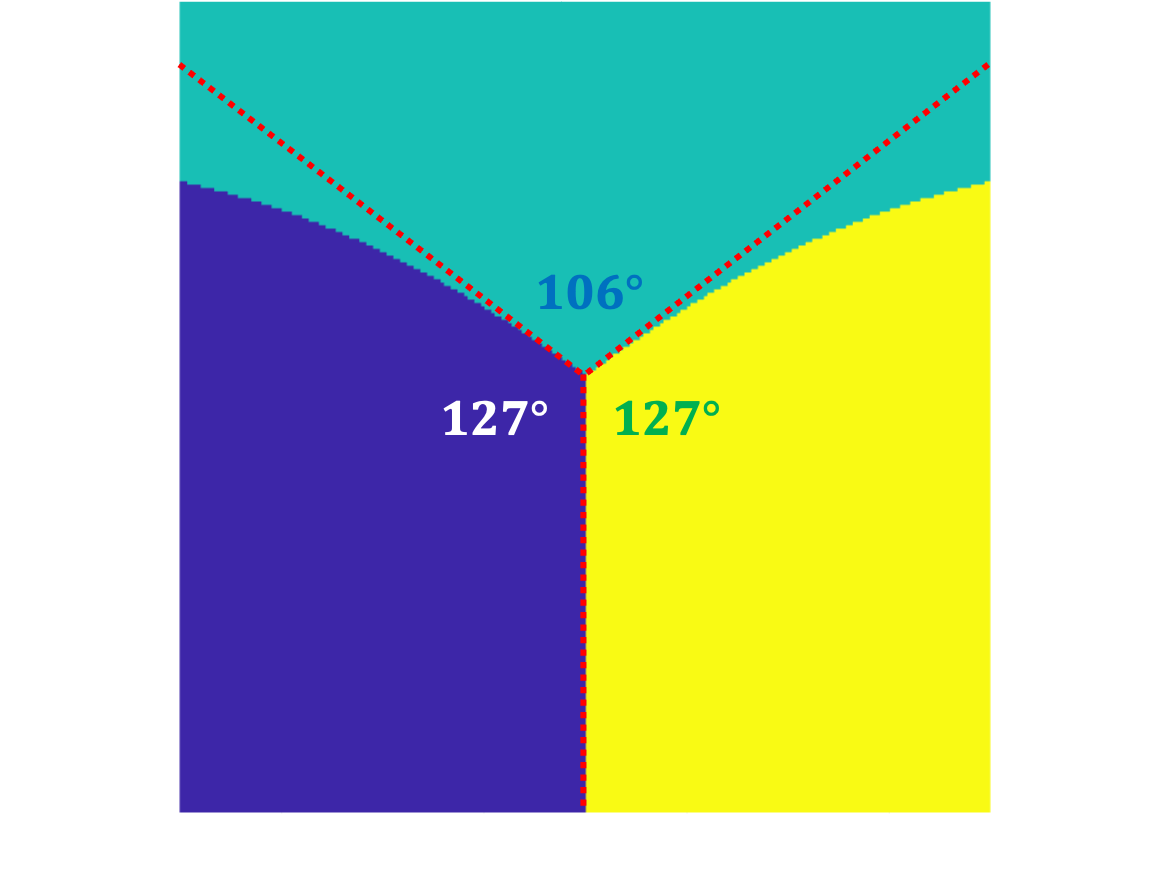}
         \caption{$\epsilon$=0.01}
         \label{subfig:Herring_b}
     \end{subfigure}
     \par\bigskip
      \begin{subfigure}{0.45\textwidth}
         \centering
         \includegraphics[trim= 0 35 0 0 clip=true, width=\textwidth]{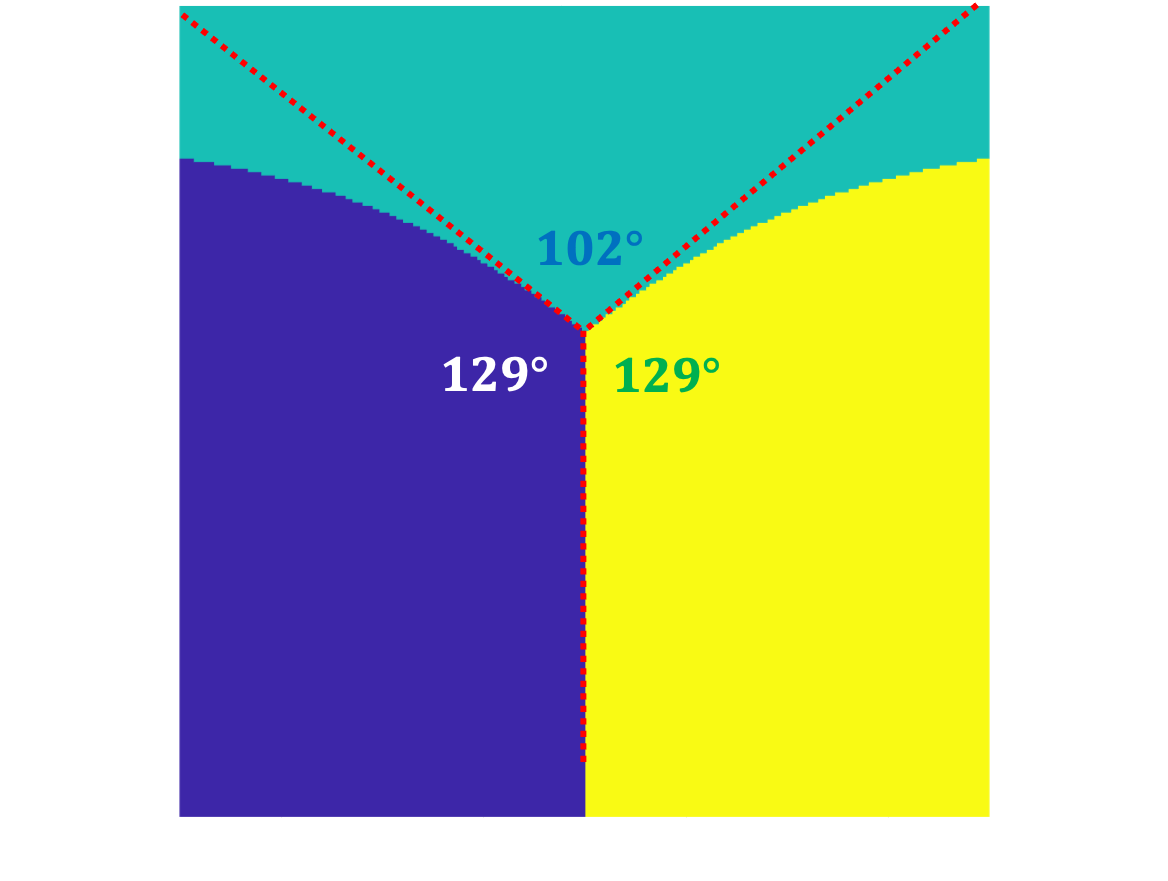}
         \caption{$\epsilon$=0.006}
         \label{subfig:Herring_c}
     \end{subfigure}
      \begin{subfigure}{0.45\textwidth}
         \centering
          \includegraphics[trim= 0 35 0 0 clip=true, width=\textwidth]{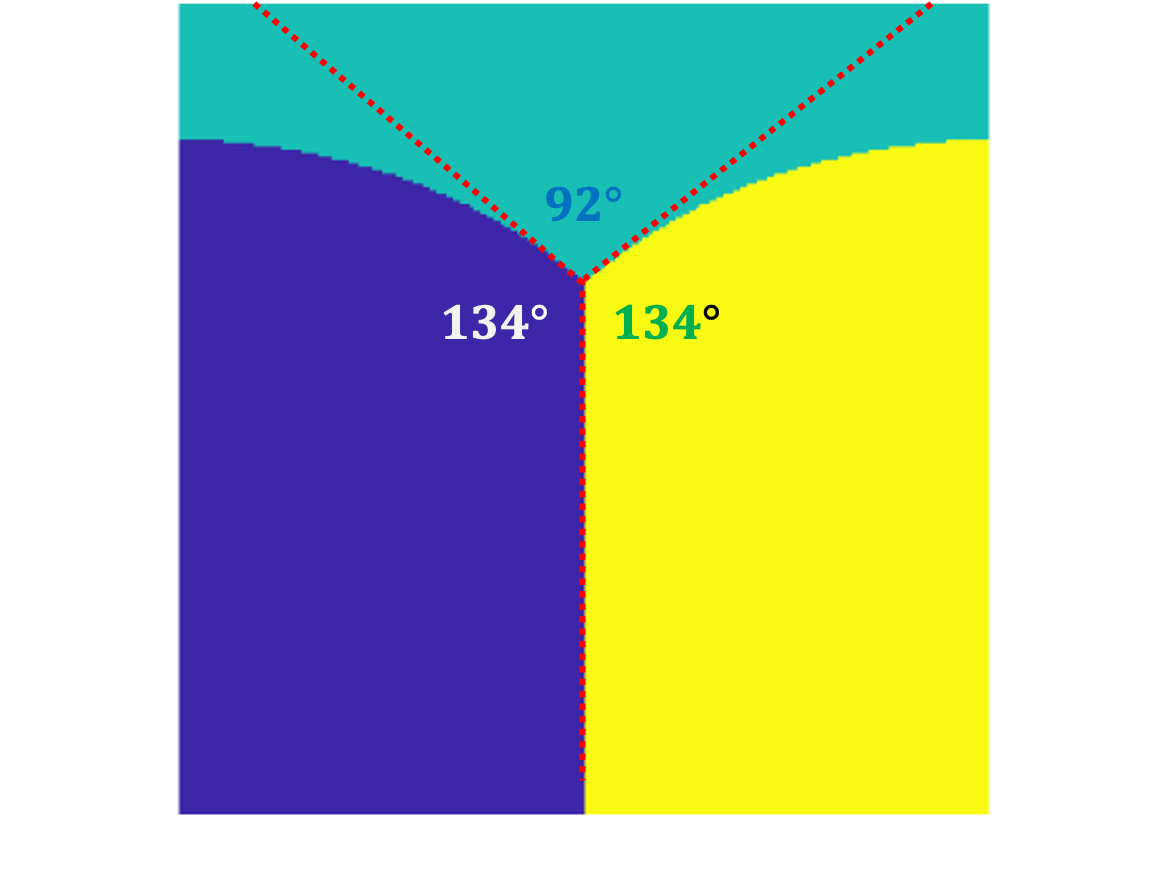}
         \caption{$\epsilon$=0.003}
          \label{subfig:Herring_d}
     \end{subfigure}
     \caption{\protect\subref{subfig:Herring_init}) The orientation distribution in an initial tricrystal under
         periodic boundary conditions with a
         triple junction at $(x,y)=(0.25.0.75)$, and dihedral angles
         $(\Theta_1,\Theta_2,\Theta_3)=(90^{\circ},180^{\circ},90^{\circ})$. The
         polycrystal is evolved using the thresholding algorithm with
         $\epsilon=0.01$, $0.006$ and $0.003$.
         \protect\subref{subfig:Herring_b})-\protect\subref{subfig:Herring_d})
         Closeups of an evolving triple junction (red box) clearly
         show that the dihedral angles converge to 
        $(\Theta_1,\Theta_2,\Theta_3)=(134.56^{\circ},90.89^{\circ},134.56^{\circ})$
        predicted by the Herring angle condition \eref{eqn:HerringCondition}, as
        $\epsilon$ converges to zero. 
}\label{fig:Herring}
\end{figure}

A triple junction is a line where three grains meet, and it is represented as a
point in two dimensions. The equilibrium of a triple junction is guaranteed if it satisfies
the Herring relation \citep{Herring} given by
\begin{equation}
    \frac{\gamma^{12}}{\sin \Theta_3}=
    \frac{\gamma^{23}}{\sin \Theta_1}=
    \frac{\gamma^{31}}{\sin \Theta_2},
\label{eqn:HerringCondition}
\end{equation}
where $\Theta_i$ is the dihedral angle of grain $i$, and $\gamma^{ij}$ is the
energy density of the grain boundary shared by grains $i$ and $j$. While the
Herring relation is derived in the sharp-interface framework, it is also
seen to hold for a triple junction governed by the original KWC model through
\eref{eqn:evolution}. This is not surprising since the
KWC model converges to the Mullins model in the sharp-interface limit \emph{and}
the evolution in \eref{eqn:evolution} has a variational structure in the form
of a gradient descent of the functional in \eref{eqn:KWC_energy}. On
the other hand, it is not clear if our 
approach to evolve the generalized KWC model arises from a
variational formulation. Therefore, it is necessary to examine the Herring
relation using our thresholding algorithm. We will now demonstrate that the
Herring relation indeed holds provided the parameter $\epsilon$ is
chosen appropriately.

We study the evolution of a triple junction in a tricrystal with orientations
$\theta_1=0$, $\theta_2=\pi/6$, and $\theta_3=\pi/3$ in $\Omega=[0,1]\times
[0,1]$. Using the Read--Shockley core energy $\mathcal J = \jump\theta$, we note from
\fref{fig:energy_bicrystal} that the energy density of the three grain boundaries
are $\gamma^{12}=0.62$, $\gamma^{23}=0.62$, and $\gamma^{13}=0.87$. 
From the Herring relation in \eref{eqn:HerringCondition}, it follows that the steady
state dihedral angles are 
$\Theta_1=134.56^{\circ}$, $\Theta_2=90.89^{\circ}$, and $\Theta_3=134.56^{\circ}$.
In order to examine the Herring relation, we consider a tricrystal under
periodic boundary conditions, with an
initial orientation distribution given by
\begin{equation}
    \theta(\bm x,t=0) =
    \begin{cases}
        \theta_2 & \text{if } x_2 \leq0.25  \text{ or }  x_2>0.75, \\
        \theta_3 & \text{if } 0.25<x_2\leq 0.75 \text{ and } 0.25\leq x_1<0.75, \\
        \theta_1 & \text{if } 0.25<x_2\leq 0.75 \text{ and } x_1>0.25 \text{ or
        } x_1>0.75,
    \end{cases}
\label{eqn:Herring_initial_condition}
\end{equation}
resulting in four triple junctions at
$(x_1,x_2)=(0.25,0.75),(0.75,0.75),(0.75,0.25)$, and $(0.25,0.25)$. The 
initial dihedral angles of the triple junctions are
$90^\circ$, $180^\circ$, and $90^\circ$. \fref{subfig:Tricrystal_init} shows a plot of the
initial orientation distribution of the tricrystal.

We begin by comparing the evolution of a triple junction predicted by the KWC
model implemented using our thresholding scheme with that obtained using the
Gaussian kernel method \citep{Esedoglu:2015}. The grain boundary energies 
$(\gamma^{12},\gamma^{23},\gamma^{13})=(0.62,0.62,87)$, pre-computed using the
KWC model, are inputs to the Gaussian kernel method, and the respective
mobilities are set to the inverse of the grain boundary energies.
The parameter $\epsilon$ of the KWC model is taken as $0.01$.
Both schemes are simulated on a $1024\times1024$ grid.
As shown in \fref{fig:Tricrystal}, the evolution dynamics of both schemes are 
qualitatively similar. As expected, the triple junction adjusts at a faster time
scale to satisfy the Herring angle condition compared to the
curvature-driven motion of grain boundaries \citep{Selim:2010}.
The motion of triple junctions induces a curvature in the grain boundaries,
which drives the shrinking of the embedded grains (blue and yellow), while
maintaining constant dihedral angles. 
The dihedral angles predicted by the Gaussian kernel method are
$(93^\circ,133.5^\circ,133.5^\circ)$,
while the generalized KWC model with $\epsilon=0.01$ yields 
$(106^\circ,127^\circ,127^\circ)$.

To further investigate the dependence of the triple junction angles on $\epsilon$, 
we implement the thresholding algorithm 
with $\epsilon=0.01$, $0.006$ and $0.003$ on a $3000\times3000$ grid.
As shown in \fref{subfig:Herring_b} to \fref{subfig:Herring_d}, 
as $\epsilon$ decreases, the stabilized triple junction angles converge 
to those predicted by the Herring relation. 
This test suggests that the Herring relation is satisfied in the limit $\epsilon \to 0$.

\subsection{The Von Neumann-Mullins Theory of Grain Growth }

\begin{figure}[t]
     \centering
     \begin{subfigure}[b]{0.45\textwidth}
         \centering
         \includegraphics[trim=150 50 150 50 mm, clip=true,width=\textwidth]{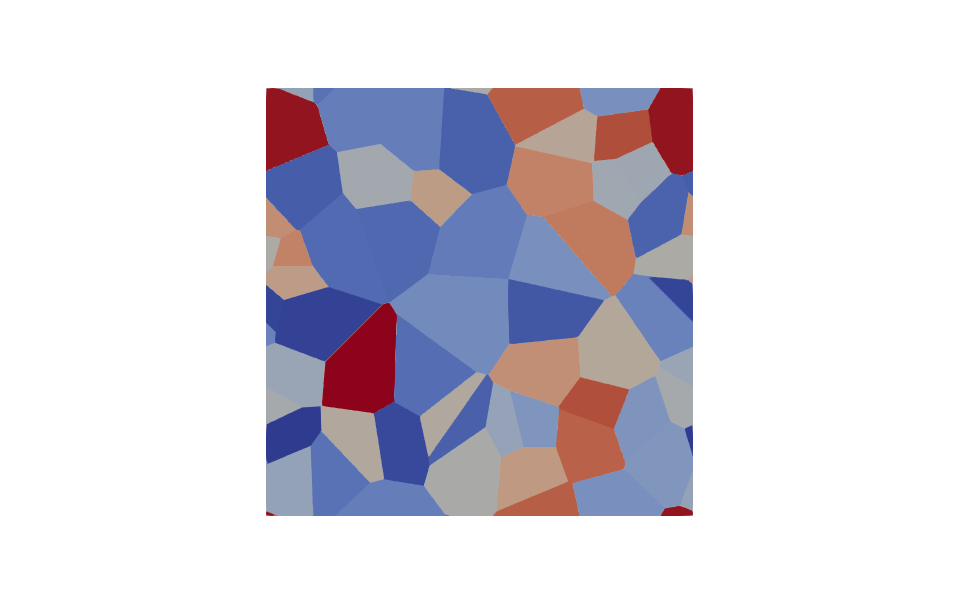}
         \caption{Initial condition}
         \label{fig:idealgrain_init}
     \end{subfigure}
     \hfill
     \begin{subfigure}[b]{0.45\textwidth}
         \centering
         \includegraphics[trim=150 50 150 50 mm, clip=true,width=\textwidth]{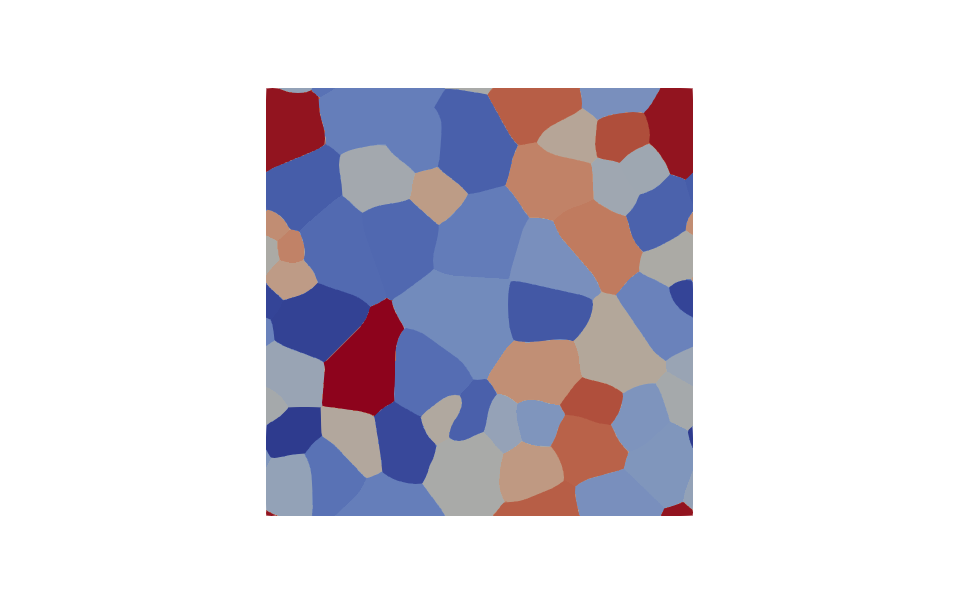}
         \caption{$t=5\times10^{-3}$}
         \label{fig:idealgrain_final}
     \end{subfigure}
     \caption{The ideal grain growth simulate by the current scheme. The constant core energy function $\mathcal J(\jump\theta) = 0.5$ results an isotropic grain boundary energy. }
     \label{fig:idealGraingrowth}
\end{figure}

\begin{figure}[t]
\begin{center}
\includegraphics[width=0.6\textwidth]{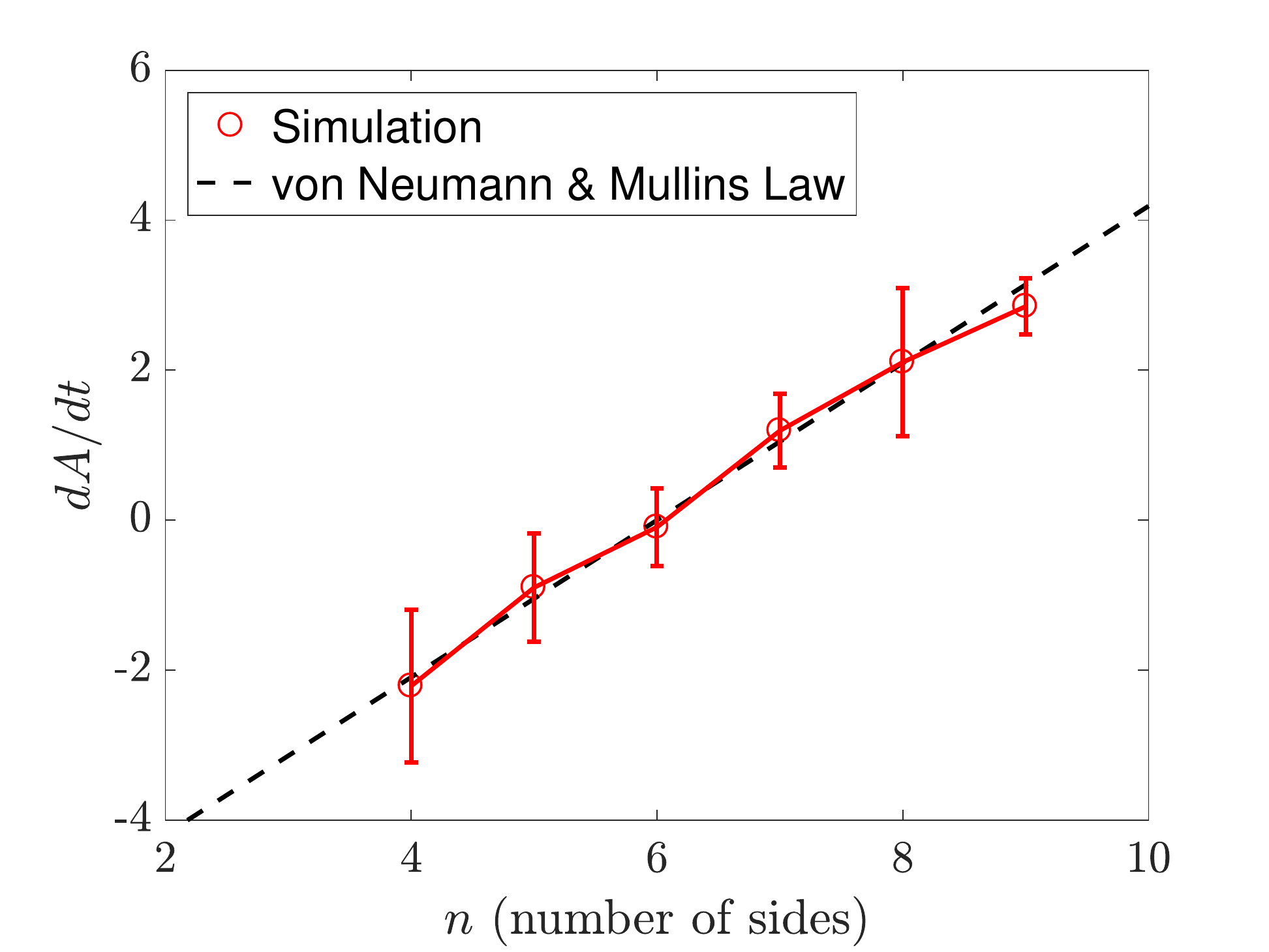}
\end{center}
\caption{
The mean and standard deviation of area change-rate for $n$-sided grain during the ideal grain growth shown in 
\fref{fig:idealGraingrowth}.
The black dashed line $\dot{A}(n)=(\pi/3)n-2\pi$ indicates the von Neumann-Mullins equation \eref{eqn:vonNeumann}
with the unit reduced mobility $m \gamma =1$. 
}
\label{fig:vonNeumannAnalysis}
\end{figure}

In this section, we validate our thresholding scheme by testing the von
Neumann--Mullins relation for a polycrystal with  uniform grain boundary
energies ($\gamma$) and mobilities ($m$). In two dimensions, the von Neumann-Mullins law
\citep{vonNeumann,Mullins} states
\begin{equation}
\frac{dA}{dt}=\frac{\pi}{3}m \gamma (n-6),
\label{eqn:vonNeumann}
\end{equation}
where $A$ is the area of a grain with $n$ sides, and $\gamma$ . In other words,
grains with more than six sides grow, while those with less than six sides will
shrink. 

To test the relation given in \eref{eqn:vonNeumann}, 
we select the core energy function as a constant ($\mathcal J(\jump\theta) =
0.5$), and consider an initial polycrystal consisting of $\mathcal N=50$ grains. 
The initial configuration, shown in \fref{fig:idealgrain_init}, is generated using a Voronoi tessellation of
uniformly distributed random points.
We implemented Algorithm~\ref{algorithm:KWC}  with parameters
$\epsilon=0.01$, $\mathsf e=10^{-6}$, and $\xi=0.05$,
on a $1024\times1024$ grid. 

As described in \sref{subsec:herring}, the evolution of the polycrystal begins with the
motion of triple junctions to attain the dihedral angles
$(120^{\circ},120^{\circ},120^{\circ})$ predicted by \eref{eqn:HerringCondition}
for constant $\gamma$. Subsequently, grain boundary motion by curvature follows.  
A snapshot of an evolving grain microstructure at $t=5\times10^{-3}$, 
simulated using our thresholding scheme, is shown in \fref{fig:idealgrain_final}.
In \fref{fig:vonNeumannAnalysis}, we plot the mean rate of change of area
of grains --- along with standard deviation --- measured during the time interval
$[2.5\times10^{-3} , 5.0\times10^{-3}]$, as a function of the number of sides.  
Noting that the plot in red is close to the theoretically predicted black dashed
line, we confirm that our thresholding scheme accurately predicts the von Neumann-Mullins law.

\subsection{Comparison with the finite element implementation of the KWC model}
\label{subsec:polycrystal}
\begin{figure}[t]
\centering
\begin{subfigure}[b]{0.36\textwidth}
\centering
\includegraphics[trim=100 70 230 60 mm, clip=true,width=\textwidth]{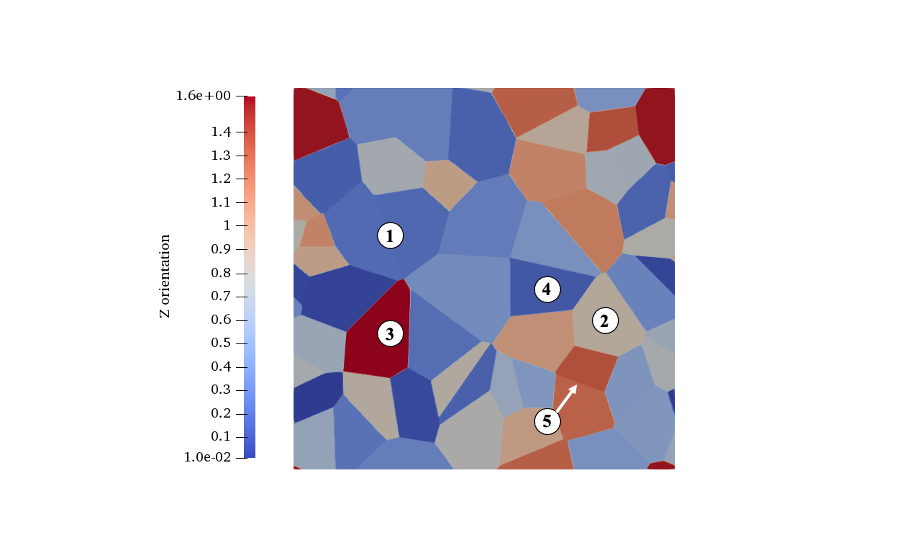}
\caption{}
\label{subfig:compare_fem_a}
\end{subfigure}
\hfill
\begin{subfigure}[b]{0.26\textwidth}
\centering
\includegraphics[trim=350 120 350 100 mm, clip=true,width=\textwidth]{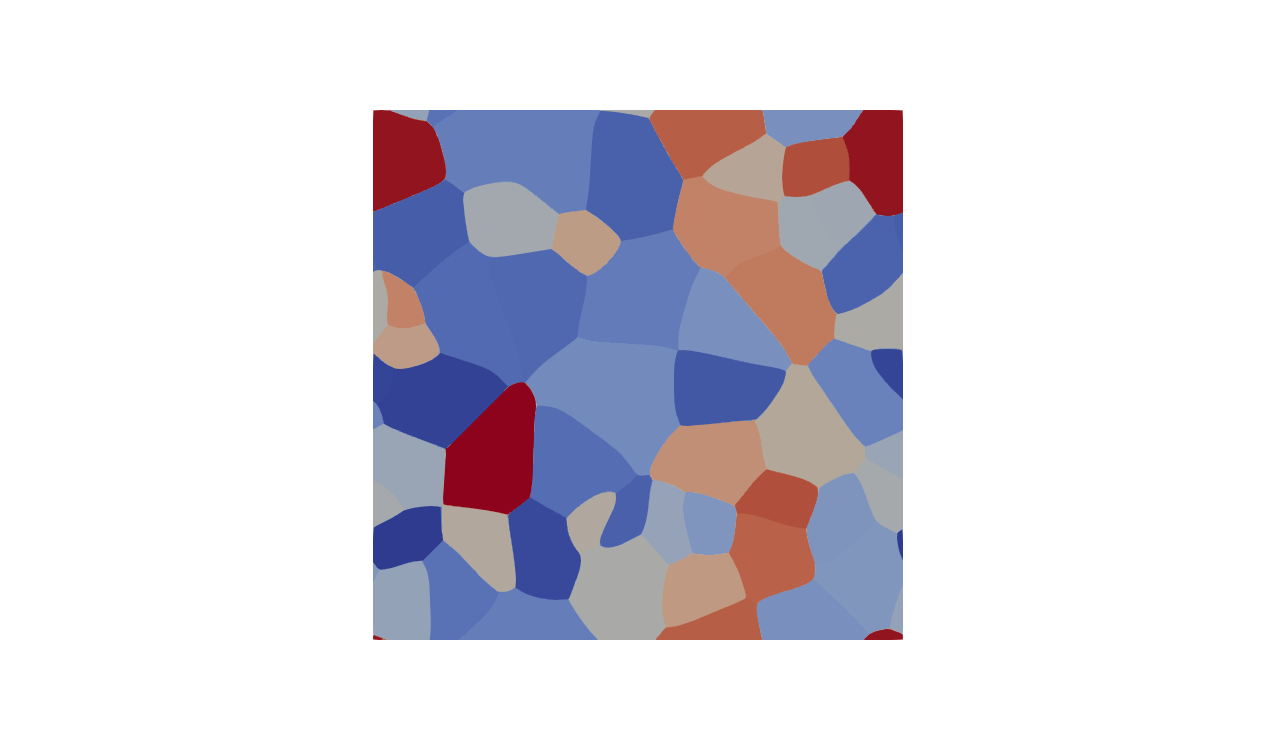}
\caption{}
\label{subfig:compare_fem_b}
\end{subfigure}
\hfill
\begin{subfigure}[b]{0.26\textwidth}
\centering
\includegraphics[trim=350 120 350 100 mm, clip=true,width=\textwidth]{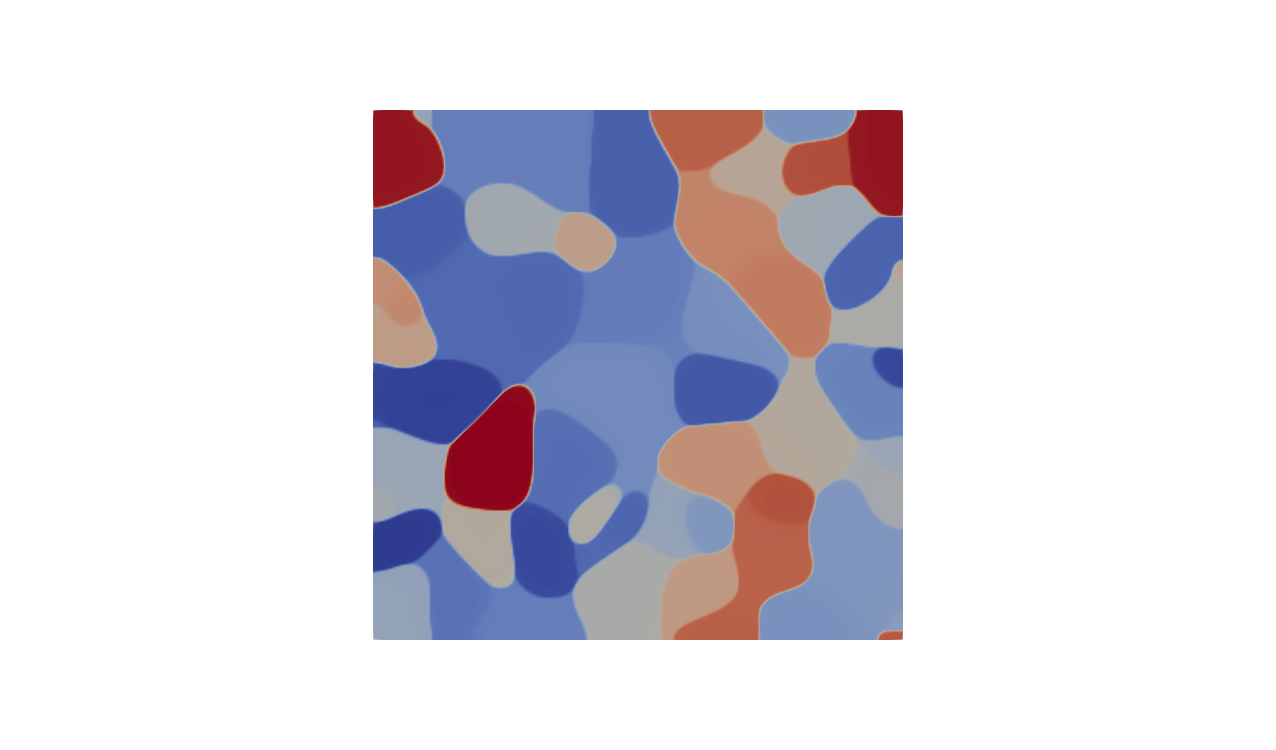}
\caption{}
\label{subfig:compare_fem_c}
\end{subfigure}
\caption{An initial polycrystal, shown in
    (\protect\subref{subfig:compare_fem_a}), is evolved using the thresholding
    and the finite element methods resulting in polycrystals shown in
    (\protect\subref{subfig:compare_fem_b}) 
and (\protect\subref{subfig:compare_fem_c}) respectively.
The two methods are consistent in predicting the growth (e.g.,
\textcircled{\scriptsize{1}}, \textcircled{\scriptsize{2}}) and shrinkage (e.g.,
\textcircled{\scriptsize{3}}, \textcircled{\scriptsize{4}}) in various grains. 
The differences in the evolution is attributed 
to the mobility function introduced in \eref{eqn:FEM_mobility} to prevent
grain rotation.}
\label{fig:compare_fem}
\end{figure}
In this section, we compare the evolutions of a polycrystal resulting form our
thresholding method and a finite element implementation of the KWC model, which
we will refer to as \emph{FE-KWC}.
An initial polycrystal consisting of $\mathcal N=50$ grains, as shown in
\fref{subfig:compare_fem_a}, is generated using a Voronoi tessellation of
uniformly distributed random points. 
The orientations of the grains are randomly chosen from the interval $[0,
\pi/2]$. The core energy is chosen to be of the Read--Shockley-type, i.e.
$\mathcal J = \jump \theta$.

The thresholding algorithm is implemented on a $1024\times1024$ grid, with parameters
$\epsilon=0.01$, $\mathsf e=10^{-6}$, and $\xi=0.05$. A snapshot of an
evolving grain microstructure at $t=5\times 10^{-3}$, simulated using our thresholding
scheme, is shown in \fref{subfig:compare_fem_b}.

\begin{figure}[t]
\centering
\begin{subfigure}[b]{0.36\textwidth}
\centering
\includegraphics[trim=100 70 230 60 mm, clip=true,width=\textwidth]{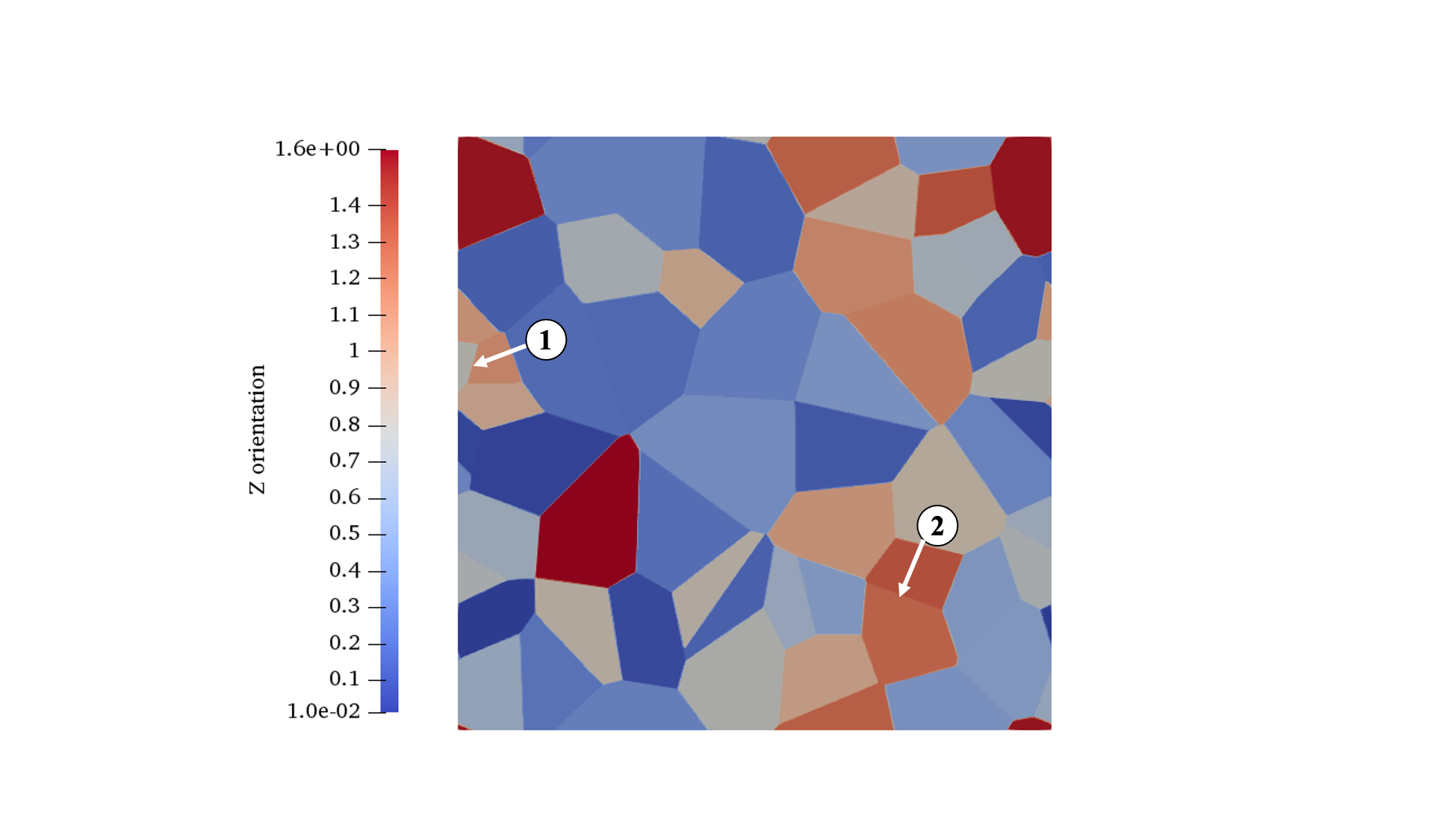}
\caption{}
\label{sub_fig:compare_rho_a}
\end{subfigure}
\hfill
\begin{subfigure}[b]{0.26\textwidth}
\centering
\includegraphics[trim=350 120 350 100 mm, clip=true,width=\textwidth]{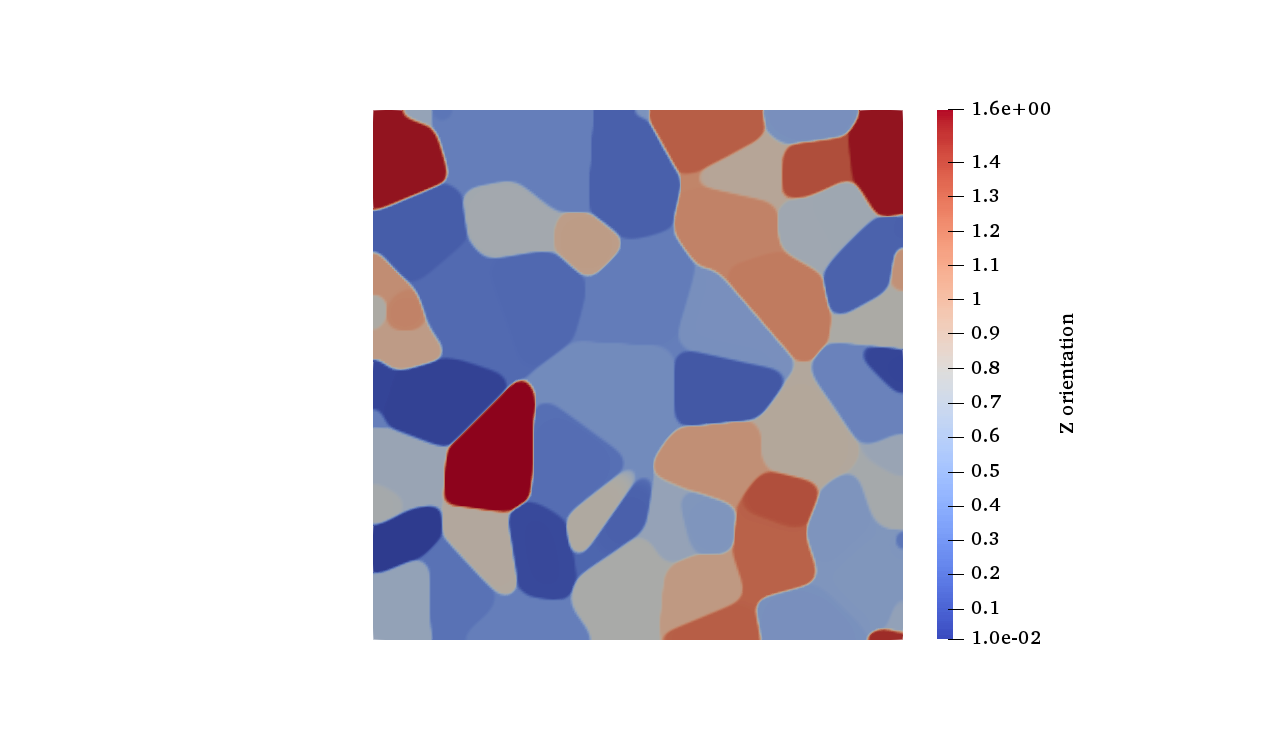}
\caption{}
\label{sub_fig:compare_rho_b}
\end{subfigure}
\hfill
\begin{subfigure}[b]{0.26\textwidth}
\centering
\includegraphics[trim=350 120 350 100 mm, clip=true,width=\textwidth]{Figures/Subfig_large_rho.png}
\caption{}
\label{sub_fig:compare_rho_c}
\end{subfigure}
\caption{An initial
polycrystal, shown in (\protect\subref{sub_fig:compare_rho_a}),
is evolved using the finite element method.
(\protect\subref{sub_fig:compare_rho_b}) and
(\protect\subref{sub_fig:compare_rho_c}) show the resulting polycrystals with regularization
parameters $\rho_0=2\times10^{-4}$ and $\rho_0=2\times10^{-3}$ respectively.
When $\rho_0$ is not sufficiently small, 
grains with small misorientation (e.g., \textcircled{\scriptsize{1}}, \textcircled{\scriptsize{2}})
blend out and grain boundaries easily become rounded.
However, the decrease in $\rho_0$ for simulating sharp interfaces, comes with significant 
computational cost contrasts to the suggested scheme. }
\label{fig:compare_rho}
\end{figure}

We note that FE-KWC, using continuous Lagrange finite elements, cannot be
carried out on our model since the solution for $\theta$ is discontinuous.
Therefore, we proceed with a finite element implementation of the regularized KWC model given
in \eref{eqn:evolution}. We use second-order quadrilateral Lagrange finite
elements to interpolate the order parameters.
Since the regularized model allows grain rotation, we inhibit
rotation using the following $\eta$-dependent mobility for $\theta$
\begin{equation}
b_{\theta}^{-1}(\eta)=10^{-5}\epsilon+ \left(1-\eta^3(10-15\eta+6\eta^2)\right)(1-10^{-5})\epsilon,
\label{eqn:FEM_mobility}
\end{equation}
as suggested by \citet{Dorr:2010}. On the other hand,
$(b_{\eta})^{-1}=\epsilon$ is chosen to be constant. To address the singularity
due to the $|\nabla\theta|$ term in \eref{eqn:theta_evolution_orig}, we use the
approximation
\begin{equation}
g(\eta)|\nabla \theta| \approx g(\eta)\sqrt{\rho_0+|\nabla \theta|^2},
\label{eqn:femRegularization}
\end{equation}
where $\rho_0=2\times10^{-3}$ is a constant. The manifestation of $\rho_0$ on
the solution will be discussed below.
FE-KWC is performed on the open-source computing
platform \texttt{Fenics}~\cite{Fenics}. 
We take an implicit time step with $dt=0.012$.
In order to compare the numerical efficiency,
we ensure that the number of degrees of freedom is the same in the thresholding and the finite
element simulations. 
The grain microstructure at $t = 1.71$, simulated
using FE-KWC, is shown in \fref{subfig:compare_fem_c}.

Comparing \frefs{subfig:compare_fem_b}{subfig:compare_fem_c}, we note that
both the methods are consistent in predicting growth (see \textcircled{\footnotesize{1}}, \textcircled{\footnotesize{2}}) 
and shrinkage (see \textcircled{\footnotesize{3}}, \textcircled{\footnotesize{4}}) 
in various grains.
It is observed that grain boundaries become rounded in the finite element
simulation,
because of the diffusive nature of orientation field.
In addition, disparities are more clear for small misorientation grain boundaries, 
e.g. \textcircled{\footnotesize{5}}, which are highly diffused. This is a
manifestation of $\rho_0$, which results in a non-zero gradient in $\theta$ in the
grain interiors. In \fref{fig:compare_rho}, we compare two finite element
simulations with $\rho_0=2\times 10^{-3}$ and $2\times 10^{-4}$, which shows
that for a smaller $\rho_0$, the grain boundaries retain their characteristic
width.\footnote{Recall that the characteristic width of a grain boundary in the
regularized KWC model is a function of $\epsilon$.} 
Thus, to simulate sharp grain interfaces comparable to the our scheme, a small 
enough $\rho_0$ is required for FE-KWC.
However, we note that the decrease in $\rho_0$ increases the stiffness of the equations, which
significantly affects the computational time as discussed below.

\begin{figure}[t]
\begin{center}
\includegraphics[width=0.55\textwidth]{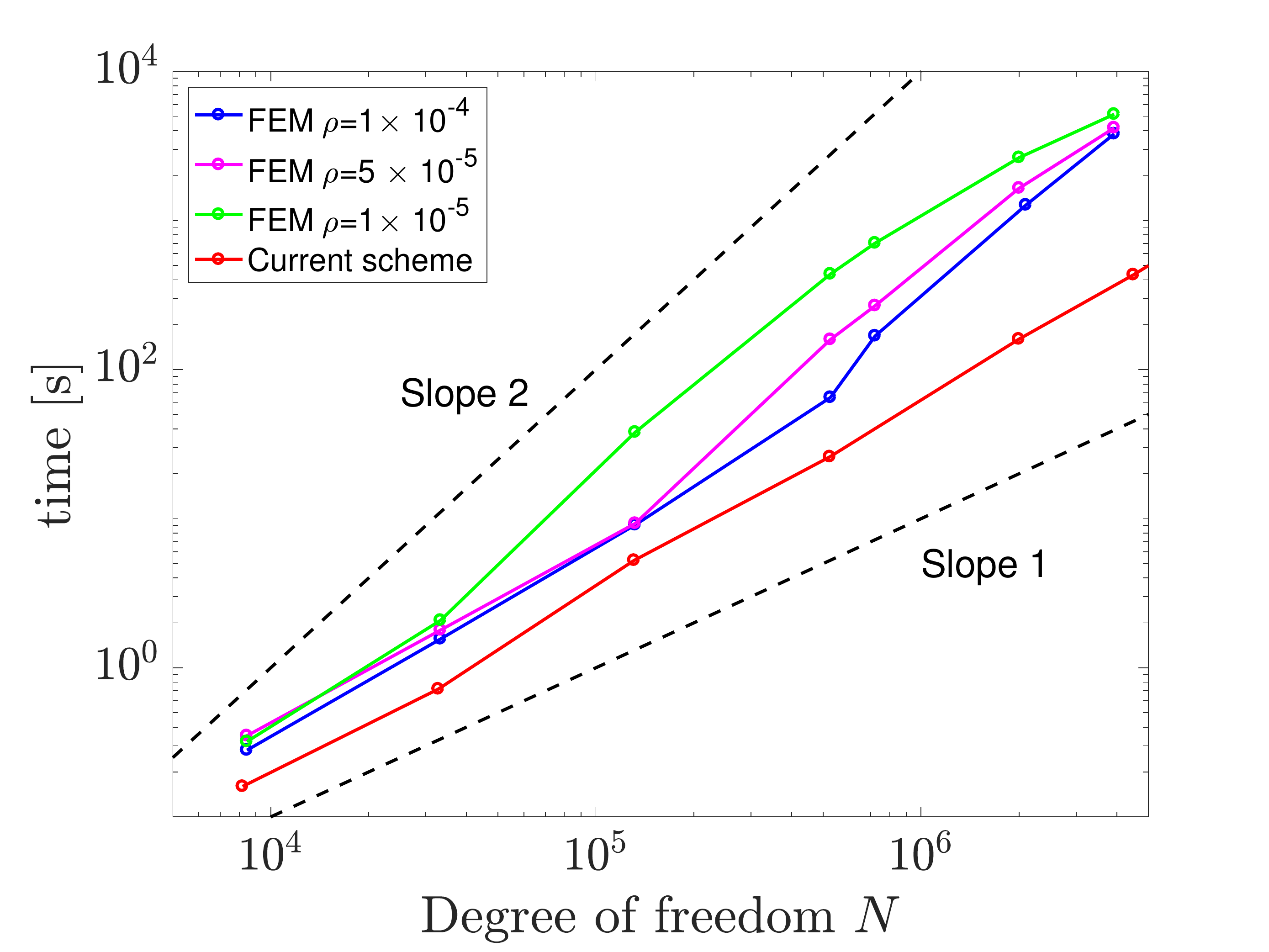}
\end{center}
\caption{A comparison of the complexity of the thresholding and the finite
    element methods. The dashed lines represent reference slopes in the log-log scale plot. 
    Slope 1 and 2 represent $\mathcal O(N)$ and $\mathcal O(\log N)$ respectively.
     While both methods have a complexity of at most $\mathcal{O}(N\log{N})$, the cost of the finite
element method depends on the choice of $\rho$.}
\label{fig:timeStudy}
\end{figure}
 
Computational time study clearly highlights the advantage of the our thresholding scheme.
Performance tasks are executed on a single 1.6 GHz core with 8 GB RAM,
and we measured the wall-clock time to complete one-full time step for the two methods. 
For our method, this includes solving for $\eta$
using the primal dual algorithm, and executing the fast marching based
thresholding algorithm to update $\theta$.
In \fref{fig:timeStudy}, we plot the dependence of the wall-clock time,
as a function of the number of degrees of freedom $N$.   
The computational complexity of the current scheme is $\mathcal{O}(N\log{N})$,
with a dominant contribution from FFT used in the primal dual algorithm to solve \eref{eqn:psi_update_form}.
On the other hand, the asymptotic computational cost of FE-KWC is estimated to
be in between $\mathcal O (N)$ and $\mathcal O (N^2)$ 
as shown in \fref{fig:timeStudy}.
The computational bottleneck of FE-KWC is in solving --- using a GMRES iterative solver~\citep{GMRES} --- a linear system of
equations formed by an $N\times N$-sized sparse matrix.
Though the asymptotic costs of the two schemes are similar in terms of $N$, we
note that the computational cost of FE-KWC also depends on the choice of the
regularization parameter $\rho_0$, which increases the stiffness of the equations in the
limit $\rho\to 0$. Therefore, as demonstrated in \fref{fig:timeStudy}, the
current scheme can be orders of magnitude faster than FE-KWC.
Both, FE-KWC and the implementation of our method, can be well-parallelized using
the current generation of graphics cards, 
which have the power, programmability and precision to implement FFT and
iterative matrix solvers \citep{GPUSOLVER,GPUFFT} respectively.

\subsection{Grain growth in an fcc copper polycrystal}
\label{subsec:fcc110}

\begin{figure}[t]
\begin{center}
\includegraphics[width=0.65\textwidth]{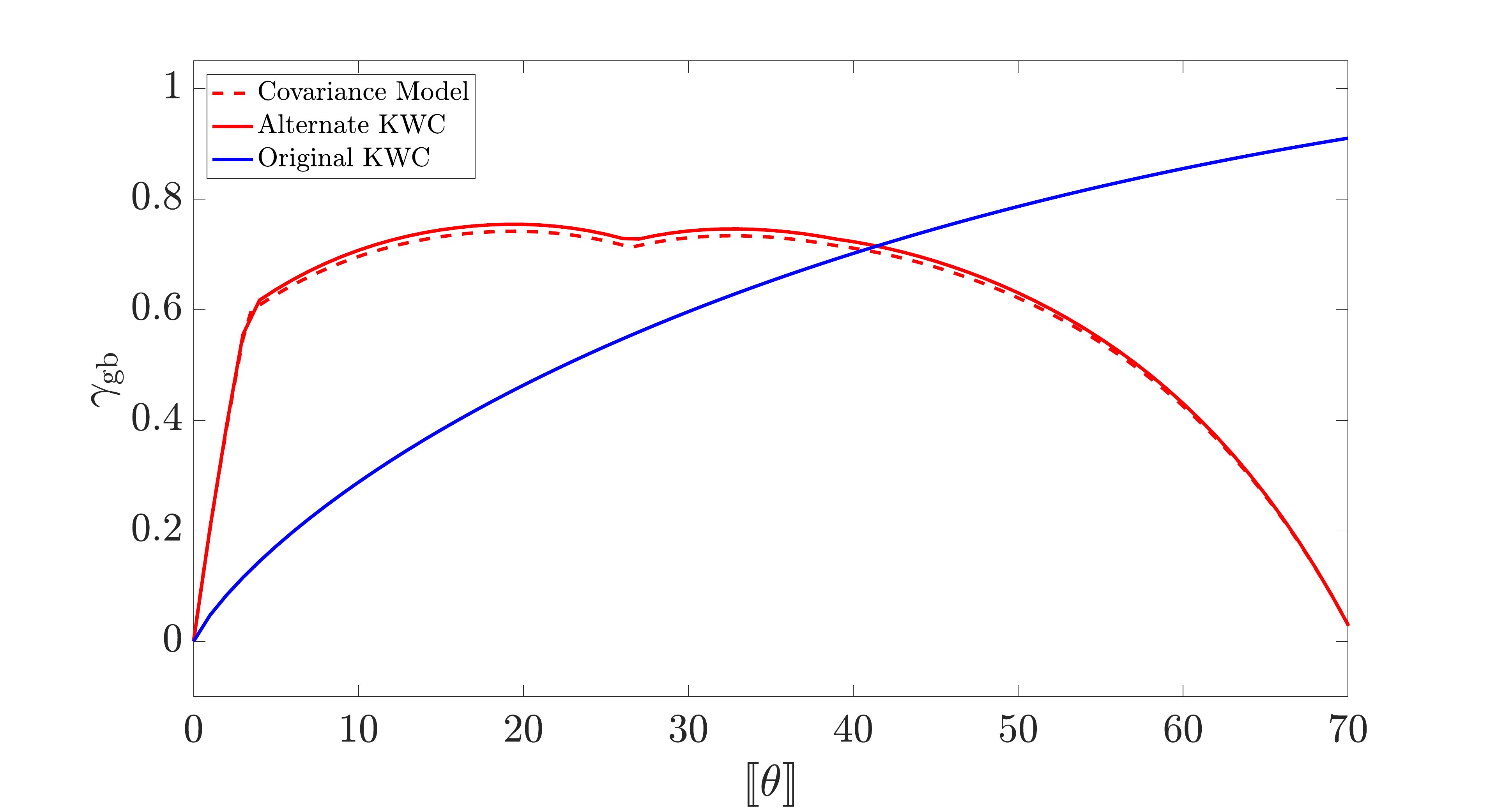}
\end{center}
\caption{
Grain boundary energies used for the polycrystal simulation in \sref{subsec:fcc110}. 
Using a core energy $\mathcal{J}([\![ \theta ]\!])$ designed in
Section~\ref{subsec:alternateKWC}, we obtain a crystal symmetry-invariant KWC
model with energy that matches the covariance model. In order to compare the
original and the new KWC models, we scale the function $g$ of the original KWC model in 
\eref{eqn:KWC_energy} to $g=-0.93\ln(1-\eta)$ such that
the averages of the grain boundary energies (with respect to misorientation) are
identical in the two models. In other words, the areas under the above plots are
equal.
}
\label{fig:simulation_110}
\end{figure}
\begin{figure}
\centering
\begin{subfigure}[b]{0.4\textwidth}
\centering
\includegraphics[trim=100 70 230 60 mm, clip=true,width=\textwidth]{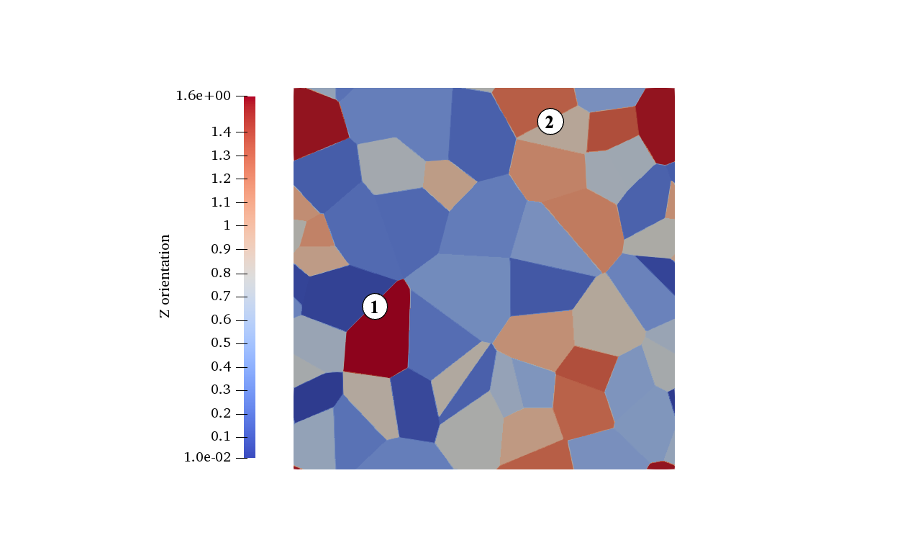}
\caption{}
\label{subfig:compare_cov_a}
\end{subfigure}
\hfill
\begin{subfigure}[b]{0.26\textwidth}
\centering
\includegraphics[trim=350 120 350 100 mm, clip=true,width=\textwidth]{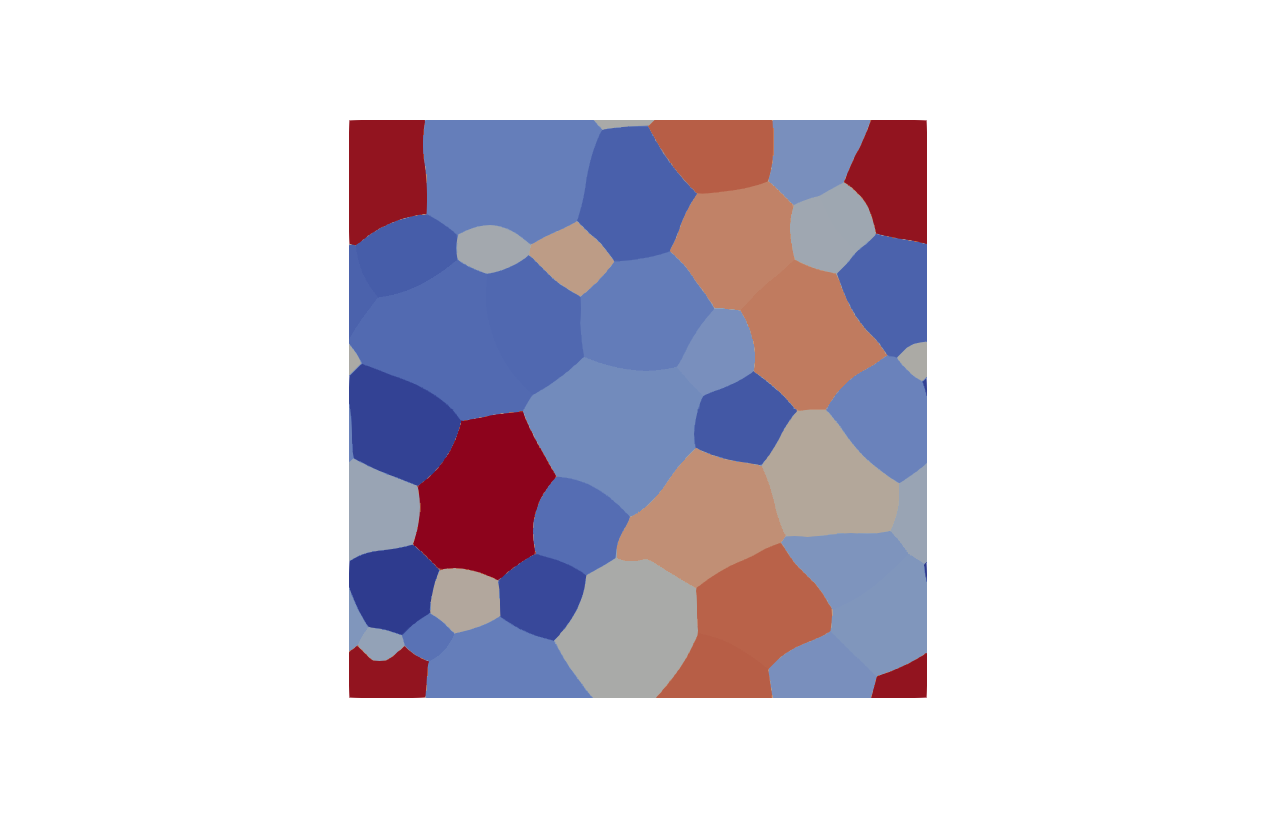}
\caption{}
\label{subfig:compare_cov_b}
\end{subfigure}
\hfill
\begin{subfigure}[b]{0.26\textwidth}
\centering
\includegraphics[trim=350 120 350 100 mm, clip=true,width=\textwidth]{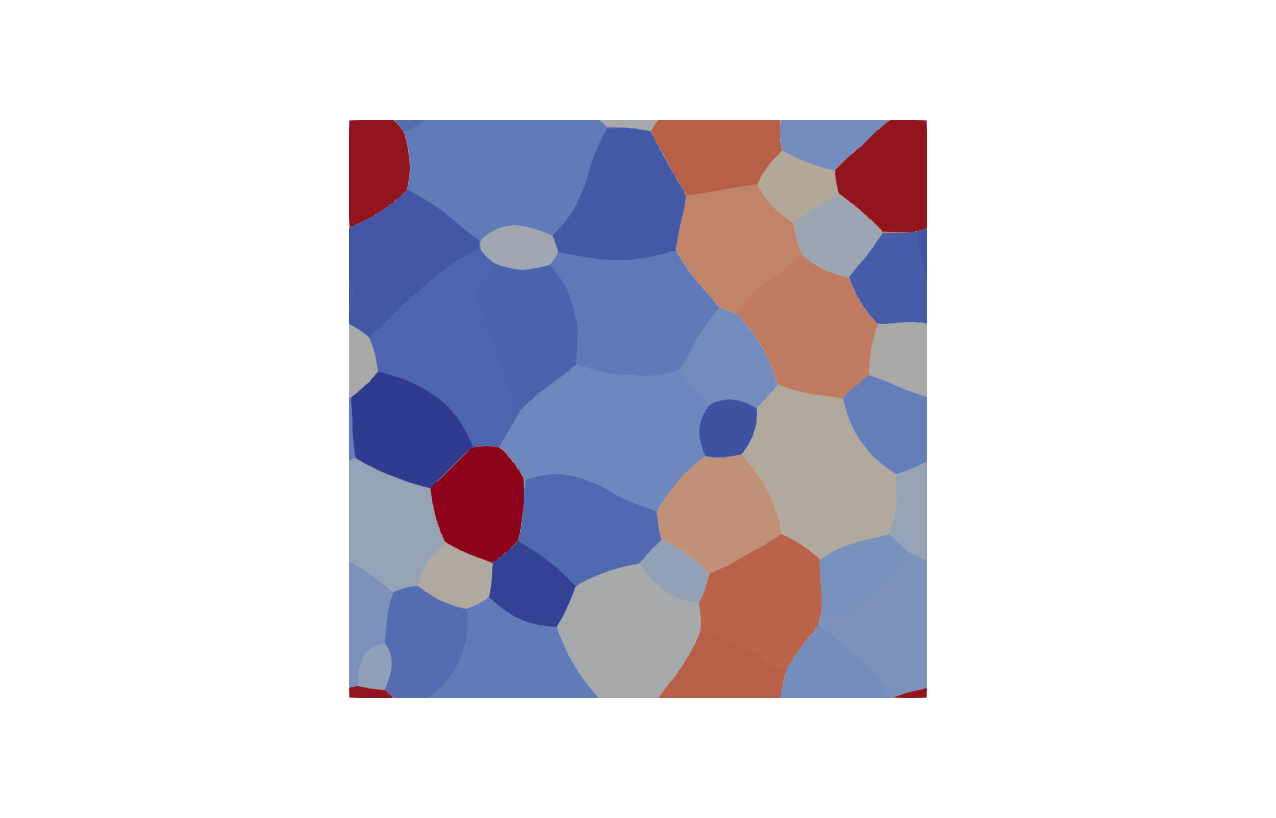}
\caption{}
\label{subfig:compare_cov_c}
\end{subfigure}
\caption{(\protect\subref{subfig:compare_cov_a}) A polycrystal with $\mathcal N=50$ grains, and an initial
    orientation distribution. (\protect\subref{subfig:compare_cov_b}) and
    (\protect\subref{subfig:compare_cov_c}) show evolved polycrystals 
    using the new and the original KWC models respectively. Grains $1$ and $2$ show
    opposite growth/shrinkage trends in the two models due to the deviation of
the grain boundary energy from the Read--Shockley-type in the new formulation.
The blue and red colors
represent the maximum and minimum orientation angles of $0^{\circ}$ and
$70.6^{\circ}$ respectively.}
\label{fig:compare_cov}
\end{figure}

\begin{figure}
\centering
\begin{subfigure}[t]{0.45\textwidth}
\centering
\includegraphics[trim=80 50 50 85 mm, clip=true,width=\textwidth]{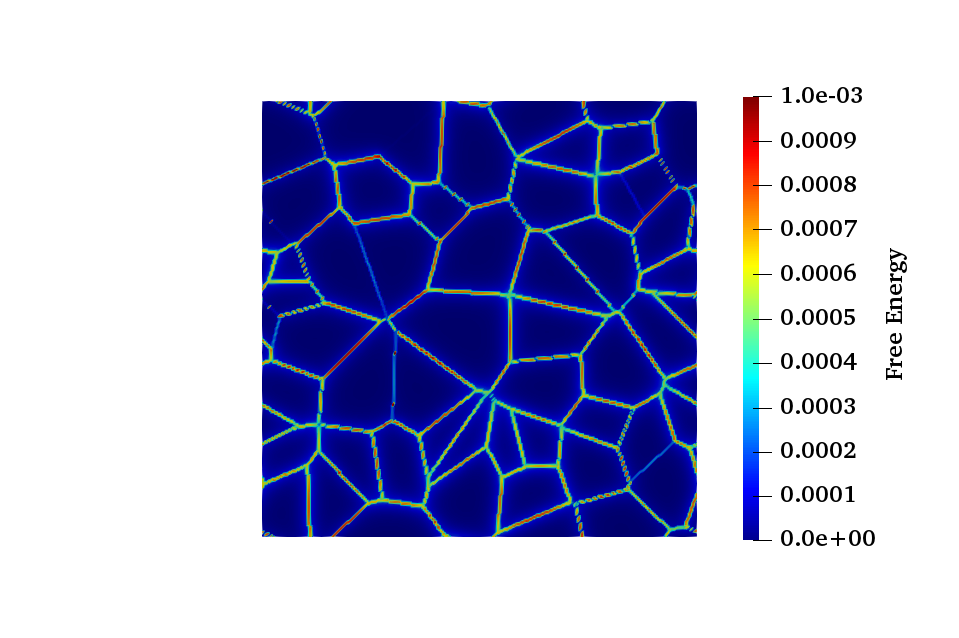}
\caption{}
\label{subfig:Cov_poly_freeEnergy}
\end{subfigure}
\hfill
\begin{subfigure}[t]{0.45\textwidth}
\centering
\includegraphics[trim=80 50 50 85 mm clip=true,width=\textwidth]{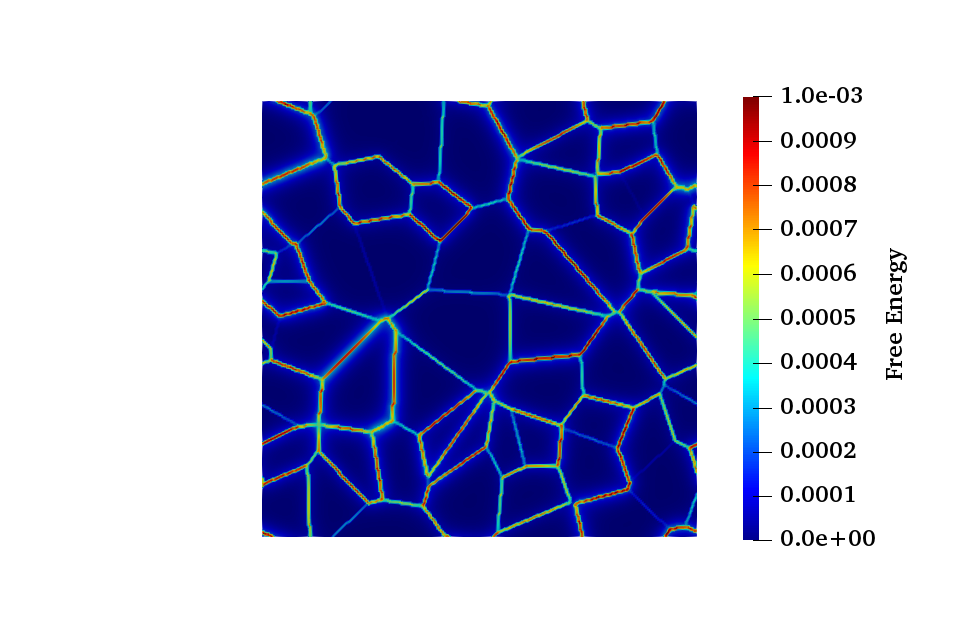}
\caption{}
\label{subfig:Orig_poly_freeEnergy}
\end{subfigure}
\caption{Initial distributions of grain boundary energies for the
    (\protect\subref{subfig:Cov_poly_freeEnergy}) generalized, and 
(\protect\subref{subfig:Orig_poly_freeEnergy}) the original KWC models. 
The grain boundary energy in the original KWC model is of the
Read--Shockley-type. On the other hand, the grain boundary energy in the
generalized KWC model reflects the crystal symmetry of copper.}
\label{fig:compare_cov_energy}
\end{figure}

In this section, we examine grain growth in a two-dimensional fcc copper
polycrystal with $[110]$-type grain boundaries simulated using the generalized
KWC model,\footnote{The $[110]$ direction of each grain is is aligned with the
$z$-axis (out of the plane).} with crystal symmetry-invariant grain boundary
energy. We compare the results with the predictions of the original
KWC model.

A two-dimensional polycrystal consisting of $\mathcal N=50$ grains,
with orientations in the range $[0, 70.6^{\circ}]$ is
generated using a Voronoi tessellation of random points. 
\fref{fig:compare_cov} shows
the initial orientation distribution in the polycrystal. 
We assume that the grain boundary energy density is
independent of inclination. We use the core energy $\mathcal J(\jump\theta)$ 
constructed in \sref{subsec:alternateKWC} (see \fref{fig:cu_110_covariance}). In
order to compare the generalized model to the
original KWC model, we scale the function $g$ of the original KWC model in 
\eref{eqn:KWC_energy} to $g=-0.93\ln(1-\eta)$ such that
the mean of the grain boundary energies
as functions of misorientation in the range $[0,70.6^\circ]$ are identical for
the two models. \fref{fig:simulation_110} shows a comparison of the grain
boundary energy densities of the two models.

The orientation
distributions of the polycrystal at the end of 200 time steps for the generalized and the
original KWC models are shown in \frefs{subfig:compare_cov_b}{subfig:compare_cov_c} respectively. Comparing the resulting polycrystals with the initial polycrystal
in \fref{subfig:compare_cov_a}, we note that the generalized KWC model predicts a growth for red grains while
the original model results in their shrinkage. 
This can be attributed to the difference in the grain boundary energies
of the two models, as shown in \fref{fig:compare_cov_energy}.
For example, the grain boundary  \textcircled{\footnotesize{1}},
which has a misorientation of $\approx 70.6^\circ$ has a relatively smaller energy 
in the generalized model due to crystal symmetry. 

On the other hand, we note an opposite trend for light blue grains for which the
generalized model predicts shrinkage while the original model results in a
growth. This is a result of relatively larger 
energy of grain boundary \textcircled{\footnotesize{2}} in the generalized
model compared to the original model. The above observations suggest that 
the generalized model can result in the growth of certain grains with
large misorientation, highlighting the importance of crystallography in grain
growth.

\section{Conclusion and future work}

In this work, we generalized the two-dimensional KWC model for grain boundaries such that it
can incorporate arbitrary misorientation-dependent grain boundary energies that
respect the bicrystallography of grain boundaries. In addition, we address the
computational challenge of solving a singular diffusive equation of the KWC
model by developing an $\mathcal O(N \log N)$
thresholding algorithm. Below, we summarize the construction of our model, its
implementation, and research directions for future work.

First, we eliminate the 
$|\nabla \theta|^2$ term in the original KWC model, which is responsible for regularizing the
orientation order parameter, and rendering a non-zero mobility to
the grain boundaries. The lack of a regularizing term results in a discontinuous
orientation order parameter with jump across the grain boundaries, and grain
boundaries with no mobility. In the presence of a piecewise-constant orientation
order parameter, the modified KWC functional can be separated into a
bulk contribution that depends on $\eta$, and a surface contribution, called the
core energy, that depends linearly on the jump $\jump\theta$ across the grain
boundaries. Next, we show that by generalizing the core energy from a linear
function to an arbitrary function $\mathcal J$ of $\jump\theta$, the 
model can incorporate arbitrary dependence of grain boundary energies on
misorientation angles.

Since the absence of the regularizing $|\nabla \theta|^2$ term renders the grain
boundaries immobile, we design an $\mathcal{O}(N\log N)$ thresholding algorithm
to evolve grain boundaries by curvature, where $N$ is the number of grid points.
The algorithm, which employs a primal-dual and the fast marching methods, is
shown to be an order of magnitude faster than the finite element implementation of the
original KWC model. We validate our implementation by predicting the Herring angle
relation, and simulate a two-dimensional polycrystal consisting of $[110]$ tilt grain boundaries.
The computational efficiency and flexibility of our approach opens 
the door to a number of exiting directions for future work. 
\begin{itemize}
\item The present framework will enable us to carry out a statistical study of
    large scale simulations of various ensembles of polycrystals to characterize
    abnormal grain growth in terms of the grain boundary energy landscape and
    crystal symmetry.

\item While arbitrary grain boundary energies can be incorporated into our model,
    its implementation is restricted to grain boundary mobility equal to the
    inverse of the energy. An extension of our algorithm to include mobilities
    independently will be explored in a future work.
    
\item The present algorithm does not allow grain rotation, which is
   another important phenomenon during recrystallization of polycrystalline materials.\footnote{We note that grain rotation
may sometimes play an important role during the transition from recovery to
continuous dynamic recrystallization. Dislocations agglomerate and form cell
walls/subgrains at the end of the recovery stage. In a phenomenon, commonly referred to as 
\emph{subgrain rotation recrystallization}, few subgrains --- aided by
bulk dislocations -- increase their misorientation and transform to
grains/nuclei which grow \citep{li1962possibility}. From this perspective, grain rotation plays an
important role during the nucleation of recrystallized grains.} We plan to augment the current scheme with a step that models grain rotation. 
\item A recent work by \citet{Admal:2019} extended the two-dimensional KWC model
    to a three-dimensional fully anisotropic (both misorientation and inclination dependent) model, 
    wherein the dependence of grain boundary energy on the misorientation angle was restrictive to a
    Read--Shockley-type.
    Due to the high computational cost of the finite element method,
    the implementation of the three-dimensional model was restricted to simple
    bicrystals.  It is envisaged that the efficiency of our thresholding algorithm will enable
    us to explore large three-dimensional polycrystals with fully anisotropic grain boundary
    energy.
\item Finally, we recall from the introduction that surface tension is not the
    only dominant driving force on a grain boundary due to grain boundary
    plasticity. Adapting our thresholding algorithm into existing unified frameworks
    \citep{Admal:2018}, wherein grain microstructure and deformation evolve
    contemporaneously, will enable us to quantify the role of grain boundary
    plasticity, and study phenomena such as dynamic recrystallization,
    superplasticity and severe plastic deformation \citep{Srolovitz:2017,Srolovitz:2020,Runnels:2020}. 
\end{itemize}

\appendix

\section*{Data availability}
A \texttt{C++} template library that implements
Algorithm~\ref{algorithm:KWC} is available at 
\url{ https://github.com/admal-research-group/GBthresholding}.
\section*{CRediT author statement}
\textbf{Jaekwang Kim:} Formal analysis, Investigation, Software, Validation,
Writing-Original draft, Visualization.
\textbf{Matt Jacobs:} Conceptualization, Methodology, Software, Formal analysis.
\textbf{Stanley Osher:} Conceptualization.
\textbf{Nikhil Admal:} Conceptualization, Methodology, Writing-Review and
Editing, Resources, Data Curation, Supervision, Project Management.

\section{Results on the 1D KWC model}
\label{app:analytic}
In this section, we collect results on the
one-dimensional KWC model which describes an infinite bicrystal
with a grain boundary at the origin. In particular, we present the derivation of
the steady-state
analytical solution under Dirichlet boundary conditions, and the resulting grain
boundary energy as a function of misorientation.

Consider the following KWC energy functional without the $|\nabla\theta|^2$
regularizing term
\begin{equation}
    \mathcal W[\eta,\theta] = \int_{-\infty}^\infty \left [
        \frac{\epsilon }{2}|\nabla \eta|^2 + \frac{(1-\eta)^2}{2\epsilon} + g(\eta) |\nabla \theta| 
    \right ] \, dV.
    \label{eqn:energy}
\end{equation}
The Euler--Lagrange equation associated with the above functional is
\begin{equation}
    \label{eqn:el}
        \epsilon \triangle \eta - \frac{\eta-1}{\epsilon} - g,_\eta
        |\nabla \theta| = 0,
\end{equation}
where $g,_\eta$ is used to denote $\partial g/\partial \eta$.
In what follows, we derive a steady-state solution of \eref{eqn:el} under
Dirichlet boundary conditions
\begin{equation}
    \eta(\pm \infty) = 1, \quad \theta(\infty) = -\theta(-\infty) = \theta/2.
    \label{eqn:bc}
\end{equation}

We begin with the ansatz that $\theta(x)$ is a step function satisfying
\eref{eqn:bc} with a discontinuity at the origin. 
Multiplying \eref{eqn:el} by $\eta'$, and integrating with respect to
$x$ in a region away from the origin, we obtain
\begin{equation}
    \frac{\epsilon}{2} \eta,^2_x - \frac{(1-\eta)^2}{2\epsilon} = 0,
    \implies \eta,_x = \pm \frac{(1-\eta)}{\epsilon},
    \label{eqn:etax}
\end{equation}
On the other hand, multiplying \eref{eqn:el} with $\eta'$, and integrating
over an arbitrarily small neighborhood of $0$ results in the jump condition
\begin{equation}
    \epsilon  \jump{\eta,_x} = g,_\eta(\bphi) \jump\theta,
    \label{eqn:jump_etax}
\end{equation}
where $\bphi:=\eta(0)$ is the value of $\eta$ at the grain boundary.
From \eref{eqn:etax} and \eref{eqn:jump_etax}, it follows that
\begin{equation}
    \epsilon \eta,_x = 
    \begin{cases}
        1-\eta & \text{ if }x>0,\\
        -(1-\eta) & \text{ otherwise},\\
    \end{cases}
    \label{eqn:etax_cases}
\end{equation}
and
\begin{equation}
    g,_\eta(\bphi) \jump\theta = 2 (1-\eta),
    \label{eqn:app_jump_condition}
\end{equation}
which relates $\bphi$ to $\jump\theta$. 
The analytical solution for $\eta$ can be obtained 
by integrating \eref{eqn:etax}. 
With our choice of $g = -\ln(1-\eta)$,  
the result can be explicitly written as
a function of misorientation $\jump \theta$: 
\begin{equation}
    \int_{\bphi}^\eta\frac{\epsilon}{1-\eta }d\eta = x,
    \implies \eta(x) = 1-\sqrt{\frac{\jump\theta}{2}} \exp
    \left(-\frac{|x|}{\epsilon}\right).
    \label{eqn:eta_analytical}
\end{equation}
The grain boundary energy $\gamma$ as a function of misorientation
is calculated by evaluating $\mathcal W[\eta,\theta]$
using the steady state solution for $\eta$ derived above. From
\eref{eqn:etax}, we have 
\begin{align}
    \gamma(\jump\theta) = \mathcal W[\eta,\theta] &= 
    \int_{-\infty}^\infty \left [ \frac{\epsilon}{2} \eta_{,x}^2 +
    \frac{(1-\eta)^2}{2 \epsilon} \right] \, dx 
    +  g(\bphi) \jump\theta \notag \\
    &= 2 \int_0^\infty \frac{(1-\eta)^2}{\epsilon} \, dx  + g(\bphi)\jump\theta \notag \\
    &= 2 \int_{\bphi}^1 (1-\eta) \, d\eta+
    g(\bphi)\jump\theta = (1-\bphi)^2 + g(\bphi)\jump\theta.
    \label{eqn:gbenergy_appendix}
\end{align}
Note that the grain boundary energy $\gamma$ and $\bphi$ are independent of
$\epsilon$, which reinforces that the model converges to its sharp interface as
$\epsilon \to 0$ while the energy remains unchanged.
Again specializing the analytical expression $\gamma$ with the choice of 
the logarithmic $g$, we obtain 
\begin{align}
    \gamma(\jump\theta) &= (1-\bphi)^2 - \jump\theta \ln(1-\bphi) \notag \\
    &= \frac{\jump\theta}{2}- \jump\theta
    \ln\left(\sqrt{\frac{\jump\theta}{2}}\right).
    \label{eqn:energy_sol}
\end{align}

\section{The covariance model of grain boundary energy}
\label{app:covariance_model}
The covariance model for grain boundary energy, developed by
\citet{Runnels:2016_1,Runnels:2016_2}, estimates grain boundary energy
using the covariance of atomic densities of the two lattices adjoining a grain boundary. 

In the covariance model, a lattice density
measure $\bar \rho$ for a given lattice\footnote{A lattice $\mathcal L$ is defined using
    three lattice vectors $\bm l_1$, $\bm l_2$, and $\bm l_3$ as
    \begin{align}
        \mathcal L = \{n_1 \bm l_1 + n_2 \bm l_2 + n_3 \bm l_3\: n_i \in \mathbb
        Z\}.
        \label{eqn:lattice}
    \end{align}
}
$\mathcal L$, defined as an infinite sum
of Dirac measures with support at the lattice points points of $\mathcal L$:
\begin{equation}
\bar\rho(\bm{x}) = \sum_{\bm{d} \in \mathcal{L}} \delta(\bm{x}- \bm{d}).
\end{equation}
A lattice density field $\rho$ is introduced as the convolution of $\bar \rho$ with
a thermalization function $\xi$, i.e. 
\begin{equation}
\rho(\bm x) = \overline{\rho}(\bm x)* \xi(\bm x),
\end{equation}
where
\begin{equation}
\xi(\bm x)=\frac{1}{\sigma^3 \pi^{3/2}} e^{-\|\bm{x}\|^2/\sigma^2},
\end{equation}
with $\sigma^2$ as the dimensionless temperature. 
The planar covariance of two thermalized lattices 
$\mathcal{L}_A$ and $\mathcal{L}_B$ with their respective density fields
$\rho_A$ and $\rho_B$, measured on $\mathbb R^2$,  is defined as 
\begin{equation}
c[\rho_A,\rho_B]=
\int_{\bm{y} \in \mathbb{R}^2}
\rho_A(P^{T}\bm{y})
\rho_B(P^{T}\bm{y}) \lambda(\bm{y})\; dA,
\label{app:eqn:covariance1}
\end{equation}
where $\lambda(\bm{x})$ is an appropriately chosen window function (see
\eref{eqn:window}, $P: \mathbb{R}^3 \to \mathbb{R}^2$ is the projection 
\begin{equation}
 P=\left(\begin{array}{@{}ccc@{}}
    1 & 0 & 0 \\
    0 & 1 & 0 \\
  \end{array}\right)
\end{equation} 
on to the plane $\mathbb R^2$. Expressing the functions $\rho_A$ and $\rho_B$ in
Fourier series, the integral in \eref{app:eqn:covariance1} simplifies as
\begin{equation}
c[\rho_{A},\rho_{B}]
=\frac{1}{\hat{\lambda} (\bm 0)}\sum_{\bm{k}_{A} \in \mathcal{L}'_A} 
\sum_{\bm{k}_{B} \in \mathcal{L}'_B}
\widehat{ \rho} (\bm{k}_A) 
\widehat{\rho}^{\; *}(\bm{k}_B)
\hat{\lambda}\left(
P(\bm{k}_B-\bm{k}_A)\right),
\end{equation}
where $\bm k_A$ and $\bm k_B$ are lattice vectors of the dual lattices $\mathcal
L'_A$ and $\mathcal L'_B$, and the window function is defined 
in terms of its Fourier transform as
\begin{equation}
\hat{\lambda}(\bm{k})=e^{-\|\bm{k}\|/\omega},
\label{eqn:window}
\end{equation}
with an adjustable parameter $\omega$.
The grain boundary energy in the covariance model is defined as 
\begin{equation}
\gamma^{\mathrm{cov}}=
E_0\left( 1- \frac{c[\rho^\mathcal{A},\rho^\mathcal{B}]}{c_{\mathrm{gs}}}\right),
\label{eqn:gamma_cov}
\end{equation}
where $c_{\mathrm{gs}}$ is the ground state covariance defined as the
supremum, over all planes, of $c[\rho_A,\rho_A]$. For example,
in fcc, $c_{\mathrm{gs}}$ corresponds to covariance measured with respect to the
$[111]$ plane. Finally, we note that 
the covariance model has three adjustable parameters $\{ E_0, \sigma,\omega \}$
that can be used to fit $\gamma^{\rm{cov}}$ to data from experiments
or molecular dynamics simulations. It is known that while \eref{eqn:gamma_cov} is a good indicator of grain
boundary energy, it over-predicts the energy for low angle grain
boundaries as the above model does not account for facet formation. \cite{Runnels:2016_1,Runnels:2016_2} have
shown that a further relaxation of the grain boundary energy, which signifies the formation of
facets, yields necessary corrections to the energy predicted by the model.

\begin{figure}[t]
\begin{center}
\includegraphics[width=0.75\textwidth]{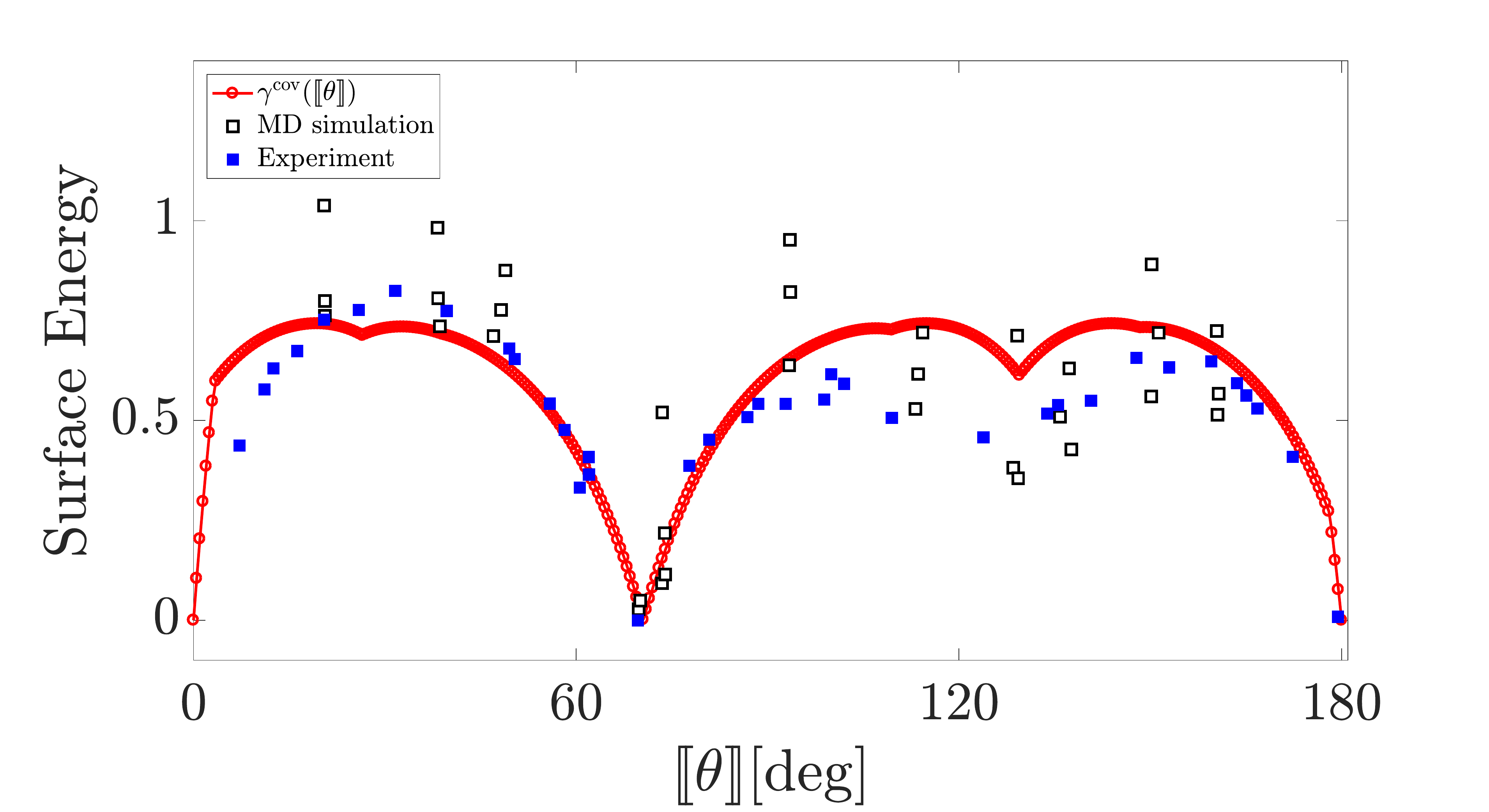}
\end{center}
\caption{A plot of the normalized grain boundary energy versus the
    misorientation angle predicted by the
    covariance model for a $[110]$ symmetric-tilt grain boundary in fcc copper,
    computed using the relaxation algorithm of \citet{Runnels:2016_1,Runnels:2016_2}. 
    For comparison, grain boundary energies obtained from experiment
    \citep{Miura}, and MD simulations \citep{Wolf:1990} are shown in blue and
    square points respectively.
}
\label{appfig:cu_110_covariance}
\end{figure}

\fref{appfig:cu_110_covariance} shows a plot of
a relaxed $\gamma^{\rm{cov}}$ computed for $[110]$ symmetric-tilt
grain boundaries in fcc copper using
$E_0=1.45 \,\mathrm{J}/\mathrm{m}^{-2}$, $\omega=0.5$, and $\sigma/a_0=0.175$,
where $a_0=3.597$ is the lattice constant of copper.
From \fref{appfig:cu_110_covariance}, it is clear that the 
grain boundary energy predicted by the covariance model
is in good overall agreement with data from molecular dynamics simulations \citep{Wolf:1990,Miura}.

\section{The primal-dual method}
\label{app:primaldual}

Primal-dual methods is a part of a class of first-order algorithms\footnote{
An algorithm that only requires the calculation of the gradient of a
functional.}
that have a long history in the context of optimization
problems \citep{addPrimaldual0,addPrimaldual1,addPrimaldual2}. 
As the name suggests, primal-dual methods proceed by concurrently solving 
a primal problem  and a dual problem.
The main benefit of primal-dual splitting is that 
it replaces an original hard problem with a set of two easy sub-problems
(primal- and dual-).
Because of this advantage, the method has been widely used in diverse fields
including compressed sensing, image processing, signal processing, 
and machine learning \citep{addPrimalDualExample0,Chambolle:2011,addPrimalDualExample1,addPrimalDualExample2}.

The motivation to use a primal-dual algorithm to solve the minimization problem
in \eref{eqn:eta_sub} for $\eta$ arises due to the presence of a highly
nonlinear term $g(\eta)\mathcal J(\jump\theta)$ along with $|\nabla \eta|^2$.
Therefore, we adopt a primal-dual method
by introducing an auxiliary dual variable
which enables us to cast \eref{eqn:eta_sub} as an equivalent optimization
problem. The choice of the dual variable is based on the observation that 
\begin{align}
    \frac{\epsilon}{2} \|\nabla \eta\|_{L^2(\Omega)}^2 &=
\epsilon \|\nabla \eta\|_{L^2(\Omega)}^2 -
\frac{\epsilon}{2} \|\nabla \eta\|_{L^2(\Omega)}^2 \notag \\
&= -\epsilon\int_\Omega \eta \triangle \eta \, dV - 
\int_\Omega \frac{\epsilon}{2}\nabla \eta \cdot \nabla \eta\, dV,
\end{align}
where we have used the divergence theorem, and the Neumann boundary condition
$\nabla \eta \cdot \bm n = \bm 0$.
Introducing an auxiliary variable $\psi$, and identifying it with $-\epsilon \triangle \eta$, we have
\begin{align}
\frac{\epsilon}{2} \|\nabla \eta\|^2 
&=\sup\limits_{\psi \in (\dot H^1(\Omega))^*} \left [
    \int_{\Omega} 
    \eta(x) \psi(x) \, dV - 
    \frac{1}{2\epsilon}\| \Delta^{-1} \nabla \psi \|_{L^2(\Omega)}^2
\right ]\notag \\
&=\sup\limits_{\psi \in (\dot H^1(\Omega))^*} \left [
    \int_{\Omega} 
    \eta(x) \psi(x) \, dV - 
    \frac{1}{2\epsilon}\|\psi\|_{(\dot H^1(\Omega))^*}^2
\right ],
\label{eqn:sup}
\end{align}
where $\dot H^1(\Omega)$ denotes the set of all functions in $H^1(\Omega)$ with
zero average, and  $(\dot H^1(\Omega))^*$ is its dual.
Substituting \eref{eqn:sup} into the KWC functional $\mathcal W^{\rm G}$, the minimization problem in \eref{eqn:eta_sub} transforms to the following saddle point problem:
\begin{equation}
    \inf_{\eta \in L^2(\Omega)} \, \sup_{\psi \in (\dot H^1(\Omega))^*} \Phi[\eta,\psi],
    \label{eqn:saddle}
\end{equation}
where
\begin{equation}
\Phi[\eta,\psi]=
-\frac{1}{2\epsilon}\|\psi\|^2_{(\dot H^1(\Omega))^*} + 
\int_{\Omega} (\eta \psi +f(\eta)) \, dV + 
\int_\mathcal S g(\eta) \mathcal{J}(\jump\theta) \, dS.
\label{eqn:PHI}
\end{equation}
The problems of minimizing $\Phi$ with respect to $\eta$, and maximizing it with respect to
$\psi$ are referred to as $\eta$ and $\psi$ sub-problems respectively. 
The advantage of using a primal-dual algorithm 
is evident from the observation that $\Psi$ \emph{does not} depend on
the gradients of $\eta$, which renders the $\eta$ sub-problem local, and the
nonlinearity in $g(\eta)$ is no longer a concern. The existence and uniqueness
of solutions to the sub-problems follows from standard convex analysis.

We solve for the saddle point of $\Phi$ using the following primal-dual update scheme
(Algorithm 2 in \citep{Chambolle:2011}):
\begin{subequations}
    \begin{align}
        \eta_{n+1}&=\operatornamewithlimits{arg\; min}_{\eta \in L^2(\Omega)}
        \left[
        \Phi(\eta, \psi_{n})+
        \frac{1}{2\tau_{n}} \|\eta - \eta_{n}\|_{L^2(\Omega)}^2
        \right],
        \label{eqn:PD_update_eta}\\
        \psi_{n+1}&=\operatornamewithlimits{arg\; max}_{\psi \in
            (\dot H^1(\Omega))^*}
        \left[
        \Phi(\tilde\eta_{n+1}, \psi)-
        \frac{1}{2\sigma_{n}} \| \psi-\psi_{n} \|^2_{L^2(\Omega)}
        \right],
        \label{eqn:PD_update_psi}
    \end{align}
    \label{eqn:PD_update}
\end{subequations}
where
\begin{equation*}
\tilde\eta_{n+1} = (1+\mu_n) \eta_{n+1} - \mu_n \eta_{n}
\end{equation*}
with
\begin{equation*}
\mu_{n} = 1/\sqrt{1+2\tau_n/\epsilon}, \quad
\tau_{n+1} = \mu_{n} \tau_n, \quad
\sigma_{n+1}=\sigma_{n}/\mu_n.
\end{equation*}
The scalars $\tau_{n}$ and $\sigma_{n}$
are the step sizes of the $\eta$- and $\psi$-update respectively. 
The stability \citep{Jacobs:2019, Chambolle:2011} of the
update scheme in \eref{eqn:PD_update} is guaranteed if $\tau_{n}\sigma_{n} \leq 1$.
We select $\tau_0=\epsilon, \sigma_0=1/\epsilon$.
The solution to \eref{eqn:PD_update} is obtained by solving the following Euler--Lagrange
equations corresponding to gradient flows of the two
functionals in \eref{eqn:PD_update}
\footnote{
    In order to obtain \eref{eqn:psi_update_form}, we note that the constrained
    gradient in $(\dot H^1(\Omega))^*$ of $\int_\Omega \tilde \eta_{n+1} \psi\, dV$ with
    respect to $\psi$ is $-\triangle \tilde \eta_{n+1}$.
}
\begin{equation}
\left( \frac{1}{\epsilon}+ \frac{1}{\tau_{n}}
\right)\eta^2(x) +
\left(
\psi_{n}(x) -\frac{2}{\epsilon}- (1+ \eta_{n})\frac{1}{\tau_{n}}
\right)\eta(x)
-  \mathcal{J}^{\star} \left( [\![  \theta ]\!] \right)
+ \frac{1}{\epsilon} -\psi_{n} + \frac{1}{\tau_{n}}\eta_{n} 
=0,
\label{eqn:PD_my}
\end{equation}
\begin{equation}
( 1/\epsilon- \Delta /\sigma_{n+1})\psi_{n+1} = - \Delta (\bar\eta_{n+1}+
\psi_{n} / \sigma_{n+1}),
\label{eqn:psi_update_form}
\end{equation}
where the surface measure $\mathcal J(\jump\theta) \, dS$ has been replaced by a
volume measure $\mathcal J^\star \, dV= \mathcal J(\jump\theta) \exp(-
x^2/2\epsilon^4)\, dV$ that depends on the distance $x$ from the grain boundary.
From \eref{eqn:PD_my}, we note that the primal dual algorithm along
with the choice $g(\eta)=-\log(1-\eta)$ not only renders the $\eta$ sub-problem local but
also analytically solvable.

We solve \eref{eqn:PD_my} and \eref{eqn:psi_update_form}
on a uniform grid of size $N=N_x \times N_y$. 
Since \eref{eqn:PD_my} is 
solved analytically at each grid point, its cost remains $\mathcal{O}(N)$.
We solve for $\psi_{n+1}$ in \eref{eqn:psi_update_form} using the fast Fourier
transform (FFT), resulting in an $\mathcal{O}(N\log{N})$ complexity for the primal
dual algorithm.
We use the following stopping criterion for the update scheme in 
\eref{eqn:PD_update},
\begin{equation}
\|\eta_{n+1}-\eta_{n}\|_{\infty} = 
\max_{1\leq j \leq N} |(\eta_{n+1})_j-(\eta_{n})_j| \leq \mathsf{e},
\label{eqn:primal_tol}
\end{equation}
where $\mathsf{e}$ is the tolerance of the iterative scheme.  
Finally, we note that the use of FFT to solve \eref{eqn:psi_update_form}
necessitates periodic boundary conditions on $\eta$. On the other hand, for Neumann boundary
conditions, we use the 
discrete cosine transform given by 
\begin{equation}
\hat{\psi}_{pq}=\lambda_p \lambda_q \sum^{N_x -1}_{i=0} \sum^{N_y-1}_{j=0}
\psi\left(\frac{i}{N_x},\frac{j}{N_y}\right) \cos{\left(\frac{\pi(2i+1)p}{2N_x}  \right)}
\cos{\left(\frac{\pi(2j+1)q}{2N_y}  \right)}, \quad
\begin{aligned}
&0\leq p\leq N_x-1\\
&0\leq q \leq N_y-1,
\end{aligned}
\end{equation}
with 
\begin{equation}
\lambda_p= \left\{
\begin{aligned}
& 1/\sqrt{N_x}, \quad p=0, \\
& \sqrt{2/N_x}, \quad 1\leq p \leq N_x-1,
\end{aligned}
\right.
\text{ and }
\quad 
\lambda_q = \left\{
\begin{aligned}
& 1/\sqrt{N_y}, \quad q=0, \\
& \sqrt{2/N_y}, \quad 1\leq q \leq N_y-1.
\end{aligned}
\right. 
\end{equation}

\section{Note on the derivations of thresholding scheme}
\label{app:derivThreshold}
In this section, we describe the steps to obtain
\eref{eqn:metric_mean_curvature} from \eref{eqn:suggested_metric}.
We begin by separating the domain of integration in \eref{eqn:suggested_metric} as
\begin{equation}
\int^{l_0}_{-\infty} (1-u(l/\epsilon))^2 \; dl
= \int^{0}_{l_0} (1-u(l/\epsilon))^2 \; dl + \int^{+\infty}_{0} (1-u(l/\epsilon))^2 \; dl. 
\label{eqn:app_derivation1}
\end{equation}
Substituting the solution in \eref{eqn:eta_profile} into
\eref{eqn:app_derivation1}, we have
\begin{equation}
\frac{(u(0)-1)^2}{ (\frac{2}{\epsilon}-\kappa)} \exp\left[
    \left( \frac{2}{\epsilon}-\kappa \right)l_0 
\right ]=
\frac{(u(0)-1)^2}{ (\frac{2}{\epsilon}-\kappa)} 
\left( 
    1 - \exp \left [\left(
        \frac{2}{\epsilon}-\kappa \right)l_0  
    \right ]
\right)
+\frac{(u(0)-1)^2}{ (\frac{2}{\epsilon}+\kappa)}.
\end{equation}
Dividing both sides by $(u(0)-1)^2$ and collecting the $l_0$ terms, we obtain
\begin{equation}
    \frac{2\exp\left[\left(\frac{2}{\epsilon}-\kappa \right)l_0\right] }{ (\frac{2}{\epsilon}-\kappa)} =
\frac{1}{ (\frac{2}{\epsilon}-\kappa)}
+ \frac{1}{ (\frac{2}{\epsilon}+\kappa)} 
=\frac{\frac{4}{\epsilon}}{ (\frac{4}{\epsilon^2}-\kappa^2)}.
\end{equation}
Taking a logarithm, we have 
\begin{equation}
\left(\frac{2}{\epsilon}-\kappa \right)l_0 =
\log{\left(\frac{2/\epsilon}{2/\epsilon + \kappa}\right)} = 
\log{\left(\frac{1}{1 + (\epsilon\kappa)/2}\right)}.
\label{appeqn:split}
\end{equation}
A Taylor expansion of the right-hand-side of \eref{appeqn:split} with respect to
$\epsilon\kappa/2$ results in 
\begin{equation}
\left(\frac{2}{\epsilon}-\kappa \right)l_0 =
- \frac{\epsilon \kappa}{2} + \frac{\epsilon^2 \kappa^2}{4}
- \frac{\epsilon^3 \kappa^3}{8} + O(\epsilon^4\kappa^4).
\end{equation}
Multiplying by $\epsilon$ on both sides, we have
\begin{equation}
\left(2 - \kappa\epsilon \right)l_0 =
-\frac{\epsilon^2 \kappa}{2} + \frac{\epsilon^3 \kappa^2}{4}
+\frac{\epsilon^4 \kappa^3}{8} + O(\epsilon^4\kappa^4).
\label{eqn:d0_infinity}
\end{equation}
Finally, using the approximation $2-\kappa \epsilon \approx 2$,
we get \eref{eqn:metric_mean_curvature}.

As mentioned in \sref{sec:fmm}, in practice, the infinite bounds of the integral in
\eref{eqn:app_derivation1} are replaced by finite bounds of magnitude $l_{\rm b}$. Under this
change, \eref{eqn:d0_infinity} modifies as 
\begin{equation}
\left(\frac{2}{\epsilon} -\kappa\right)l_0 
=\log{\left(\frac{\frac{2}{\epsilon}}{\frac{2}{\epsilon}+\kappa}
+\boxed{
\frac{1}{2}\exp{\left(-\left( \frac{2}{\epsilon}-\kappa \right) l_b\right) }
-\frac{\left( \frac{2}{\epsilon}-\kappa\right)}{2\left(\frac{2}{\epsilon}+\kappa \right)}
\exp{\left(-\left(\frac{2}{\epsilon}+\kappa \right)l_b \right)} 
} \; 
\right) }.
\label{eqn:thresholidngError}
\end{equation}
It can be easily shown that the boxed terms resulting from a finite value of
$d_b$ decay exponentially as $\epsilon \to 0$, which leaves
\eref{eqn:metric_mean_curvature} unchanged.

\section{Fast marching method}
\label{app:fastmarching}

The fast marching method (FMM), developed by \citet{tsitsiklis1995efficient} is used to evolve a surface in the outward unit normal
direction with a speed $V(\bm x)>0$. The fast marching method
reformulates a time-dependent initial value problem describing the evolution of
a surface into an equivalent boundary value formulation. In this section, we
summarize the FMM algorithm as described in \citet{Sethian}. For illustration,
let $\bm s(t)$ describe a surface evolving with speed
$\mathcal V$ from a given initial surface $\bm s(0) = \Gamma$. Instead of
solving a time-dependent problem for $\bm s(t)$, the
fast marching method solves for a function $\zeta(\bm x)$ which represents the
time it takes for the surface to reach $\bm x$. By the definition of $\zeta$, we
have
\begin{equation}
    \zeta(\bm s(t))=t,
    \label{eqn:surface}
\end{equation}
with $\zeta = 0$ on $\Gamma$. Differentiating \eref{eqn:surface} with respect to $t$, and noting that 
$\nabla \zeta$ is normal to the surface, we arrive at the following boundary
value problem
\begin{equation}
    |\nabla \zeta | V=1, \quad \zeta = 0 \text{ on } \Gamma,
\label{eqn:eikonal_equation}
\end{equation}
commonly referred to as the Eikonal equation.

Next, we describe the algorithm to solve \eref{eqn:eikonal_equation} on a
two-dimensional grid. In order to compute $|\nabla \zeta|$, an operator
$D^{-x}_{ij}$, representing the standard backward finite difference operation on the grid point
$ij$, is defined as
\begin{equation}
D^{-x}_{ij}\zeta=\frac{\zeta_{ij} - \zeta_{(i-1)j}}{\Delta x}.
\end{equation}
Similarly, $D^{+x}$, $D^{-y}$, and $D^{+y}$ denote
forward in $x$, backward and forward in $y$ finite difference operators
respectively.
To guarantee a unique viscosity solution\footnote{See \citet{Sethian} on the reason
behind seeking a viscosity solution.} of the evolving surface,
one should consider an upwind finite difference scheme to compute the gradient, 
which is conveniently written as
\begin{equation}
\begin{aligned}
|\nabla \zeta | & \approx 
\left[
(\mathrm{max} (D^{-x}_{ij}\zeta,0)^2 + \mathrm{min} (D^{+x}\zeta_{ij},0)^2
+ (\mathrm{max} (D^{-y}_{ij}\zeta,0)^2 + \mathrm{min} (D^{+y}\zeta_{ij},0)^2
\right]^{1/2}\\
& = 
\left[(\mathrm{max} (D^{-x}_{ij}\zeta,0)^2 + \mathrm{max} (-D^{+x}\zeta_{ij},0)^2
+ (\mathrm{max} (D^{-y}_{ij}\zeta,0)^2 + \mathrm{max} (-D^{+y}\zeta_{ij},0)^2
\right]^{1/2}.
\end{aligned}
\label{eqn:intermediate_upwind}
\end{equation}
Using \eref{eqn:intermediate_upwind}, we rewrite
\eref{eqn:eikonal_equation} in an algebraic form 
\begin{equation}
\left[(\mathrm{max} (D^{-x}_{ij}\zeta, D^{+x}\zeta_{ij}, 0)^2 
+ (\mathrm{max} (D^{-y}_{ij}\zeta,-D^{+y}\zeta_{ij}, 0)^2 
\right]^{1/2}=\frac{1}{V(x,y)}.
\label{eqn:eikonal_update}
\end{equation}
Note that if the neighboring values of $\zeta_{ij}$ are known, 
then \eref{eqn:eikonal_update} is a quadratic equation for $\zeta_{ij}$
that can be solved analytically.
 
The fast marching method begins with the following initialization step 
\begin{enumerate}
\item Assign $\zeta(x)=0$ for grid points in the area enclosed by the initial
    surface, and tag them as \textit{accepted}.
\item Assign $\zeta(x)=+\infty$ for the remaining grid points, and tag them as \textit{far}.
\item Among the \textit{accepted} points, identify the points that are 
in the neighborhood of points tagged as \textit{far}, and tag them as \textit{considered}.
\end{enumerate}
The key step in the fast marching method is to update $\zeta$ with a trial value
using \eref{eqn:eikonal_update} for grid points tagged as
\textit{considered} , but only accept the update with the smallest value. In
order to identify the smallest value efficiently, the grid
points tagged as \textit{considered} are stored in a min-heap\footnote{
    A min-heap structure is a complete binary tree with a property that the value
    at any given node is less than or equal to the values at its children.
}
structure \citep{Heap} borrowed from discrete network algorithms. The fast
marching method then proceeds as follows.
\begin{enumerate}
    \item Construct a min-heap structure for the \textit{considered} points.
    \item Access the root (minimum value) of the heap.
    \item Find a trial solution $\tilde{\zeta}$ on the neighbors of the root using \eref{eqn:eikonal_update}.
            If the trial solution is smaller than the present values, then update $\zeta(x)=\tilde{\zeta}$.
    \item If a point, previously tagged as \textit{far}, is updated using a trial value, 
         relabel it as \textit{considered}, and add it to the heap structure.
    \item Tag the root of the heap as \textit{accepted}, and delete it from the heap.
    \item Repeat steps 2 to 5, until every grid point is tagged as \textit{accepted}.
\end{enumerate}

\begin{figure}[t]
\begin{center}
\includegraphics[width=0.5\textwidth]{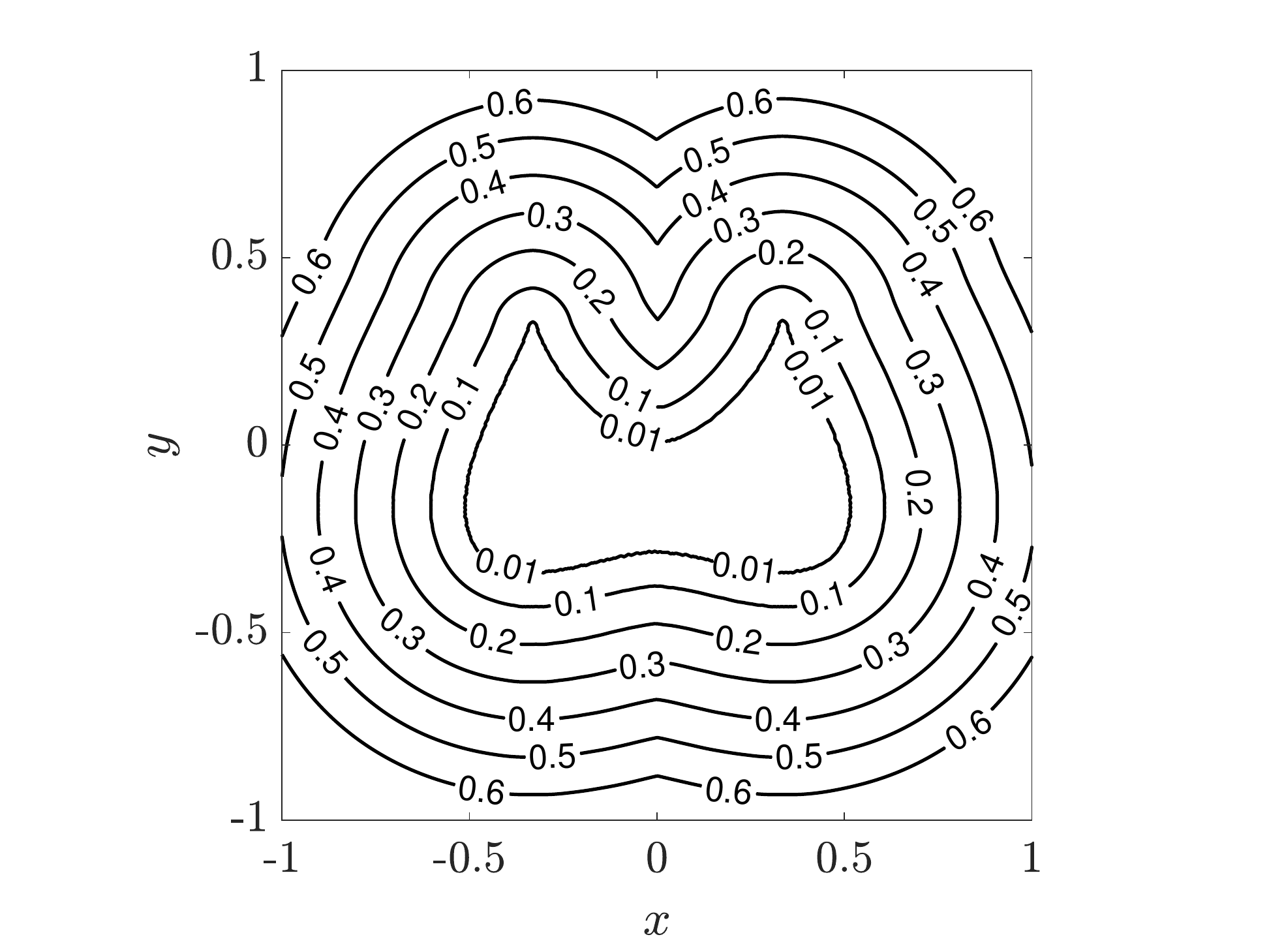}
\end{center}
\caption{The level sets of the solution to the Eikonal equation
    \eref{eqn:eikonal_equation}, computed using the fast marching method,
    describe a surface evolving with outward normal velocity $V(x,y)=1$.
}
\label{fig:fast_marching_example}
\end{figure}
\fref{fig:fast_marching_example} demonstrates the fast marching method used to track an initial surface 
\begin{equation}
(9x^2 - 1)^2 - (3y+1)(1-3y)^3=0,
\end{equation}
growing with a uniform outward normal velocity $V(x)=1$.

\section*{References}
\bibliographystyle{model1-num-names}
\bibliography{references}

\end{document}